\documentclass[aps,prl,twocolumn,floatfix]{revtex4}
\usepackage{bm}
\usepackage[german,english]{babel}
\usepackage[dvips]{epsfig}
\usepackage{amsmath}

\begin{document} 
{\small To appear in International Reviews in Physical
Chemistry (October 2007) \hfill}
\title{Dynamical Tunneling in Molecules: Quantum Routes to Energy Flow}

\author{Srihari Keshavamurthy}
\affiliation{Department of Chemistry, Indian Institute
of Technology, Kanpur, India 208 016}

\begin{abstract}
Dynamical tunneling, introduced in the molecular context,
is more than two decades old and refers to phenomena
that are classically forbidden but allowed by quantum mechanics. 
The barriers for dynamical tunneling, however, 
can arise in the momentum or more generally in the full
phase space of the system.  
On the other hand the phenomenon of 
intramolecular vibrational energy redistribution (IVR)
has occupied a central place in the field of chemical physics for
a much longer period of time.
Despite significant progress in understanding IVR {\it a priori}
prediction of the pathways and rates is still a difficult task.
Although the two phenomena seem to be unrelated
several studies indicate that
dynamical tunneling, in terms of its mechanism and timescales,
can have important implications for IVR.
It is natural to associate dynamical tunneling with a purely quantum
mechanism of IVR.
Examples include the observation of local mode doublets, clustering of
rotational energy levels, and extremely narrow vibrational features
in high resolution molecular spectra.
Many researchers have demonstrated the usefulness of a phase space
perspective towards understanding the mechanism of IVR.
Interestingly dynamical tunneling is also
strongly influenced by the nature of the underlying
classical phase space. Recent studies show that 
chaos and nonlinear resonances in the
phase space can enhance or suppress dynamical tunneling by many orders of
magnitude. Is it then possible that both the 
classical and quantum mechanisms of
IVR, and the potential competition between them, 
can be understood within the phase space perspective? 
This review focuses on addressing the question
by providing the current state of understanding of dynamical tunneling
from the phase space perspective and the consequences for intramolecular
vibrational energy flow in polyatomic molecules.
\end{abstract}

\maketitle

\tableofcontents

\section{Introduction}
\label{intro}

The field of chemical physics is full of phenomena that occur quantum
mechanically, with observable consequences, 
despite being forbidden by classical mechanics\cite{bmw94}. 
All such
processes are labeled as tunneling with the standard elementary example
being that of a particle surmounting a potential barrier despite insufficient
energy. However the notion of tunneling is far more general in the sense
that the barriers can arise in the phase space\cite{lc79,dh81,hp84,Ozo84}. 
In other words the barriers 
are dynamical and arise due to the existence of one or several
conserved quantities. Barriers due to exactly conserved quantities
{\it i.e.,} constants of the motion are usually easy to identify. 
A special case is that of a particle in a one dimensional double well
potential wherein the barrier is purely due to the conserved energy.
On the other hand it is possible, and frequently observed,
that the dynamics of the system can result in one or more approximate
constants of the motion which can manifest as barriers. The term approximate
refers to the fact that the relevant quantities,
although strictly nonconserved, are constant over
timescales that are long compared to certain 
system timescales of interest\cite{kuzstubook}.
Such approximate dynamical barriers are not easy to identify and 
in combination with the other exact constants of the motion can give
rise to fairly complex dynamics. As usual one anticipates that the dynamical
barriers will act as bottlenecks for the classical dynamics whereas 
quantum dynamics will `break free' due to tunneling through the 
dynamical barriers {\it i.e.,} {\em dynamical tunneling}. 
However, as emphasized in this review,
the situation is not necessarily that straightforward since
the mere existence of a finite dynamical barrier does not guarantee that
dynamical tunneling will occur. 
This is especially true in multidimensions since a variety of other
dynamical effects can localize the quantum dynamics. 
In this sense the mechanism of dynamical tunneling is far more subtle
as compared to the mechanism of tunneling through nondynamical barriers.

\begin{figure} [htbp]
\begin{center}
\includegraphics[width=80mm]{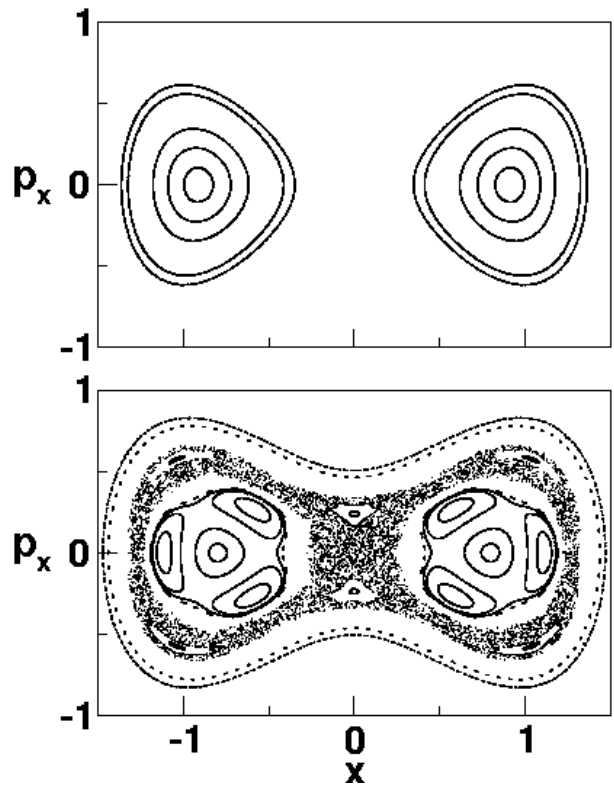}
\caption{Poincar\'{e} surface of section for the example two-dimensional
potential in Eq.~(\ref{dwellham}). The parameter values are chosen to
be $V_{0}=0.25$, $a=\omega=\gamma=1$ and the static barier height is therefore
$V_{b} = 0.25$. The surface of section are defined by $y=0$ and $p_{y} > 0$.
The top panel shows the phase space for $E=0.2$ (below the static barrier
height) and one observes two disconnected regular regions separated by
energetic barrier. The bottom panel corresponds to $E=0.4$ (above the
static barrier height) and shows the left and right regular regions
separated now only by dynamical barriers. Model system and parameters
taken from ref.~\onlinecite{Schl06}.}
\label{fig1}
\end{center}
\end{figure}

The distinction between energetic and dynamical
barriers can be illustrated by considering the two-dimensional
Hamiltonian
\begin{eqnarray}
H(x,y,p_{x},p_{y})&=&\frac{1}{2}(p_{x}^{2}+p_{y}^{2}) + V(x,y) 
                          \label{dwellham}\\
V(x,y) &=& V_{0}(x^{2}-a^{2})^{2}+\frac{1}{2}\omega^{2} y^{2} +
\gamma x^{2} y^{2} \nonumber
\end{eqnarray} 
which represents a one-dimensional double well (in the $x$ degree of freedom)
coupled to a harmonic oscillator (in the $y$ degree of freedom). For energies
$E$ below the barrier height $V_{b}=V_{0}a^{4}$, the $y=0$ Poincar\'{e}
surface of section in Fig.~\ref{fig1} shows two disconnected regions
in the phase space. This reflects the fact that the corresponding
isoenergetic surfaces are also disconnected.
Thus one speaks of an {\em energetic barrier} separating
the left and the right wells. On the other hand Fig.~\ref{fig1} shows
that for $E > V_{b}$ the Poincar\'{e} surface of section again
exhibits two regular regions related to the motion in
the left and right wells despite the absence of the energetic barrier. 
In other words the two regions are now part of the same singly connected
energy surface but the left and the right regular regions are separated
by {\em dynamical barriers}. The various nonlinear resonances and the
chaotic region seen in Fig.~\ref{fig1} should be contrasted with the phase
space of the one-dimensional double well alone which is integrable.
Later in this review examples will be shown wherein there are only
dynamical barriers and no static potential barriers.
Note that in higher dimensions, see discussions towards the end of this
secion, one does not even have the luxury of
visualizing the global phase space, as in Fig.~\ref{fig1}, let alone
identifying the dynamical barriers!

In the case of multidimensional tunneling through potential
barriers it is now well established that valuable insights can
be gained from the phase space perspective\cite{bm72,Mil74,Wil86,Hel99,sc06}
It is not necessary that tunneling is associated with transfer of an atom
or group of atoms from one site to another site\cite{Hel01}. 
One can have, for instance, 
vibrational excitations transferring from a particular mode of a molecule
to a completely different mode\cite{Hel99,Hel95}. 
Examples like Hydrogen atom transfer
and electron transfer belong to the former class whereas 
the phenomenon of 
intramolecular vibrational energy redistribution (IVR) is associated with
the latter class\cite{Hel95}. 
In recent years dynamical tunneling has been realized in
a number of physical systems. 
Examples include driven atoms\cite{zdb98,bdz02},
microwave\cite{dghhrr00} or optical cavities\cite{ns97}, 
Bose-Einstein condensates\cite{hetal01,sor01},
and in quantum dots\cite{bafmli03}. 
Thinking of dynamical tunneling as a close cousin\cite{Hel99,Hel01} 
of the ``above barrier
reflection" (cf. example shown in Fig.~\ref{fig1})
a recent paper\cite{gutk05} by Giese {\it et al.} suggests the
importance of dynamical tunneling in order to understand the
eigenstates of the dichlorotropolone molecule.
This review is concerned with the molecular
manifestations of dynamical tunneling and specifically 
on the relevance of dynamical tunneling 
to IVR and the corresponding signatures in molecular spectra.
One of the aims is to reveal the detailed
phase space description of dynamical tunneling. This, in turn,
leads to the identification of the key structures in phase space and a
universal mechanism for dynamical tunneling
in all of the systems mentioned above.

In IVR studies, which is of paramount interest to chemical
dynamics, one is interested in the
fate of an initial nonstationary excitation in terms
of timescales, pathways and 
destinations\cite{lsp94,gb98,Gru00,mq01,gw04}.
Will the initial energy ``hot-spot'' redistribute throughout
the molecule statistically? Alternatively, to borrow a term from
a recent review\cite{Gru03}, 
is the observed statisticality only ``skin-deep''?
The former viewpoint is at the heart of one of the most
useful theories for reaction rates - the RRKM (Rice-Rampsperger-Kassel-
Marcus) theory\cite{baerhasebook}. 
On the other hand, recent studies\cite{gsh03} seem to be
leaning more towards the latter viewpoint.
As an aside it is worth mentioning that IVR in
molecules is essentially the FPU\cite{fpu55,crz05} (Fermi-Pasta-Ulam) 
problem; only
now one needs to worry about a multidimensional network of
coupled nonlinear oscillators.
In a broad sense, the hope is that a mechanistic understanding of IVR
will yield important insights into mode-specific chemistry and
the coherent control of reactions.
Consequently, substantial experimental\cite{Gru00,nf96,kp00} and 
theoretical efforts\cite{Rice81,Uzer91,Ezra98}
have been directed towards understanding IVR in both time and
frequency domains.
Most of the studies, spanning many decades, have focused on
the gas phase. More recently, researchers have studied
IVR in the condensed phase and it appears that the gas phase
studies provide a useful starting point.
A detailed introduction to the literature on IVR is beyond the
scope of the current article. A brief description is provided in the
next section and the reader is refered to the literature below
for a comprehensive account of the recent advances. 
The review\cite{nf96} by Nesbitt and Field gives
an excellent introduction to the literature. The review\cite{Gru04}
by Gruebele highlights the recent advances and
the possibility of controlling IVR.
The topic of molecular energy flow in solutions has been recently
reviewed\cite{aka03} by A\selectlanguage{german}"s\selectlanguage{english}mann,
Kling and Abel. 

Tunneling leads to quantum mixing between states
localized in classically disconnected regions of the
phase space. 
In this general setting
barriers arise due to exactly or even approximately conserved quantities.
Thus, for example, it is possible for two or more localized quantum
states to mix with each other despite the absence of any energetic
barriers separating them. 
This has significant 
consequences for IVR in isolated molecules\cite{Hel95}
since energy can flow from an initially excited
mode to other, qualitatively different, modes;
classically one would predict
very little to no energy flow. Hence it is appropriate to
associate dynamical tunneling with a purely
quantum mechanism for energy flow in molecules.  
In order to have detailed insights into IVR pathways and rates
it is necessary to study both the classical and quantum routes. The
division is artificial since both mechanisms coexist and compete with
each other. However, from the standpoint of control of IVR 
deciding the importance of one route over the other can be useful.
In molecular systems the dynamical barriers to IVR are related to
the existence of the so called {\em polyad numbers}\cite{fe87,Kell90}.
Usually the polyad numbers are quasiconserved quantities and act
as bottlenecks for energy flow between different polyads. Dynamical
tunneling, on the otherhand, could lead to interpolyad energy flow.
Historically the time scales for IVR via dynamical tunneling have
been thought to be of the order of hundreds of picoseconds. However
recent advances
in our understanding of dynamical tunneling suggest
that the timescales could be much smaller in the mixed phase space
regimes. 
This is mainly due to the
observations\cite{Tom01} that chaos in the underlying phase space 
can enhance the tunneling by several orders of magnitude. 
Conversely, very early on it was thought that the extremely long timescales
would allow for mode-specific chemistry. 
Presumably the chaotic enhancement would spoil the mode specificty and hence
render the initially prepared state unstable.
Thus it is important
to understand the effect of chaos on dynamical tunneling in order to
suppress the enhancement.

Obtaining detailed insights into the phenomenon of dynamical
tunneling and the resulting consequences for IVR requires
us to address several questions.
How does one identify the barriers? Can one explicitly
construct the dynamical barriers for a given system? What is the role of
the various phase space structures like resonances, partial barriers due
to broken separatrices and cantori, and chaos? Which, if any, of the
phase space structures provide for a universal description and mechanism
of dynamical tunneling? Finally the most important question: Do experiments
encode the sensitivity of dynamical tunneling to the nature of the
phase space? 
This review is concerned with addressing most of the questions
posed above in a molecular context. Due to the nature of the issues involved
considerable work has been done by the nonlinear physics community 
and in this work some of the recent
developements will be highlighted since they can have significant impact
on the molecular systems as well. 
Another reason for bringing together developements in different fields
has to do with the observation that the literature on dynamical tunneling
and its consequences are, curiuosly enough, disjoint with hardly any overlap
between the molecular and the `non' molecular areas.

At this stage it is important and appropriate, given the 
viewpoint adopted in this review, to highlight certain crucial issues
pertaining to the nature of the classical phase space for systems with
several ($N \geq 3$) degrees of freedom. A significant portion 
of the present review
is concerned with the phase space viewpoint of dynamical tunneling in
systems with $N \leq 2$. The last section of the review
discusses recent work on a model with three degrees of freedom. 
The sparsity of work in $N \geq 3$ is not entirely surprising and
parallels the situation that prevails in the classical-quantum correspondence
studies of IVR in polyatomic molecules. Indeed it will become
clear from this review that classical phase space structures are, to
a large extent, responsible for both classical and quantum mechanisms
of IVR. 
Thus, from a
fundamental, and certainly from the molecular, standpoint it is highly
desirable to understand the mechanism of classical phase space
transport in systems with $N \geq 3$. 
In the early eighties the seminal work\cite{mmp84}
by Mackay, Meiss, and Percival on transport in Hamiltonian systems with
$N=2$ motivated researchers in the chemical physics community
to investigate the role of various phase space structures in IVR dynamics.
In particular these studies\cite{Dav85,dg86,gr87,ml89,sd88} 
helped in providing a deeper understanding of
the dynamical origin of nonstatistical behaviour in molecules. However 
molecular systems typically have atleast $N=3$ and soon the need for a
generalization to higher degrees of freedom was felt; this task 
was quite difficult since the concept of transport across
broken separatrices and cantori do not have a straightforward generalization
in higher degrees of freedom. In addition tools like the Poincar\'{e}
surface of section cease to be useful for visualizing the global phase
space structures. 
Wiggins\cite{Wigg90} provided one of the early 
generalizations based on the concept
of normally hyperbolic invariant manifolds (NHIM) which led to
illuminating studies\cite{ge91,almm90}
in order to elucidate the dynamical nature of the intramolecular
bottlenecks for $N \geq 3$. Although useful insights were gained from
several studies\cite{ge91,almm90,grd86,mde87,tr88,lmt91,Leo91} the 
intramolecular bottlenecks could not be characterized
at the same levels of detail as in the two degree of freedom cases.
It is not feasible to review the various studies
and their relation/consequences to the topic of this article. 
The reader is referred to the paper\cite{ge91} by Gillilan
and Ezra for an introduction and to the monograph\cite{wigbook} by Wiggins
for an exposition to the theory of NHIMs.
Following the initial studies, which were perhaps ahead of
their time, far fewer efforts were made
for almost a decade but
there has been a 
renewal of interest in the problem over the last few years with the
NHIMs playing a crucial role. 
Armed with the understanding
that the notion of partial barriers,
chaos, and resonances are very different in $N \geq 3$ fresh insights
on transition state and RRKM theories, 
and hence IVR, are beginning to 
emerge\cite{wwju01,ujpyw02,wbw04,sclu04,acp130,bhc05,wbw05,gkmr05,
bhc06,sblkt06,sk06,slkt07}.

To put the issues raised above in perspective for the current review note
that there are barriers in phase space through which dynamical tunneling
occurs and at the same time there are barriers that
can also lead to the localization of the quantum dynamics. 
The competition between dynamical tunneling and dynamical localization
is already important for systems with $N=2$ and this will be briefly
discussed in the later sections. 
Towards the end of the review some work on a
three degrees of freedom model are presented.
However, as mentioned above, the ideas are necessarily preliminary
since to date one does not have a detailed understanding of
either the classical barriers to transport nor the dynamical barriers
which result in dynamical tunneling. Naturally, the competition between
them is an open problem.
We begin with a brief overview of IVR
and the explicit connections to dynamical tunneling.

\subsection{State space model of IVR}
\label{statespace}

Consider a molecule with $N$ atoms
which has $s \equiv 3N-6$ ($3N-5$ for linear molecules) vibrational modes.
Dynamical studies, classical and/or quantum, require the
$s$-dimensional Born-Oppenheimer potential
energy surface $V(Q_{1},Q_{2},\ldots,Q_{s})$
in terms of 
some convenient generalized coordinates ${\bf Q}$ and their choice
is a crucial issue. Obtaining the global $V({\bf Q})$ by solving the
Schr\"{o}dinger equation is a formidable task even for mid-sized
molecules. Traditionally, therefore, a perturbative viewpoint is
adopted which has enjoyed a considerable degree of success.
For example, near the bottom of the well and for low
levels of excitations
the approximation of vibrations by uncoupled harmonic normal
modes is sufficient and the molecular vibrational Hamiltonian can
be expressed perturbatively as\cite{brionfieldbook,papalibook}:
\begin{equation}
H = \sum_{i=1}^{s} \frac{\omega_{i}}{2}(P_{i}^{2}+Q_{i}^{2}) +
\sum_{ijk=1}^{s} \phi_{ijk}Q_{i}Q_{j}Q_{k} + \ldots
\label{eq1}
\end{equation}
In the above $({\bf P},{\bf Q})$ are the dimensionless vibrational
momenta and normal mode coordinates respectively. The deviations from the
harmonic limit are captured by the anharmonic terms with strengths
$\{\phi_{ijk}\}$. Such small anharmonicities account for the relatively
weak overtone and combination transitions observed in experimental
spectra.
However with increasing energy the
anharmonic terms become important and doubts arise regarding the
appropriateness of a perturbative approach.
The low energy normal modes get coupled and the very concept of
a mode becomes ambiguous. There exists sufficient 
evidence\cite{iffjkbs99,jf00,jb80,kt07} for the 
appearance of new `modes', unrelated to the
normal modes, with increasing vibrational excitations.
Nevertheless detailed theoretical studies over the last couple of
decades has shown that it is still possible to understand
the vibrational dynamics via a suitably generalized perturbative
approach - the canonical 
Van-Vleck perturbation theory (CVPT)\cite{Van51,js02,ms00}.
The CVPT leads to an
effective or spectroscopic
Hamiltonian which can be written down as
\begin{equation}
\widehat{H} = \widehat{H}_{0}({\bf n}) +
V({\hat{\bf a}},
{\hat{\bf a}^{\dagger}})
\label{specham}
\end{equation}
where $\hat{\bf n}\equiv \hat{\bf a}^{\dagger}\hat{\bf a},
\hat{\bf a}$, and $\hat{\bf a}^{\dagger}$
are the harmonic number, destruction and creation operators.
The zeroth-order part {\it i.e.,} the Dunham expansion\cite{Dun32}
\begin{equation}
\widehat{H}_{0}({\bf n})=\sum_{j}\omega_{j}\hat{N}_{j} +
\sum_{i \geq j} x_{ij} \hat{N}_{i}\hat{N}_{j} + \ldots
\end{equation}
is diagonal in the number representation. In the above
expression $\hat{N}_{j} \equiv \hat{n}_{j}+d_{j}/2$ with
$n_{j}$ being the number of quanta in the j$^{\rm th}$ mode
with degeneracy $d_{j}$.
The off-diagonal
terms {\it i.e.,} anharmonic resonances have the form
\begin{equation}
V({\hat{\bf a}},{\hat{\bf a}^{\dagger}}) =
\sum_{\bf m} \prod_{j} \Phi_{\bf m}
(a_{j}^{\dagger})^{m_{j}^{+}}
(a_{j})^{m_{j}^{-}} \equiv \sum_{\bf m}
V_{\bf m}(\hat{\bf a},\hat{\bf a}^{\dagger})
\end{equation}
with ${\bf m} = \{m_{1}^{\pm},m_{2}^{\pm},\ldots\}$. These terms
represent the couplings between the zeroth-order modes and are responsible
for IVR starting from a general initial state $|\Psi \rangle =
\sum_{j}c_{{\bf n}j} |{\bf n} \rangle$.

\begin{figure} [htbp]
\begin{center}
\includegraphics[width=80mm]{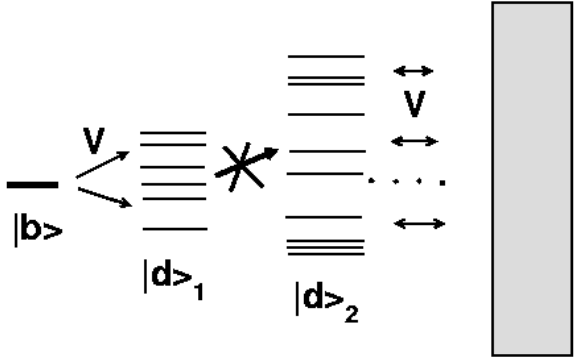}
\caption{Schematic diagram illustrating the tier model of IVR. The
bright state $|b\rangle$ is coupled to the various dark states sorted into
tiers through perturbations $V$. As an example this sketch indicates
fast IVR between $|b\rangle$ and states in $|d\rangle_{1}$ whereas
further flow of energy to states $|d\rangle_{2}$ in the second tier
is restricted due to the lack of couplings. Nevertheless it is possible that
vibrational superexchange can couple the first and second tiers and energy
can eventually flow into the final tier (grey) over long timescales.}
\label{fig2}
\end{center}
\end{figure}

A few key points regarding the effective Hamiltonians can be noted at
this juncture. There are two routes to the effective Hamiltonians.
In the CVPT method
the parameters of the effective Hamiltonian are related to the
original molecular parameters of $V({\bf Q})$.
The other route, given the difficulties associated with
determining a sufficiently accurate $V({\bf Q})$,
is {\em via}
high resolution spectroscopy. The parameters of
Eq.(\ref{specham}) are determined\cite{brionfieldbook}
by fitting the experimental data on line positions
and, to a lesser extent, intensities. The resulting effective
Hamiltonian is only a model and hence
the parameters of the effective Hamiltonian do not have
any obvious relationship to the molecular parameters of $V({\bf Q})$.
Although the two routes to Eq.(\ref{specham}) are very different, they
complement one another and each has its own advantages and drawbacks. The
reader is refered to the reviews by Sibert and Joyeux for a detailed
discussion of the CVPT approach.
The review\cite{iffjkbs99}
by Ishikawa {\it et al.} on the experimental and theoretical studies
of the HCP $\leftrightarrow$ CPH isomerization dynamics provides,
amongst other things, a detailed comparison between the two routes.

Imagine preparing a specific initial state, say an eigenstate
of $H_{0}$, denoted as $|{\bf b}\rangle$.
Theoretically one is not restricted to $|{\bf b}\rangle$ and
very general class of initial states can be considered.
However, time domain experiments with short laser pulses excite specific
vibrational overtone or combination states, called as the zeroth-order
bright states (ZOBS),
which approximately correspond to the eigenstates of $H_{0}$.
The rest of the optically inaccessible states are called
as dark states $\{|{\bf d} \rangle\}$. Here the zeroth-order eigenstates
$|{\bf n} \rangle$ are partitioned as $\{|{\bf b}\rangle,|{\bf d}\rangle_{1},
|{\bf d}\rangle_{2},\ldots\}$.
The perturbations $V({\bf a},{\bf a}^{\dagger})$ couple
the bright state with the dark states leading to energy flow and
impose a hierarchical
coupling structure between the ZOBS and the dark states. Thus one
imagines, as shown in Fig.~\ref{fig2}, the dark states
to be arranged in tiers, determined by
the order $p=\sum_{j}(m^{+}_{j}+m^{-}_{j})$ of the $V_{\bf m}$,
with an increasing density of states across the tiers.
This implies that the the local density of states around $E_{\bf b}^{0}$
is the important factor and
experiments do point to a hierarchical IVR process\cite{lsp94}.
For example, Callegari {\it et al.}
have observed seven different time scales ranging from 100 femtoseconds
to 2 nanoseconds for IVR out of the first CH stretch overtone
of the benzene molecule\cite{csmlsd97}. Gruebele and coworkers
have impressively modeled the IVR in many large molecules based
on the hierarchical tiers concept\cite{gb98}.
In a way the coupling $V({\bf a},{\bf a}^{\dagger})$ imposes a tier
structure on the IVR; similar tiered IVR flow would have been observed
if one had investigated the direct dynamics ensuing from the global
$V({\bf Q})$. Thus the CVPT and its classical analog 
help in unraveling the tier structure.
Nevertheless dominant classical and quantum energy flow routes
still have to be identified for detailed mechanistic insights.

In the time
domain the survival probability
\begin{equation}
P_{\bf b}(t) = |\langle {\bf b}(0)|{\bf b}(t) \rangle|^{2}
= \sum_{\alpha,\beta} p_{{\bf b}\alpha} p_{{\bf b}\beta}
e^{-i (E_{\alpha}-E_{\beta})t/\hbar}
\end{equation}
gives important information on the IVR process. The eigenstates of
the full Hamiltonian have been denoted by $|\alpha \rangle,
|\beta \rangle, \ldots$ with
$p_{{\bf b},\alpha} = |\langle \alpha|{\bf b} \rangle|^{2}$ being the
spectral intensities.
The long time average of $P_{\bf b}(t)$
\begin{equation}
\sigma_{\bf b} = \lim_{T \rightarrow \infty} \frac{1}{T} \int_{0}^{T}
P_{\bf b}(t) dt = \sum_{\alpha} p_{{\bf b},\alpha}^{2}
\end{equation}
is known as the inverse participation ratio (also called
as the dilution factor in the spectroscopic community).
Essentially $\sigma_{\bf b}^{-1}$ indicates the number of states
that participate in the IVR dynamics out of $|{\bf b} \rangle$.

\begin{figure} [htbp]
\begin{center}
\includegraphics[width=85mm]{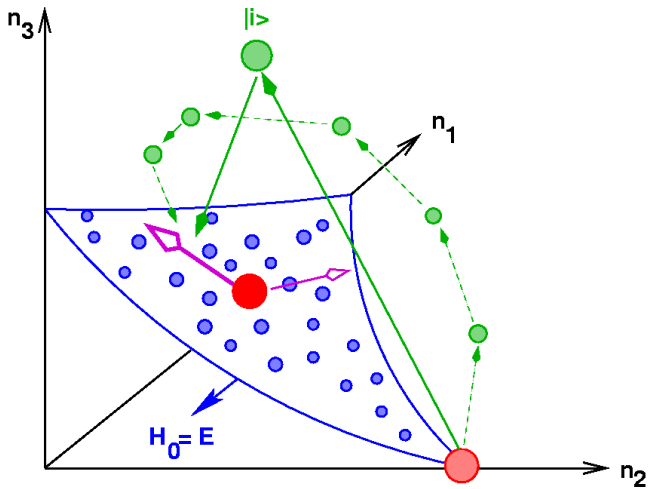}
\caption{(Colour online)
Schematic diagram illustrating the state space model of IVR.
Zeroth order states on the energy shell (blue circles) and
off the energy shell (green circles) are indicated. An interior state
(large red circle) can couple to nearby states via several anharmonic
resonances indicated by purple arrows. 
Edge states (shaded red circle) can couple
via off-resonant virtual states $|i_{\alpha}\rangle$ to states
on the energy shell. Examples of such one-step and multistep superexchange
paths in the state space are shown.}
\label{fig3}
\end{center}
\end{figure}

Although the tier picture has proved to be useful in analysing IVR,
recent studies\cite{Gru00,sw92,sw95,sww95,sww96} 
emphasize the far greater utility in visualising
the IVR dynamics of $|{\bf b}(t) \rangle$ as a
diffusion in the zeroth-order quantum number space or {\em state space}
denoted by $(n_{1},n_{2},\ldots,n_{s})$.
The state space picture, shown schematically in Fig.~\ref{fig3},
highlights the local nature and the directionality
of the energy flow due to the various anharmonic resonances.
In the ``one-dimensional" tier scheme shown in Fig.~\ref{fig2}
the anisotropic nature of the IVR dynamics is hard to discern.
Tiers in the state space are formed by zeroth-order states within
a certain distance from the bright state and hence still organized by
the order of the coupling resonances. In the long time limit most states
that participate in the IVR dynamics are confined, due to the 
quasi-microcanonical nature of the laser excitation, 
to an energy shell
$E^{0}_{b} \pm \delta E$ with $E^{0}_{b}$ being the bright state energy.
The IVR dynamics occurs on a $D$-dimensional subspace, refered to as
the IVR ``manifold'',
of the state space due to the local nature of the couplings
between $|{\bf b}\rangle$ and ${|{\bf d}_{j}\rangle}$ with
$H \approx E^{0}_{b} \approx E^{0}_{d_{j}}$.
One of the key prediction of the state space model is that the
survival probability exhibits a 
power law at intermediate times\cite{gw04,wg99}
\begin{equation}
P_{\bf b}(t) \sim \sigma_{\bf b} + (1-\sigma_{\bf b})\left[1+
\frac{2t}{\tau D} \right]^{-D/2}
\label{powerlaw}
\end{equation}
with $D \ll 3N-6$, the number of vibrational modes of the molecule. 
Thus the effective dimensionality $D$ of the IVR manifold in the state space
can be fairly small and a large body of work in the recent years
support the state space viewpoint\cite{Gru00,gw04,aka03}. 
The effective dimensionality itself is
crucially dependent on the extent to which classical and quantum 
mechanisms of IVR manifest themselves at specific energies.
One possible interpretation of $D$ is as follows.
If $D \approx 3N-6$ then the dynamics in the state space
can be thought of as a normal diffusive process and thus ergodic over
the state space. 
In such a limit
for large $N$ (molecules) the survival can be well
approximated by an exponential behaviour. 
For $D \ll 3N-6$ the IVR dynamics is anisotropic and the dynamics
is nonergodic with $D$ typically being nonintegral. 
Note that the terms ``manifold'' and ``effective dimensionality''
are being used a bit loosely since one does not have a very
clear idea of the topology of the IVR manifold as of yet.

Leitner and Wolynes, building upon the 
earlier work of Logan and Wolynes\cite{low90},
have provided criteria\cite{lw96jcp} 
for vibrational state mixing and energy flow
from the state space perspective. Using a local random matrix approach
to the Hamiltonian in Eq.~(\ref{specham})
the rate of energy flow out of $|{\bf b}\rangle$ is given by
\begin{equation}
k(E) = \frac{2\pi}{\hbar} \sum_{Q} K_{Q} \langle|\psi_{Q}|^{2}\rangle
D_{Q}(E)
\label{lwrate}
\end{equation}
with $Q$ being a distance in the state space. 
The term $\langle|\psi_{Q}|\rangle$
represents the average effective coupling of $|{\bf b}\rangle$ 
to the states, 
$K_{Q}$ of them with density $D_{Q}(E)$, a distance $Q$ away in state space.
The extent of state mixing is characterized by the transition parameter
\begin{equation}
T(E) = \frac{2\pi}{3} \left(\sum_{Q} K_{Q} \langle|\psi_{Q}|\rangle
D_{Q}(E) \right)^{2}
\label{lwqet}
\end{equation}
and the transition between localized and extended states is located at
$T(E)=1$. The key term is the effective coupling $\psi_{Q}$ which
involves both low and high order resonances. Applications to several
systems shows that the predictions based on Eq.~(\ref{lwrate})
and Eq.~(\ref{lwqet}) are both qualitatively 
and quantitatively\cite{lw97} accurate.

\subsection{Connections to dynamical tunneling}
\label{connections}

The perspective
of IVR being a random walk on an effective $D$-dimensional manifold
in the quantum number space is reasonable\cite{Gru00,sw92}
as long as direct anharmonic resonances exist which connect the bright
state with the dark states. It is useful to emphasize again that
the existence of direct anharmonic resonances in itself does not
imply ergodicity over state space {\it i.e.,} the random walk
need not be normal.
What happens if, for certain bright states,
there are no direct resonances available?
In other words, the coupling matrix elements
$\langle {\bf d}_{j}|V|{\bf b}\rangle \approx 0$ for all $j$.
What would be the mechanism of IVR, if any, in such cases?
Examples of such states in fact correspond
to overtone excitations which are typically prepared in experiments.
The overtone states are also called as edge states from the state space
viewpoint since most of the excitation is localized in one single mode
of the molecule. 
Thus, for example, in a system with four degrees of freedom a state 
$|{\bf b}\rangle \equiv |n_{1},0,0,0\rangle$ would be called as an edge
state whereas the state
$|n_{1},0,n_{3},n_{4}\rangle$, corresponding to a combination state, is
called as an interior state. The edge states, owing to their location in
the state space, have fewer anharmonic resonances available for IVR as
compared to the interior states. To some extent such a 
line of reasoning leads 
one to anticipate overtone excitations to have slower IVR when compared
to the combination states.
A concrete example comes from the theoretical work of
Holme and Hutchinson wherein the dynamics of overtone
excitations in a model system representing coupled CH-stretch and CCH-bend
interacting with an intense laser field was studied\cite{hh86,Hut89}. 
They found that classical dynamics
predicted no significant energy flow from the high overtone excitations.
However the corresponding quantum calculations did indicate significant
IVR. Based on their studies Holme and Hutchinson concluded that overtone
absorption proceeds via dynamical tunneling on timescales 
of the order of a few nanoseconds.

Experimental evidence for the dynamical tunneling route to IVR was provided
in a series of elegant works\cite{klmps91,gtls93} by the Princeton group.
The frequency domain experiments involved
the measurement of IVR lifetimes of the acetylinic CH-stretching
states in (CX$_{3}$)$_{3}$YC{\bf C$\equiv$H} 
molecules with X$=$H,D and Y$=$C,Si. The
homogeneous line widths, which are related to the rate of IVR out
of the initial nonstationary state, arise due to
the vibrational couplings between the CH-stretch and the various other
vibrational modes of the molecule. Surprisingly Kerstel {\it et al.}
found\cite{klmps91} extremely 
narrow linewidths of the order of 10$^{-1}$ - 10$^{-2}$
cm$^{-1}$ which translate to timescales of the order of thousands of
vibrational periods of the CH-stretching vibration.
Thus the initially prepared CH-stretch excitation remains localized
for extremely long times.
Such long IVR timescales were ascribed to the lack of strong/direct
anharmonic resonances coupling the CH-stretch with the rest of the
molecular vibrations. The lack of resonances, despite
a substantial density of states,
combined with IVR timescales of several nanoseconds implies
that the mechanism of energy flow is inherently quantum.
Another study\cite{glls94} by Gambogi {\it et al.} 
examines the possibility of long
range resonant vibrational energy exchange between equivalent CH-stretches
in CH$_{3}$Si(C$\equiv$CH)$_{3}$.
There have been several 
experiments\cite{fcsk88,mn89,mmp92,gp92,uctc95} that indicate
multiquantum zeroth-order state mixings due to extremely weak couplings of the
order of 10$^{-1}$ - 10$^{-2}$ cm$^{-1}$ in highly vibrationally
excited molecules.

Since the early work by Kerstel {\it et al.} other
experimental studies have revealed the existence of the dynamical
tunneling mechanism for IVR in large organic molecules. 
For instance in a recent work\cite{cpcesgls03} 
Callegari {\it et al.} performed
experimental and theoretical studies on the
IVR dynamics in pyrrole (C$_{4}$H$_{4}$NH) and 
1,2,3-triazine (C$_{3}$H$_{3}$N$_{3}$). Specifically, they chose the
initial bright states to be the edge state $2\nu_{14}$ CH-stretch
for pyrrole and
the interior state $\nu_{6}^{1}+2\nu_{7}^{2}$ corresponding
to the CH stretching-ring breathing combination for triazine.
In both cases very narrow IVR features, similar to the
observations by Kerstel {\it et al.}, were seen
in the spectrum. This pointed towards an important role
of the off-resonant states
to the observed and calculated narrow IVR features.
The analysis by Callegari {\it et al.} reveals that near the IVR
threshold it is reasonable to expect such highly off-resonant coupling
mechanisms to be operative. Another example for the possible existence of the
off-resonant mechanism comes from the experiment\cite{prb04} 
by Portonov, Rosenwaks, and
Bar wherein the IVR dynamics ensuing from 
$2\nu_{1}$, $3\nu_{1}$, and $4\nu_{1}$ CH-acetylinic stretch
of 1-butyne are studied. It was found that the homogeneous linewidth
of the $4\nu_{1}$ state, $\sim$ 0.5 cm$^{-1}$, is about a factor of two
smaller than the widths of $2\nu_{1}$ and $3\nu_{1}$. 

It is important to note that the involvement of such off-resonant states
can occur at any stage of the IVR process from a tier perspective. 
In other words it is possible that the bright state undergoes fast
initial IVR due to strong anharmonic resonances with states in the
first tier but the subsequent energy flow might be extremely slow. Several
experiments point towards the existence of such unusually long secondary
IVR time scales which might have profound consequences for the interpretation
of the high resolution spectra. 
Boyarkin and Rizzo demonstrated\cite{br96} the
slow secondary time scales in their experiments on the IVR 
from the alkyl CH-stretch overtones of CF$_{3}$H.
Upon excitation of the CH-stretch fast IVR occurs to the CH stretch-bend
combination states on femtosecond timescales. However the vibrational
energy remains localized in the mixed stretch-bend states and flows out to
the rest of the molecule on timescales of the order of hundred picoseconds.
Similar observations were made\cite{lbspr96} 
by Lubich {\it et al.} on the IVR dynamics
of OH-stretch overtone states in CH$_{3}$OH.

The interpretation of the above experimental results as due to dynamical
tunneling is motivated by the theoretical analysis by Stuchebrukhov and Marcus 
in a landmark paper\cite{sm93}.
In the initial work they explained the narrow features,
observed in the experiments\cite{klmps91} by Kerstel {\it et al.},
by invoking the coupling of the bright states with
highly off-resonant gateway states which, in turn, couple back to states
that are nearly isoenergetic with the bright state.
In Fig.~\ref{fig3} an example of the indirect coupling via a state which is
off the energy shell in the state space is illustrated.
More importantly Stuchebrukhov and Marcus argued that the
mechanism, described a little later in this review and anticipated
earlier by Hutchinson, Sibert, and Hynes\cite{hsh84},
involving high order coupling chains is essentially
a form of generalized tunneling\cite{sm932}.
Hence one imagines dynamical barriers which act to prevent classical
flow of energy out of the initial CH-stretch whereas quantum dynamical
tunneling does lead to energy flow, albeit very slowly.
Similar arguments had been put forward by
Hutchinson demonstrating\cite{Hut84} the importance of the off-resonant
coupling leading to the mixing of nearly degenerate 
high energy zeroth-order states in cyanoacetylene HCC$\equiv$CN.
It was argued that such purely quantum relaxation pathway would explain
the observed broadening of the overtone bands in various substituted
acetylenes.

The above examples highlight the possible connections
between IVR and dynamical tunneling. 
However it was realized very early on
that observations of local mode doublets\cite{lc80,lc81,lc82}
in molecular spectra and the 
clustering of rotational energy sublevels with high angular momenta\cite{hp84}
can also be associated with dynamical tunneling. 
Indeed much of our understanding of the phenomenon of dynamical tunneling
in the context of molecular spectra comes from these early works. 
One of the key feature of the analysis was the focus on a phase
space description of dynamical tunneling {\it i.e.,} identifying the
disconnected regions in the phase space and hence classical structures
which could be identified as dynamical barriers. Thus the mechanism of
dynamical tunneling, confirmed by computing the splittings based on
specific phase space structures, could be understood in exquisite detail.
However such detailed phase space analysis are exceedingly difficult, if
not impossible, for large molecules. For example in the case of the
(CX$_{3}$)$_{3}$YCCH system there are $3N-6=42$ vibrational degrees of freedom.
Any attempt to answer the questions on the origin and location of the
dynamical barriers in the phase space is seemingly a futile excercise.
It is therefore not surprising that Stuchebrukhov and Marcus, although aware of
the phase space perspective, provided a purely quantum explanation for
the IVR in terms of the chain of off-resonant states\cite{sm93}. 

Thus one cannot help but ask if the phase space viewpoint
is really needed. The answer is affirmative and many reasons can be provided
that justify the need and utility of a phase space viewpoint.
First, as noted by Heller\cite{Hel95,Hel99}, 
the concept of tunneling is meaningless without
the classical mechanics of the system as a baseline. 
In other words, in order to label a process as purely quantum it is
imperative that one establishes the absence of the process within
classical dynamics. Note that for dynamical tunneling 
it might not be easy to {\it a priori} make such a distinction.
There are mechanisms of classical transport, especially in systems with
three or more degrees of freedom, involving long timescales which might
give an impression of a tunneling process. It is certainly not possible
to differentiate between classically allowed and forbidden
mechanisms by studying the spectral intensity and splitting patterns alone
due to the nontrivial role played by the stochasticity in the classical
phase space.
Secondly,
the quantum explanation is in essence a method to calculate the contribution
of dynamical tunneling to the IVR rates in polyatomic molecules. In order
to have a complete qualitative picture of dynamical tunneling it is
necessary, as emphasised by Stuchebrukhov and Marcus\cite{sm932}, 
to find an explicit form of the effective dynamical potential
in the state space of the molecule. Third reason, mainly due to
the important insights gained from recent studies, has to do with the
sensitivity of dynamical tunneling to the stochasticity in the phase
space\cite{Tom01}. 
Chaos can enhance as well as supress dynamical tunneling and
for large molecules the phase space can be mixed regular-chaotic  
even at energies corresponding to the first or second overtone levels.
Undoubtedly the signatures should be present in the splitting and intensity
patterns in a high resolution spectrum. However the precise mechanism
which governs the substantial enhancement/suppression
of dynamical tunneling,
and perhaps IVR rates is not yet clear. 
Nevertheless recent developements indicate
that a phase space approach is capable of providing
both qualitative and quantitative insights into the problem.

A final argument in favor of a phase space perspective
needs to be mentioned.
It might appear that the dimensionality issue
is not all that restrictive for the quantum studies {\it a la}
Stuchebrukhov and Marcus. 
However it will become clear from the discussions in
the later sections that in the case of large
systems even calculating, for example local mode splittings via
the minimal perturbative expression in Eq.~(\ref{minimal}) can be
prohibitively difficult. One might argue that a clever basis would
reduce the number of perturbative terms that need to be considered or
alternatively one can look for and formulate criteria that would
allow for a reliable estimate of the splitting. Unfortunately,
{\it a priori} knowledge of the clever basis or the optimal set
of terms to be retained in Eq.~(\ref{minimal}) implies a certain level
of insight into the dynamics of the system. The main theme of this
article is to convey the message that such insights truly originate
from the classical phase space viewpoint.
With the above reasons in mind the next section provides brief reviews of
the earlier approaches to dynamical tunneling from the phase space
perspective. This will then set the stage to
discuss the more recent advances and the issues involved in the molecular
context.

\section{Dynamical tunneling: early work from the phase space perspective}
\label{dtew}

In a series of pioneering papers\cite{lc80,lc81,lc82}, 
nearly a quarter of a century ago,
Lawton and Child showed that the experimentally observed local mode
doublets in H$_{2}$O could be associated with a generalized tunneling
in the momentum space. In molecular spectroscopy it was appreciated from
very early on that highly excited spectra associated with X-H stretching
vibrations are better described in terms of local modes rather than the
conventional normal modes\cite{Hal98,Jen00,hk02}. 
In a zeroth-order description this corresponds to
very weakly coupled, if not uncoupled, anharmonic oscillators. Every
anharmonic oscillator, modelled by an appropriate Morse function,
represents a specific local stretching mode of the molecule. 
The central question
that Lawton and Child asked was that to what extent are the molecular
vibrations actually localized in the individual bonds (local modes)?
Analyzing the classical dynamics on the Sorbie-Murrell 
potential energy surface for H$_{2}$O, focusing on the two local O-H
stretches, it was found that a clear distinction between normal mode and local
mode behaviour could be observed in the phase space. 
The classical phase
space at appropriate energies revealed\cite{lc79} two equivalent but classically
disconnected regions as a signature of local mode dynamics. Lawton and Child
immediately noted the topological similarity between their two dimensional
Poincar\'{e} surface of section and the phase space of a one dimensional
symmetric double well potential. However they also noted that the barrier
was in momentum space and hence the lifting of the local mode degeneracy
was due to a generalized tunneling in the momentum space.
Detailed classical\cite{lc79}, quantum\cite{lc80}, 
and semiclassical\cite{lc81} investigations of the system
allowed Lawton and Child to provide an 
approximate formula for the splitting $\Delta E_{nm}$
between two symmetry related local mode states $|nm^{\pm}\rangle$ as:
\begin{equation}
\Delta E_{nm} = \frac{2\hbar \bar{\omega}_{nm}}{\pi} 
       \exp \left(-\frac{1}{\hbar} 
               \int_{{\cal C}} |{\bf P}| \cdot d{\bf Q}\right)
\label{lcsplit}
\end{equation}
The variables $({\bf P},{\bf Q})$ correspond to the mass-weighted momenta
and coordinates\cite{lc79,lc80}. 
It is noteworthy that Lawton and Child realized, correctly,
that the frequency factor $\hbar \bar{\omega}_{nm}$ should be evaluated
at a fixed total quantum number $v=n+m$. However the choice of the
tunneling path ${\cal C}$ proved to be more difficult.
Eventually ${\cal C}$ was taken to be a one dimensional, nondynamical
path in the two dimensional coordinate space of the symmetric and
antisymmetric normal modes. Although Eq.~(\ref{lcsplit}) correctly predicts the
trend of decreasing $\Delta E_{nm}$ for increasing $m$ and fixed $n$,
the origin and properties of the dynamical barrier remained unclear.

\begin{figure} [htbp]
\begin{center}
\includegraphics[width=90mm]{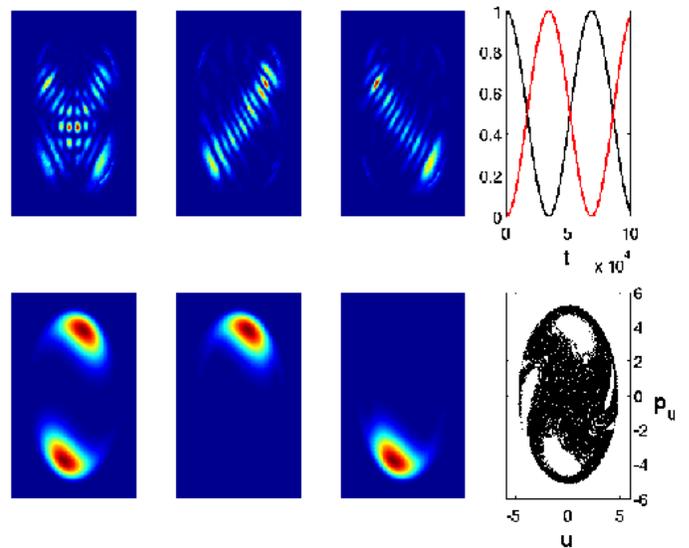}
\caption{(Colour online) An example of the local mode doublet 
in the Davis-Heller system.
In (a) the $(s,u)$ space representation of
the even state ($E = 13.589216$) is shown. In (b) and (c) the linear
combination $\psi_{\pm}=(\phi_{1} \pm \phi_{2})/\sqrt{2}$
are shown. Right below
the configuration space representations the corresponding phase
space Husimi representations in the $(u,p_{u})$ Poincar\'{e}
surface of section are shown. The rightmost figure on the top panel
shows the dynamical tunneling occuring when starting with the initial
nonstationary state $\psi_{+}$. The period of the oscillations
is consistent with the splitting $\Delta E \approx 9.125 \times 10^{-5}$.
The rightmost figure in the bottom panel shows the $(u,p_{u})$ surface
of section.}
\label{fig4}
\end{center}
\end{figure}

\begin{figure} [htbp]
\begin{center}
\includegraphics[width=85mm]{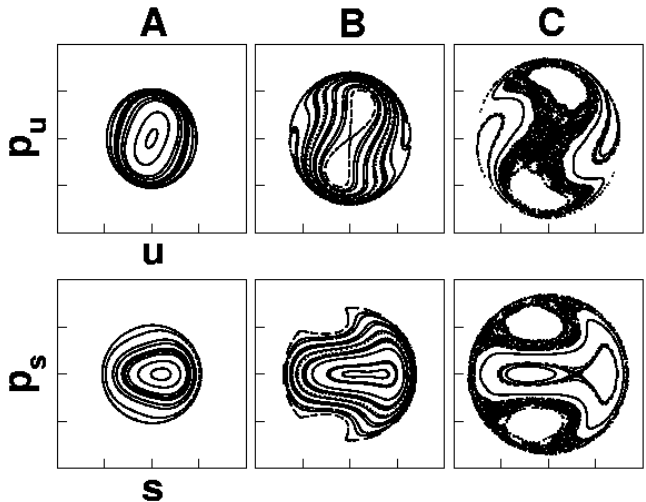}
\caption{Poincar\'{e} surface of sections for the Davis-Heller
Hamiltonian. The top panel shows the $(u,p_{u})$ section and the bottom
panel shows the $(s,p_{s})$ surface of section for
increasing total energy $E=$ 5.0 (A), 9.0 (B), and
14.0 (C).
Note that the symmetric stretch periodic orbit ($u=p_{u}=0$) becomes
unstable around $E \approx 6.5$ whereas the asymmetric stretch periodic orbit
($s=p_{s}=0$) becomes unstable around $E \approx 8.5$.}
\label{fig5}
\end{center}
\end{figure}

The analysis by Lawton and Child, although specific to the vibrational
dynamics of H$_{2}$O, suggested that the phenomenon of dynamical tunneling
could be more general. This was confirmed in an influential paper by
Davis and Heller wherein it was argued that dynamical tunneling could have
significant effects on bound states of polyatomic molecules\cite{dh81}. 
Incidentally,
the word {\em dynamical tunneling} was first introduced by Davis and Heller.
This work is remarkable in many respects and provided a fairly detailed
correspondence\cite{dh84} between dynamical 
tunneling and the structure of the
underlying classical phase space for the Hamiltonian
$H(s,p_{s},u,p_{u}) = (p_{s}^{2}+p_{u}^{2})/2 + V(s,u)$. Specifically, the
analysis was done on the following two-dimensional potential:
\begin{equation}
V(s,u) = \frac{1}{2}(\omega_{s}^{2} s^{2} + \omega_{u}^{2} u^{2})
+ \lambda u^{2} s 
\end{equation}
The system has a discrete symmetry $V(s,-u)=V(s,u)$.
For the specific choice of parameter values $\omega_{s} = 1.0, \omega_{u}=1.1$,
$\lambda=-0.11$, and $\hbar=1$ the
dissociation energy is equal to $15.125$ and the potential supports about
$115$ bound states. At low energies no doublets are found.
However above a certain energy, despite the lack of an energetic barrier, 
near-degenerate eigenstates $\phi_{1},\phi_{2}$ were observed with
small splittings $\Delta E$. 
In analogy with the symmetric double well system, linear combinations
$\psi_{\pm}=(\phi_{1} \pm \phi_{2})/\sqrt{2}$ yielded states localized in the
configuration space $(s,u)$. One such example is shown in Fig.~\ref{fig4}.
In the same figure the probabilities 
$|\langle \psi_{+}(0)|\psi_{+}(t)\rangle|^{2}$ and 
$|\langle \psi_{-}(0)|\psi_{+}(t)\rangle|^{2}$ are also shown confirming the
two state scenario.
The main issue once again had to do with the origin and nature of
the barrier. Davis and Heller showed\cite{dh81,dh84} that the
doublets could be associated with the formation of two classically
disconnected regions in the phase space - very similar to the
observations by Lawton and Child\cite{lc79}. 
The symmetric stretch periodic orbit, stable
at low energies, becomes unstable leading to the topological change in
the phase space; a separatrix is created which separates the normal and
local mode dynamics. In Fig.~\ref{fig5} the phase space are shown for
increasing total energy and the creation of the separatrix can be clearly
seen in the $(u,p_{u})$ surface of section.
The near-degenerate eigenstates correspond to the
local mode regions of the phase space and several such pairs were identified
upto the dissociation energy of the system. The corresponding phase space
representation shown in Fig.~\ref{fig4} (bottom panel) confirms the
localized nature of $\psi_{\pm}$. Thus a key structure in the phase
space, a separatrix, arising due to a $1$:$1$ resonance between the two modes
is responsible for the dynamical tunneling. Although Davis and Heller
gave compelling arguments for the connections between phase space
structures and dynamical tunneling, explicit determination of the
dynamical barriers and the resulting splittings was not attempted.

\begin{figure} [htbp]
\begin{center}
\includegraphics[width=80mm]{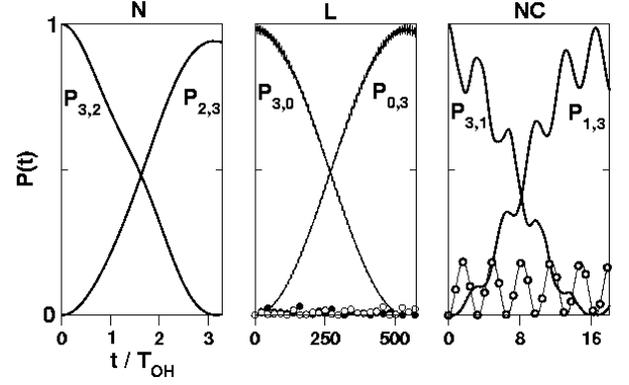}
\caption{Three classes of energy flow dynamics, normal (N),
local (L), and nonclassical (NC) illustrated for the
coupled Morse system by computing the survival probabilities
$P(t)$ for the states 
$|3,2\rangle$, $|3,0\rangle$, and
$|3,1\rangle$ respectively. 
The cases L and NC are examples of dynamical
tunneling whereas case N occurs classically. Note the difference in
the timescales for complete energy transfer between modes for L and NC.
In the case of NC significant probability builds up in the normal
mode state $|2,2\rangle$ (symbols). Very little probability buildup in the
intermediate states $|2,1\rangle$ (open circles) and $|1,2\rangle$ (filled
circles) is seen in case L.}
\label{fig6}
\end{center}
\end{figure}

The observation by Davis and Heller, implicily present in the work by Lawton
and Child, regarding the importance of the
$1$:$1$ resonance to the dynamical 
tunneling was put to test in an elegant
series of papers\cite{srh82c,srh82q} 
by Sibert, Reinhardt, and Hynes. These authors investigated
in great detail the classical\cite{srh82c} and quantum dynamics\cite{srh82q} 
of energy transfer between
bonds in ABA type triatomics and gave explicit expressions for
the splittings. 
In particular they studied the
model H$_{2}$O Hamiltonian:
\begin{equation}
H({\bf x},{\bf p}) = \sum_{j=1,2} \left(\frac{1}{2} G_{jj} p_{j}^{2} +
U(x_{j}) \right) + G_{12} p_{1} p_{2}
\label{srhham}
\end{equation}
with $(x_{1},x_{2})$ and $(p_{1},p_{2})$ denoting the two OH bond 
stretches (local modes) and their corresponding momenta respectively.
The bending motion was ignored and the stretches were modelled by
Morse oscillators
\begin{equation}
U(x) = D\left[1-e^{-a x}\right]^{2}
\end{equation}
with $D$ being the OH bond dissociation energy. Due to the equivalence of the
OH stretches $G_{11} = G_{22}$ and
the coupling strength $G_{12}$
is much smaller than $G_{11}$. The authors studied 
the dynamics of initial states $|n_{1},n_{2}\rangle$ where $|n_{j}\rangle$
are eigenstates of the unperturbed j$^{th}$ Morse oscillator. Due to the
coupling the initial states have nontrivial dynamics and thus mix with
other zeroth-order states. An initial state $|n_{1},n_{2}\rangle$
implies a certain energy distribution in the molecule and the nontrivial
time dependence signals flow of energy in the molecule. 
Later studies by Hutchinson, Sibert, and Hynes showed\cite{hsh84}
that based on the dynamics the various initial states could be classified
as normal modes or local modes. Morever at the border between the two classes
were initial states that led to a `nonclassical' energy flow. 
A representative example, using a different but equivalent system,
for the three classes is shown in Fig.~\ref{fig6}.
The normal mode regime illustrated with $|n_{1},n_{2}\rangle = |3,2\rangle$
shows complete energy flow between the modes on the time scale of a few
vibrational periods - something that occurs classically as well. The
local mode behaviour for the initial state $|3,0\rangle$ exhibits complete
energy transfer resulting in the state $|0,3\rangle$ but the timescale is
now hundreds of vibrational periods. Despite the large timescales involved
it is important to note that such a process does not happen classically. Thus
this is an example of dynamical tunneling. The so called nonclassical
case illustrated with an initial state $|3,1\rangle$ is also
an example of dynamical tunneling. However the associated timescale is 
nowhere as large when compared to the local mode regime\cite{hsh84}.

The vibrational energy transfer process illustrated through the initial
state $|3,0\rangle$ and $|3,1\rangle$ are examples of pure quantum
routes to energy flow.  
Hutchinson, Sibert, and Hynes proposed\cite{hsh84} 
that the mechanism for this quantum
energy flow can be understood as an indirect state-to-state flow of
probability involving normal mode intermediate states. For instance, in
the case involving the initial state $|3,0\rangle$ the following mechanism
was proposed:
\begin{equation}
|3,0\rangle \rightarrow |2,1\rangle \rightarrow |1,2\rangle 
\rightarrow |0,3\rangle
\label{hshpert}
\end{equation}
The reason for this indirect route has to do with the fact that
estimating the splitting directly (at first order)
$\Delta = 2 \langle 3,0|G_{12} p_{1} p_{2}|0,3\rangle$
yields a value which is more than an order of magnitude smaller than
the the actual value\cite{hsh84}. Indeed note that the indirect route
corresponds to a third order perturbation in the coupling and hence
it is possible to estimate the contribution to the splitting as:
\begin{equation}
\frac{\Delta}{2} = \frac{\langle 3,0|V|2,1\rangle \langle 2,1|V|1,2\rangle
\langle 1,2|V|0,3\rangle}{(E^{0}_{3,0}-E^{0}_{2,1})^{2}} 
\label{effcouphrh}
\end{equation}
with $V \equiv G_{12} p_{1} p_{2}$ and $E^{0}_{n_{1},n_{2}}$ 
are the zeroth-order
energies associated with the states $|n_{1},n_{2}\rangle$.

Once again a clear interpretation of the mechanism comes from taking
a closer look at the underlying classical phase space\cite{srh82c,hsh84}. 
Using the classical
action-angle variables $({\bf I},{\bm \theta})$ for Morse oscillators
it is possible to write the original Hamiltonian in Eq.~(\ref{srhham}) as:
\begin{equation}
H({\bf I},{\bm \theta}) = \sum_{i=1,2} \left(\Omega I_{i} -
\frac{\Omega^{2}I_{i}^{2}}{4D}\right) + V({\bf I},{\bm \theta})
\end{equation}
with the harmonic frequency $\Omega = \sqrt{2DG_{11}a^{2}}$. 
The coupling term $V({\bf I},{\bm \theta})$ has infinitely many terms of
the form $V_{pq}({\bf I}) \cos(p\theta_{1}-q\theta_{2})$ which
represent nonlinear resonances $p\omega_{1} \approx q\omega_{2}$ 
involving the nonlinear mode frequencies 
$\omega_{i} = \Omega - \Omega^{2}I_{i}/2D$. 
Arguments were provided\cite{srh82c} for the dominance of the
$p=q=1$ resonance and hence to an
excellent approximation
\begin{eqnarray}
H({\bf I},{\bm \theta}) &\approx& \sum_{i=1,2} \left(\Omega I_{i} -
\frac{\Omega^{2}I_{i}^{2}}{4D}\right) + 
V_{11}({\bf I}) \cos(\theta_{1}-\theta_{2}) \nonumber \\
&\approx& \Omega P - \frac{\Omega^{2} P^{2}}{8D} -
\left(\frac{\Omega^{2} p^{2}}{8D} + V_{0}(P,p)\cos(2\psi)\right) \nonumber \\
&\equiv& H_{M}(P) - H_{R}(p,\psi)
\label{hindrot}
\end{eqnarray}
The form $H_{M}-H_{R}$ originates via a canonical transformation
from $({\bf I},{\bm \theta}) \rightarrow (P,\phi,p,\psi)$ with
$I_{1}+I_{2}=P$,$I_{1}-I_{2}=p$,$\theta_{1}+\theta_{2}=2\phi$, and
$\theta_{1}-\theta_{2}=2\psi$. 
Thus the key term in the above Hamiltonian
is the hindered rotor part denoted by $H_{R}(p,\psi)$ 
and the analysis now focuses on a one dimensional Hamiltonian due to
the fact that $P$ is a conserved quantity. Note that this is consistent with
the observation\cite{lc81} by Lawton and Child regarding the evaluation of the
frequency factor $\hbar \bar{\omega}_{nm}$ in eq.~\ref{lcsplit}. The
classical variables $(P,p)$ are quantized as 
$P=(n_{1}+n_{2}+1)\hbar \equiv (m+1)\hbar$ and
$p=(n_{1}-n_{2})\hbar \equiv r\hbar$;thereby the rotor barrier is different 
for different states $|m,r\rangle$ {\it i.e.,} $V_{0}=V_{0}(m,r)$. 
In this rotor representation, motion
below and above the rotor barrier correspond to normal and local mode
behaviour respectively. 
Exploiting the fact that $H_{R}$ is a one
dimensional Hamiltonian a semiclassical (WKB) expression for the local
mode splitting can be written down as
\begin{equation}
\frac{\Delta}{2} = 
\frac{\omega_{0}}{2\pi} e^{-\theta}
\end{equation}
The frequency factor is $\omega_{0} = 2 \partial H_{R}/\partial p$ and the
the tunneling action integral $\theta$ is taken between the two turning points
$p(\psi*)=0$. Since the local modes are above the rotor barrier the turning
points are purely 
imaginary $\psi* \equiv i \eta* = \cosh^{-1}(E^{0}_{R}/V_{0})/2$ and thus
\begin{equation}
\theta \equiv \int d\psi p(\psi) = 
\int_{-\eta*}^{\eta*} d\eta \sqrt{8D(E^{0}_{R}-V_{0}\cosh(2\eta))}
\label{tunact}
\end{equation}
where, $E^{0}_{R} = \Omega^{2}r^{2}/8D$ and 
a crucial assumption has been made
that $V_{0}$ does not depend on $p$. Although such an asusmption is
not strictly valid the estimates for the splittings agreed fairly well with
the exact quantum values. 

The analysis thus implicates
a $\omega_{1}$:$\omega_{2}=1$:$1$ nonlinear resonance in the phase space for
the dynamical tunneling between local modes and several interesting
observations were made.
First, the rotor Hamiltonian explains the coupling scheme outlined in
Eq.~(\ref{hshpert}). Second, although there is a fairly 
strong coupling between $|3,0\rangle$ and
the intermediate state $|2,1\rangle$ 
no substantial probability build-up was noticed in
the state $|2,1\rangle$ (cf. Fig.~\ref{fig6}L). 
On the other hand substantial probability
does accumulate, as shown in Fig.~\ref{fig6} (NC), 
in the intermediate state $|2,2\rangle$ involved in
the energy transfer process $|3,1\rangle \leftrightarrow |1,3\rangle$.
It was also commented\cite{hsh84} that the double well 
analogy provided by Davis and
Heller\cite{dh81,dh84} was very different from the one 
that emerges from the hindered
rotor analysis - local modes in the former case would be trapped
below the barrier whereas they are above the barrier in the latter case.  
Finally, a very important observation\cite{hsh84} 
was that small amounts of asymmetry between
the two modes quenched the $|3,0\rangle \leftrightarrow |0,3\rangle$ process
whereas there was little effect on the 
$|3,1\rangle \leftrightarrow |1,3\rangle$ process.
The different effects of the asymmetry was one of the primary reasons
for distinguishing between local and nonclassical states.

Several questions arise at this stage. Is the rotor analysis applicable
to the Lawton-Child and Davis-Heller systems? 
Why is there a difference
in the interpretation of the local modes, and hence the mechanism of dynamical
tunneling, between the phase space and hindered rotor pictures? 
Is it reasonable to
neglect the higher order $\omega_{1}$:$\omega_{2}=p$:$q$ resonances solely
based on the relative strengths? If multiple resonances 
do exist then one has the possibility of their overlap
leading to chaos. Is it still possible to use the rotor formalism in
the mixed phase space regimes?
Some of the questions were answered in a work by Stefanski and Pollak
wherein a periodic orbit quantization technique was proposed to calculate
the splittings\cite{sp87}. 
Stefanski and Pollak pointed out\cite{sp87} an important difference
between the Davis-Heller and the Sibert-Reinhardt-Hynes Hamiltonians in
terms of the periodic orbits at low energies. 
Nevertheless, they were able to show that
a harmonic approximation to the tunneling action integral Eq.~(\ref{tunact})
yields an expression for the splitting which is identical to the
one derived by assuming that the symmetric stretch periodic (unstable)
orbit gives rise to a barrier separating
the two local mode states - an interpretation that Davis and Heller
provided in their work\cite{dh81}. 
Thus Stefanski and Pollak resolved an apparent
paradox and emphasized the true phase space nature of dynamical tunneling.

In any case it is worth noting that 
irrespective of the representation the key feature is the
existence of a dynamical barrier separating two qualitatively
different motions; the corresponding structure in the phase space
had to do with the appearance of a resonance. Further support for the
importance of the resonance to tunneling between tori in the phase space
came from a beautiful analysis by Ozorio de Almeida\cite{Ozo84}
which, as seen later, forms an important basis for the recent advances. 
A different viewpoint, using group theoretic arguments, 
based on the concept of dynamical symmetry breaking was advanced\cite{Kell82}
by Kellman. Adapting Kellman's arguments, originally applied to the local
mode spectrum of benzene, imagine placing one quantum in one
of the local O-H stretching mode of H$_{2}$O. Classially the energy remains
localized in this bond and thus there is a lowering or breaking of
the $C_{2v}$ point group symmetry of the molecule. However quantum
mechanically, dynamical tunneling restores the broken symmetry.
Invoking the more general permutation-inversion group\cite{bunkjenbook} 
and its feasible
subgroups provides insights into the pattern of local mode splittings
and hence insights into the energy transfer pathways via dynamical tunneling.
The dominance of a pathway of course cannot be established within a
group theoretic approach alone and requires additional analysis.
Note that recently\cite{bwf03} Babyuk, Wyatt, and Frederick have reexamined the
dynamical tunneling in the Davis-Heller and the coupled Morse 
systems via the Bohmian approach to quantum mechanics.
Analysing the relevant quantum trajectories 
Babyuk {\it et al.} discovered that there were several
regions, at different times, wherein the potential energy exceeds the 
total energy (which includes the quantum potential). 
In this sense, locally, one has a picture that is similar to tunneling
through a one dimensional potential barrier.
Interestingly such regions
were associated with the so-called quasinodes which arise during the
dynamical evolution of the density. Therefore the dynamical nature of
the barriers is clearly evident, but more work is needed to understand
the origin and distributions of the quasinodes in a given system. Needless
to say, correlating the nature of the quasinodes to the underlying phase space
structures would be a useful endeavour. 

\section{Quantum mechanism: vibrational superexchange}
\label{qmvs}

The earlier works 
established the importance of the phase space perspective
for dynamical tunneling using systems that possesed
discrete symmetries and therefore implicitly invoked an analogy to
symmetric double well models. That is not to say that the earlier studies
presumed that dynamical tunneling would only occur in symmetric systems. Indeed
a careful study of the various papers reveal several insightful comments on
asymmetric systems as well. However mechanistic details and
quantitative estimates for dynamical tunneling rates were lacking.
Important contributions in this regard were made by Stuchebrukhov, Mehta, and
Marcus nearly a decade ago\cite{sm93,sm932,smm93,msm95}. 
The experiments\cite{klmps91} that motivated these studies
have been discussed earlier in the introduction. 
In this section the key aspects of the
mechanism are highlighted.

The inspiration comes from the well known
superexchange mechanism of long distance 
electron transfer in molecules\cite{New91,sm92}.
Stuchebrukhov and Marcus argued\cite{sm93,sm932} 
that since the initial, localized (bright)
state is not directly coupled by anharmonic resonances to the other
zeroth-order states it is necessary to invoke off-resonant virtual states
to explain the sluggish flow of energy. Specifically, it was noted that
at any given time very little probability accumulates in the
virtual states. In this sense the situation is very similar to the
mechanism proposed by Hutchinson, Sibert, and Hynes\cite{hsh84} as illustrated
by Eq.~(\ref{hshpert}). However Stuchebrukhov and Marcus extended the
mechanism, called as vibrational superexchange\cite{sm932}, 
to explain the flow of energy between inequivalent bonds in the molecule
and noted the surprising accuracy despite the large number of
virtual transitions involved in the process. 
As noted earlier, Hutchinson's work\cite{Hut84} 
on state mixing in cyanoacetylene
also illustrated the flow of energy between inequivalent modes.
In order to illustrate the essential idea consider the hindered rotor
Hamiltonian (cf. Eq.~(\ref{hindrot})):
\begin{equation}
H_{R} = \frac{p^{2}}{2I} + V_{0}(P,p) \cos(2\psi)
\end{equation}
where we have denoted $I \equiv 4D/\Omega^{2}$. 
The free rotor energies and the associated eigenstates are known to be
\begin{subequations}
\begin{eqnarray}
E_{R}^{0}(k) &=& \frac{k^{2}}{2I} \\
\langle \psi|k \rangle &=& \frac{1}{\sqrt{2 \pi}} \exp(i k \psi)
\end{eqnarray}
\end{subequations}
The perturbation to the free rotor connects states that differ by
two rotational quanta {\it i.e.,} $\langle k|V_{0}(P,p) \cos(2\psi)|k'\rangle
\approx V_{0}(P)/2$ with $|k-k'| = 2$. Clearly the local mode
states $|m,r\rangle$, and $|m,-r\rangle$ with $r \gg 1$ are not
directly connected by the perturbation. 
Nevertheless the local mode states can be coupled through a sequence of
intermediate virtual states with quantum numbers $-(r-2) \leq k \leq (r-2)$.
The effective coupling matrix element
\begin{equation}
V_{\rm eff} = \frac{V_{0}}{2} \prod_{l=-(r-2)}^{(r-2)}
           \frac{V_{0}/2}{E_{R}^{0}(r)-E_{R}^{0}(l)}
\label{effcoupsm}
\end{equation}
can be obtained via a standard application of high-order perturbation
theory involving the sequence
\begin{equation}
|m,r\rangle \rightarrow |m,r-2\rangle \rightarrow \ldots \rightarrow
|m,-(r-2)\rangle
\rightarrow |m,-r\rangle
\label{coup1dchain}
\end{equation}
Note that the polyad quantum number $m$ in the above sequence is fixed
and is consistent with the single resonance approximation.
In addition $V_{0}/2 \ll E_{R}(r)-E_{R}(r-2)$ must be satisfied for
the perturbation theory to be valid.
Thus the splitting between
the symmetry related local mode states $|m,\pm r \rangle$ is given by
$\Delta_{m,r} = 2 V_{\rm eff}$. Interestingly, Stuchebrukhov and Marcus
showed\cite{sm932} that $V_{\rm eff}$ could be derived by analysing the
semiclassical action integral 
\begin{equation}
\theta = \int_{-p_{m}}^{p_{m}} dp \, \eta(p)
\end{equation}
with $\eta = -i \psi$ and
$|p_{m}| = \sqrt{2I(E_{R}^{0}(r)-V_{0})}$ being the minimum classical
value of the momentum of the rotor excited above the barrier $V_{0}$.
This suggests 
that the high-order perturbation theory, if valid, is the correct
approach to calculating dynamical tunneling splittings in multidimensions.
An additional consequence is
that the nonclassical mechanism of IVR is equivalent to dynamical tunneling
as opposed to the earlier 
suggestion\cite{hsh84} of an activated barrier crossing.
As a cautionary note it must be mentioned that the above statements
are based on insights afforded by near-integrable classical dynamics
with two degrees of freedom. 

Although the example above corresponds to a symmetric case the arguments
are fairly general. In fact the vibrational
superexchange mechanism is appropriate for describing quantum pathways for
IVR in molecules. For example consider the acetylinic CH-stretch ($\nu_{1}$)
excitations in propyne, H$_{3}$CCCH. 
Experiments by Lehmann, Scoles, and
coworkers\cite{gtls93} indicate that the overtone 
state $3\nu_{1}$ has a faster IVR rate
as compared to the nearly degenerate $\nu_{1}+2\nu_{6}$ combination state.
Similarly Go, Cronin, and Perry in their study\cite{gcp93} found evidence for
a larger number of perturbers for the $2\nu_{1}$ state than for the
$\nu_{1}+\nu_{6}$ state. The spectrum corresponding to both the
first and the second overtone states implied a lack of direct low-order
Fermi resonances. It was shown\cite{msm95} that the slow
IVR out of the $2\nu_{1}$ and $3\nu_{1}$ states of propyne can be
explained and understood via the 
vibrational superexchange mechanism.

There is little doubt that the vibrational superexchange mechanism, as long
as one is within the perturbative regimes, is applicable to fairly
large molecules. However several issues still remain unclear. 
The main issue
has to do with the effective coupling 
between the initial bright state 
$|{\bf b}\rangle \equiv |v_{1}^{b},v_{2}^{b},\ldots \rangle$ and
a nearly degenerate zeroth-order dark state 
$|{\bf d} \rangle \equiv |v_{1}^{d},v_{2}^{d},\ldots \rangle$ 
in the multidimensional state space. 
In the case of a lone anharmonic resonance there is only a single
off-resonant sequence of states connecting $|{\bf b}\rangle$ and
$|{\bf d} \rangle$ as in Eq.~(\ref{coup1dchain}). 
This sequence or chain of states translates to
a path in the state space. Note that the `paths' being refered to here
are nondynamical, although it might be possible to provide a dynamical
interpretation by analysing 
appropriate discrete action functionals\cite{sm93}.
In general there are several anharmonic
resonances of different orders and in such instances the number of state space
paths can become very large and the issue of the relative
importance of one path over the other arises. For instance it could very
well be the case that $|{\bf b} \rangle$ and $|{\bf d} \rangle$
can be connected by a path with many steps involving low order resonances
and also by a path involving few steps using higher order resonances. 
It is not {\it a priori} clear as to which path should be considered.
An obvious choice is to use some perturbative criteria to decide between
the many paths. Such a procedure was used by Stuchebrukhov and Marcus
to create the tiers in their analysis\cite{sm93}. 
More recently\cite{pg98} Pearman and Gruebele
have used the perturbative criteria to estimate the importance of
direct high/low order couplings and low order coupling chains to the IVR
dynamics in the state space. Thus consider the following possible coupling
chains, shown in Fig.~\ref{fig3}, assuming that the 
states $|{\bf b}\rangle$ and $|{\bf d}\rangle$
are separated by $n$ quanta in the state space
{\it i.e.,} $\sum |v_{k}^{b}-v_{k}^{d}|=n$
\begin{eqnarray*}
|{\bf b}\rangle &\stackrel{n}{\longrightarrow}& |{\bf d}\rangle \\
|{\bf b}\rangle &\stackrel{n_{1}}{\longrightarrow}& |{\bf i}_{1} \rangle
\stackrel{n_{2}}{\longrightarrow} |{\bf d}\rangle \\
|{\bf b}\rangle &\stackrel{n_{1}}{\longrightarrow}& |{\bf i}_{1} \rangle
\stackrel{n_{2}}{\longrightarrow} |{\bf i}_{2}\rangle
\stackrel{n_{3}}{\longrightarrow} |{\bf d}\rangle
\end{eqnarray*} 
Note that $n$ is the distance between $|{\bf b}\rangle$ and $|{\bf d}\rangle$
in the state space and thus identical to $Q$ in Eq.~(\ref{lwqet}).
In the above the first equation indicates a direct coupling between the
states of interest by an anharmonic resonance coupling 
$V_{bd}^{(n)}$ of order $n$. In the second case the coupling is mediated by
one intermediate state $|{\bf i}_{1}\rangle$ involving two anharmonic
resonances $V_{bi_{1}}^{(n_{1})}$ and $V_{i_{1}d}^{(n_{2})}$ with
$n=n_{1}+n_{2}$. Each step of a given state space path, coupling
two states $|{\bf i}\rangle$ and $|{\bf j}\rangle$ at order $m$,
is weighted by the
perturbative term 
\begin{eqnarray}
{\cal L}_{ij}^{(m)}&=&\left\{1 + \left(\frac{\Delta E_{ij}^{0}}{V_{ij}^{(m)}}
\right)^{2} \right\}^{-1/2} \\
&\approx& \frac{V_{ij}^{(m)}}{\Delta E_{ij}^{0}} \qquad 
{\rm for} \quad \Delta E_{ij}^{0} \gg V_{ij}^{(m)} \nonumber \\
&\approx& 1 \qquad \qquad {\rm for} \quad \Delta E_{ij}^{0} \ll V_{ij}^{(m)}
\nonumber
\end{eqnarray}
with $\Delta E_{ij}^{0} \equiv E_{i}^{0} - E_{j}^{0}$ and
$V_{ij}^{(m)} \equiv \langle {\bf i}|V^{(m)}|{\bf j}\rangle$.
A specific path in the state space is then associated with the
product of the weightings for each step along the path. For example the
state space path above involving two intermediate states is associated with
the term ${\cal L}_{bi_{1}}^{(n_{1})} {\cal L}_{i_{1}i_{2}}^{(n_{2})}
{\cal L}_{i_{2}d}^{(n_{3})}$. It is not hard to see that such products of
${\cal L}$ correspond to the terms contributing to the effective coupling
as shown in Eq.~(\ref{effcouphrh}) and Eq.~(\ref{effcoupsm}).
The various terms also contribute to the effective coupling
$\psi_{Q}$ (cf. Eq.~(\ref{lwqet})) 
in the Leitner-Wolynes criteria\cite{lw96jcp}
for the transition from localized to extended states.

The crucial issue is wether or not one can identify dominant paths,
equivalently the key intermediate states, based on the perturbative
criteria alone. In a rigorous sense the answer is negative since
such a criteria ignores the dynamics. Complications can also
arise due to the fact that each segment of the path contributes with
a $\pm$ phase\cite{pg98}. 
Furthermore one would like to
construct the explicit dynamical barriers in terms of the molecular
parameters and conserved or quasi-conserved quantities.
The observation that the superexchange 
$V_{\rm eff}$ can be derived\cite{sm93} from
a semiclassical action integral provides a clue to some of the issues.
However an explicit demonstration of such a correspondence has
been provided only in the single resonance (hindered rotor) case. In the
multidimensional case the multiplicity of paths obscures this
correspondence. 
The next few sections highlight the recent advances
which show that clear answers to the
various questions raised above come from viewing
the phenomenon of dynamical tunneling in the most natural 
representation - the classical phase space.

\section{Dynamical tunneling: recent work}
\label{dtrw}

Nearly a decade ago Heller wrote a brief review\cite{Hel95} on
dynamical tunneling and its spectral consequences which was
motivated in large part
by the work of Stuchebrukhov and Marcus\cite{sm93}. 
Focusing on systems with two degrees of freedom in the
near-integrable regimes, characteristic of polyatomic molecules at low
energies,  
Heller argued that the $10^{-1}$-$10^{-2}$ cm$^{-1}$ {\em broadening
of lines is due to dynamical tunneling between remote regions of 
phase space facilitated by distant resonances}.
This is an intriguing statement which could perhaps be interpreted in
many different ways. Moreover the meaning of the words 'remote' and
'distant' are not immediately clear and fraught with conceptual
difficulties in a multidimensional phase space setting. 
In essence Heller's statement, more appropriately called as a conjecture,
is an effort to provide a phase
space picture of the
superexchange mixing between
two or more widely separated zeroth-order
states in the state space. Some of the examples in the later part of the
present review, hopefully, demonstrate that the conjecture is reasonable.
Interestingly though in the same review it is mentioned\cite{Hel95} that in the
presence of widespread chaos the issue of tunneling as a mechanism for IVR
is of doubtful utility. Similar sentiments were echoed by Stuchebrukhov
{\it et al.} who, in the case of high dimensional phase spaces, envisaged
partial rupturing of the invariant tori and hence domination of
IVR by chaotic diffusion rather than dynamical tunneling. 
This too has to be considered as a conjecture at the present time since
the timescales and the competition between chaotic diffusion (classically
allowed) and dynamical tunneling (classically forbidden) in systems with
three or more degrees of freedom has not been studied in any detail.
The issues involved are subtle and extrapolating the insights from
studies on two degrees of freedom to higher degrees of freedom is
incorrect. For instance, in three and higher degrees of freedom the
invariant tori, ruptured or not, do not have the right dimensionality
to act as barriers to inhibit diffusion. 
Therefore, is it possible that classical chaos could assist or inhibit
dynamical tunneling? 
Several 
studies\cite{lb91r,si95,bbem93,btu93,tu94,lgw94,ammk95,udh94,df95,lu96,si96} 
carried out during the early nineties till the present time point to
an important role played by chaos in the process of dynamical tunneling.
We start with a brief review of these
studies followed by very detailed investigations into the role played
by the nonlinear resonances. Finally, some of the
very recent work is highlighted
which suggest that the combination of resonances and chaos in the phase
space can yet play an important role in IVR.

\subsection{Chaos-assisted tunneling (CAT)}
\label{cat}

\begin{figure} [htbp]
\begin{center}
\includegraphics[width=85mm]{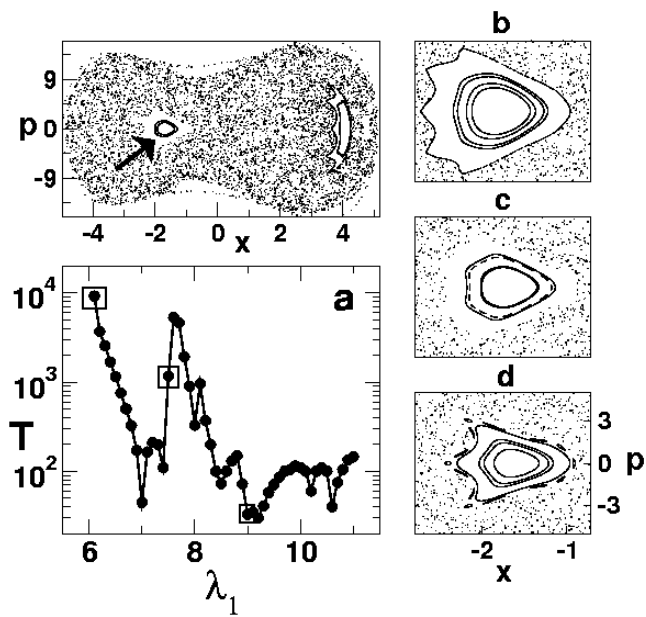}
\caption{Tunneling time for the Lin-Ballentine system as a function
of the field strength $\lambda_{1}$. 
The parameters are chosen\cite{lb90} as $m=1$,
$b=0.5$, and $d=10$. The phase space for $\lambda_{1}=7.8$ in
the upper left panel shows the two regular islands embedded in the
chaotic sea.
The initial state corresponds to
a coherent state localized on the left regular island (indicated by
an arrow).
The three phase space strobe plots at increasing values of $\lambda_{1}$
indicated in (a) show the local structure near the left island.
Despite the phase space being
very similar the tunneling times fluctuate over
many orders of magnitude. 
Figure courtesy of Astha Sethi (unpublished work).}
\label{fig7}
\end{center}
\end{figure}

One of the first studies on the possible influence of chaos on tunneling
was made by Lin and Ballentine\cite{lb90,lb91r,lb92}. 
These authors investigated the tunneling
dynamics of a driven double well system described by the Hamiltonian
\begin{equation}
H(x,p,t) = \frac{1}{2m}p^{2} + bx^{4}-dx^{2} + \lambda_{1} x \cos(\omega t)
\label{lbham}
\end{equation}
which has the
discrete symmetry $H(x,p,t) = H(-x,-p,t+T_{0}/2)$ with $T_{0}=2\pi/\omega$.
In the absence of the monochromatic field ($\lambda_{1} = 0$) the system
is integrable and the phase space is identical to that of a one
dimensional symmetric double well potential. 
However with $\lambda_{1} \neq 0$
the system is nonintegrable and the phase space is mixed regular-chaotic.
Despite the mixed phase space the discrete symmetry 
implies that any regular islands in the phase space will occur
in symmetry related pairs and the quantum floquet states will occur as
doublets with even/odd symmetries. A coherent state (wavepacket)
localized on one of the regular islands will tunnel to the other, clasically
disconnected,
symmetry related island. Thus this
is an example of dynamical tunneling. An important observation by
Lin and Ballentine was that the tunneling was 
enhanced\cite{lb90,lb92} by orders of
magnitude in regimes wherein significant chaos existed in the phase space.
For example with $\lambda_{1}=0$ the ground state tunneling 
time is about $10^{6}T_{0}$
whereas with $\lambda_{1}=10$ the tunneling time is only about $10^{2}T_{0}$.
Futhermore strong fluctuations in the tunneling times were observed.
In Fig.~\ref{fig7} a typical example of 
the fluctuations over several
orders of magnitude are shown. The crucial thing
to note, however, is that the gross phase space structures are similar over
the entire range shown in Fig.~\ref{fig7}. 
Thus although the chaos seems to influence the tunneling
the mechanistic insights were lacking {\it i.e.,} the precise role of
the stochasticity was not understood. 

Important insights came from the work\cite{gdjh91} by Grossmann {\it et al.}
who showed that it is possible to suppress the
tunneling in the driven double well by an appropriate choice of the
field parameters $(\lambda_{1},\omega)$.
The explanantion for such a coherent destruction
of tunneling and the fluctuations comes from analysing the
the floquet level motions with $\lambda_{1}$.
Gomez Llorente and Plata provided\cite{lp92} 
perturbative estimates within a two-level
approximation for the enhancement/suppression of tunneling.
A little later Utermann, Dittrich, and H\"{a}nggi
showed\cite{udh94} that there is a 
strong correlation between the splittings of the
floquet states and their overlaps with the chaotic part of the phase space.
Breuer and Holthaus\cite{bh93} highlighted the role of classical phase space
structures in the driven double well system.
Subsequently other works\cite{fm93,lgw94,ammk95,fh96,zl97,mmkgd01,om05} 
on a wide variety of driven systems established
the sensitivity of tunneling to the classical stochasticity.
An early discussion of dynamical tunneling and the influence of
dissipation can be found in the work\cite{gh87} by 
Grobe and Haake on kicked tops.
A comprehensive account of the various studies on driven systems
can be found in the review\cite{gh98} by Grifoni and H\"{a}nggi.
Such studies have provided, in recent times, important insights
into the process of coherent control of quantum processes. 
Recent review\cite{gb05}
by Gong and Brumer, for instance, discusses 
the issue of quantum control of classically chaotic systems
in detail. Perhaps it is apt to highlight a historical fact
mentioned by Gong and Brumer - {\em coherent control emerged from two
research groups engaged in studies of chaotic dynamics}. 

\begin{figure} [htbp]
\begin{center}
\includegraphics[width=85mm]{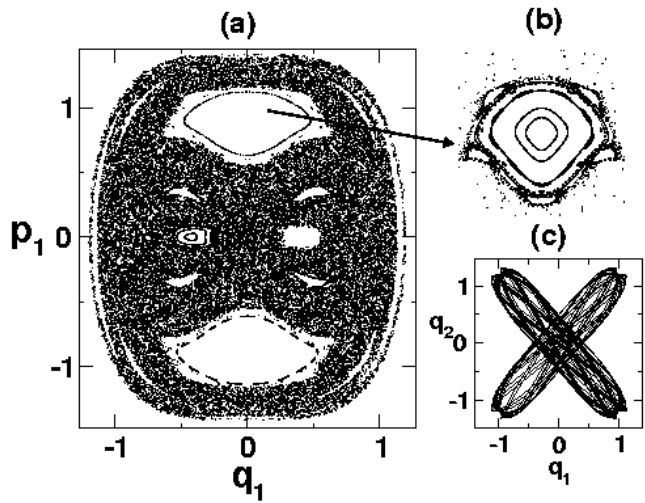}
\caption{In (a) the $q_{2}=0$ surface of section is shown for $E=1$ and
the parameter values $(\lambda,b) = (-0.25,\pi/4)$ for the Hamiltonian
in Eq.~(\ref{btuham}). 
The large islands
are part of a $1$:$1$ resonance island chain and are related to each other
by symmetry. (b) Details of the upper island and (c) shows the
configuration space plot of the symmetrically related tori.}
\label{fig8}
\end{center}
\end{figure}

In the Lin-Ballentine example above the perturbation (applied field)
not only increases the size of the chaotic region but also affects the
dynamics of the unperturbed tunneling doublet. 
Hence the enhanced tunneling, relative to the unperturbed or
the weakly perturbed case, cannot be immediately ascribed
to the increased amount of chaos in
the phase space. To observe a more direct influence of chaos on tunneling
it is necessary that the classical phase space dynamics scales with energy.
Investigations of the coupled quartic oscillator system
by Bohigas, Tomsovic, and Ullmo\cite{btu93,tu94} 
provided the first detailed view
of the CAT process. The choice of the Hamiltonian
\begin{equation}
H({\bf p},{\bf q}) = \frac{1}{2} {\bf p}^{2} + a(\lambda) \left[
\frac{q_{1}^{4}}{b} + b q_{2}^{4} + 2\lambda q_{1}^{2} q_{2}^{2} \right]
\label{btuham}
\end{equation}
with $b=\pi/4$ was made in order to study the classical-quantum correspondence
in detail. This is due to the fact that the potential is homogeneous
and hence the dynamics at a specific energy $E$ is related to
the dynamics at a different energy by a simple rescaling. 
Consequently one can fix the energy and investigate the effect of
the semiclassical
limit $\hbar \rightarrow 0$ on the tunneling process.
The classical
dynamics is integrable for $\lambda=0$ and 
strongly chaotic for $\lambda=-1$. 
Again, note that the Hamiltonian
has the discrete symmetry ${\bf q} \rightarrow -{\bf q},{\bf p} \rightarrow
-{\bf p}$ and thus the quantum eigenstates are expected to show up as
symmetry doublets. In contrast to the driven system discussed
above, and similar to
the Davis-Heller system discussed earlier, 
the system in eq.~(\ref{btuham}) does not have any
barriers arising from the two dimensional quartic potential.
The doublets therefore must correspond
to quantum states localized in the symmetry related islands in phase
space shown in Fig.~\ref{fig8}. Thus denoting the states localized in
the top and bottom regular islands by $|T\rangle$ and $|B\rangle$ respectively
the doublets correspond to the usual linear combinations
$|\pm\rangle = (|T\rangle \pm |B\rangle)/\sqrt{2}$.

Bohigas, Tomsovic, and Ullmo focused on quantum states associated with
the large regular islands seen in Fig.~\ref{fig8}. 
Specifically, they studied the variations in the
splitting $\Delta E \equiv |E_{+}-E_{-}|$
associated with the doublets $|\pm\rangle$
upon changing $\hbar$ and the coupling $\lambda$. 
Since the states are residing in a regular region of the phase space
it is expected, based on one dimensional tunneling theories, that the
splittings will scale exponentially with $\hbar$ {\it i.e.,}
$\Delta E \sim \exp(-1/\hbar)$. However the splittings
exhibit\cite{btu93} fluctuations of several orders of magnitude and no obvious
indication of a predictable dependence on $\hbar$ seemed to exist. Note that
a similar feature is seen in Fig.~\ref{fig7} showing
the tunneling time fluctuations of the
driven double well system. More precisely, it was observed\cite{btu93} that for
the range of $\lambda$ that corresponded to integrable or near-integrable
phase space $\Delta E$ roughly follows the exponential scaling. On the
other hand for $\lambda = \lambda^{*} \approx -0.2$ chaos becomes pervasive
and $\Delta E$ exhibits severe fluctuations. 
Tomsovic and Ullmo argued\cite{tu94} that
the fluctuations in $\Delta E$ could be traced to the crossing of the
quasidegenerate doublets with a third irregular state.
By irregular it is meant that when viewed from the phase space the state
density, represented by either the Wigner or the
Husimi distribution function, is appreciable in the chaotic sea. 
The irregular state, in contrast to the regular states,
do not come as doublets since there is no dynamical
partition of the chaotic region into mutually exclusive symmetric parts.
Nevertheless the irregular states do
have fixed parities with respect to the reflection operations.
Thus chaos-assisted tunneling
is necessarily atleast a three-state process\cite{tu94,bbem93}.
In other words, assuming that the even-parity irregular state $|C\rangle$ 
with energy $E_{c}$ couples to $|+\rangle$ with strength $v$
a relevant, but minimal, model Hamiltonian in the symmetrized
basis has the form:
\begin{equation}
H = \left(\begin{array}{ccc} E+\epsilon & 0 & 0 \\ 
                             0 & E-\epsilon & v\\
                             0 & v & E_{c} \end{array} \right)
\end{equation}
with $\epsilon$ being a direct coupling between $|T\rangle$ and $|B\rangle$.
If $\epsilon$ is dominant as compared to $v$ then one has the usual
two level scenario. However if $v \gg \epsilon \approx 0$ then 
the splitting can be approximated as
\begin{equation}
\Delta E \rightarrow
  \begin{cases}
   \frac{v^{2}}{E-E_{c}}&  E-E_{c} \gg v,\\
   |v|& E-E_{c} \ll v.
  \end{cases}
\end{equation}
Hence varying a parameter, for example $\lambda$, one expects a peak of
height $|v|$ as $E_{c}$ crosses the $E_{+}$ level.
Detailed theoretical studies\cite{df95,df98} 
of the dynamical tunneling process in
annular billiards by Frischat and Doron confirmed the crucial role
of the classical phase space structure. Specifically it was found that
the tunneling between two symmetry related 
whisphering gallery modes (corresponding to clockwise and counterclockwise
rotating motion in the billiard) is strongly influenced by quantum states
that inhabit the regular-chaotic border in the phase space. Such states
were termed as ``beach states'' by Frischat and Doron.
Soon thereafter Dembowski
{\it et al.} provided experimental 
evidence\cite{dghhrr00} for CAT in a microwave
annular billiard simulated by means of a two-dimensional electromagnetic
microwave resonator. Details of the experiment can be found in a recent
paper\cite{hadghhrr05} by Hofferbert {\it et al.} wherein support 
for the crucial role
of the beach region is provided.
It is, however, significant to note that  
the regular-chaotic border in case of the annular billiards is 
quite sharp.

An important point to note is that 
despite the intutively appealing three-state model,
determining the coupling $v$ between the
regular and chaotic states is nontrivial. This, in turn, is related to
the fact that accurate determination of the positions of the chaotic levels
can be a difficult task
and determining the nature of a chaotic state 
for systems with mixed phase spaces is still an open problem.
Such difficulties prompted Tomsovic
and Ullmo\cite{tu94} to adopt a statistical approach, 
based on random matrix theory,
to determine the tunneling splitting 
distribution $P(\Delta E)$ in terms of
the variance of the regular-chaotic coupling {\it i.e.,} $\overline{v^{2}}$.
In a later work
Leyvraz and Ullmo showed\cite{lu96} that $P(\Delta E)$ is a truncated Cauchy
distribution 
\begin{equation}
P(\Delta E) = \frac{2}{\pi} \frac{\overline{\Delta E}}{(\Delta E)^{2} +
(\overline{\Delta E})^{2}}
\label{trunccauchy}
\end{equation}
with $\overline{\Delta E}$ being the mean splitting.
On the other hand Creagh and Whelan\cite{cw00,cw99} showed that
the splitting between states localized
in the chaotic regions of the phase space but separated by an energetic barrier
is characterized by a specific tunneling orbit. 
Specifically, the resulting splitting distribution depends on the stability
of the tunneling orbit and is not universal. In certain limits\cite{cw00} the
splittings obey the Porter-Thomas distribution. 
Thus the fluctuations in $\Delta E$ in the CAT process are different
from those found in the usual double-well system. Note that in case
of CAT the distribution pertains to the splitting between states localized
in the regular regions of phase space embedded in the chaotic sea.
The reader is referred to the excellent discussion by Creagh
for details\cite{Crebook}. Nevertheless, in a recent
work Mouchet and Delande observed\cite{md03} that the $P(\Delta E)$ exhibits a
truncated Cauchy behaviour despite a near-integrable phase space.
Similar observations have been made by Carvalho and Mijolaro 
in their study of the splitting distribution in the annular billiard
system as well\cite{cm04}.
This indicates that the Leyvraz-Ullmo distribution Eq.~(\ref{trunccauchy})
is not sufficient to
characterize CAT. 
A first step towards obtaining a semiclassical estimate
of the regular-chaotic coupling was 
taken by Podolskiy and Narimanov\cite{pn03}. 
Assuming a regular island seperated by a structureless, on the
scale of $\hbar$, chaotic sea
they were able to show that:
\begin{equation}
\Delta E \propto \hbar \frac{\Gamma({\cal A},2{\cal A})}{\Gamma({\cal A}+1,0)}
\label{podnari}
\end{equation}
with ${\cal A} \equiv A/\pi \hbar$ and a system specific proportionality
factor which is independent of $\hbar$. 
The phase space area covered by the regular island is denoted by $A$.
Application of the theory by Podolskiy and Narimanov to the splittings
between near-degenerate optical modes, localized on a pair of
symmetric regular islands in phase space, in a non-integrable microcavity
yielded very good agreement with the exact data\cite{pn03}.
In fact Podolskiy and Narimanov have recently shown\cite{pn05} that
the lifetimes and emission patterns of optical modes in asymmetric
microresonators are strongly affected by CAT.
Encouraging results were obtained in the case of tunneling in periodically
modulated optial lattices as well. However the agreement in the splittings
displayed significant deviations in the deep semiclassical regime {\it i.e.,}
large vaues of $\hbar^{-1}$. Interestingly the critical value of $\hbar^{-1}$
beyond which the disagreement occurs correlates with the existence of
plateau regions, discussed in the next section (cf. Fig.~\ref{fig11}), 
in the plot of 
$\log(\Delta E)$ versus $\hbar^{-1}$. Such plateau regions
have been noted earlier\cite{rbiwg94} 
by Roncaglia {\it et al.}
and in several other recent studies\cite{es05,seu05,wseb06} as well. 

The qualitative picture that emerged from the numerous studies is that
CAT process is a result of competition 
between classically allowed (mixing
or transport in the chaotic sea) and classically forbidden (tunneling or
coupling between regular regions and the chaotic sea) dynamics.
Despite the significant advancements, 
the mechanism by which a state localized in the 
regular island couples to the chaotic sea continued to puzzle the
researchers.
It might therefore come as a surprise that based
on recent studies there is growing evidence that the
explanation for the regular-chaotic coupling lies in the nonlinear resonances
and partial barriers in the phase space. 
The work by Podolskiy
and Narimanov provides one possible answer but an equally strong clue is
hidden in the plateaus that occur 
in a typical $\log(\Delta E)$ versus $\hbar^{-1}$ plot.
It is perhaps reasonable to expect that the theory of Podolskiy and
Narimanov is correct  
when phase space structures like the nonlinear resonances
and partial barriers, ignored in the derivation of Eq.~(\ref{podnari}),
in the vicinity of regular-chaotic border regions
are much smaller in area as compared to the Planck constant $\hbar$.
However, for sufficiently small $\hbar$ the 
rich hierarchy of structures like
nonlinear resonances, partially broken tori, cantori, and even vague tori
in the regular-chaotic border region have to be taken into
consideration.
It is significant to note that Tomsovic and Ullmo had
already recognized the importance of accounting for mechanisms which tend
to limit classical transport\cite{tu94} and proposed generalized random
matrix ensembles as a possible approach.
There also exist a number of studies\cite{grr86,grracp89,khsw00,rp88,mh00}, 
involving less than
three degrees of freedom, which highlight
the nature of the regular-chaotic border 
and their importance to tunneling between KAM tori. 

A different perspective on the influence of chaos on dynamical tunneling
came from the detailed investigations by Ikeda and coworkers
involving the study\cite{si95,si96,si98,osit01,ti03,ti05}
of semiclassical propagators in the complex
phase space. In particular, deep insights into CAT were gained by examining the
so called ``Laputa chains" which contribute dominantly to the propagator in
the presence of chaos\cite{si95,si96}. For a detailed review the 
paper\cite{si98} by Shudo and Ikeda is highly reccommended.
Hashimoto and Takatsuka\cite{ht98}
discuss a situation wherein dynamical tunneling
leads to mixing between states localized about unstable fixed points
in the phase space. However in this case the transport is energetically
as well as dynamically allowed and it is not very clear whether the term
`tunneling' is an appropriate choice.

In the following subsections recent progress towards quantitative
prediction of the average tunneling rates, and hence the
coupling strengths between the regular and chaotic states, are described. 
The key ingredient to the success of the theory are the nonlinear resonances in
the phase space.
In fact, depending on the relative
size of $\hbar$, one needs to take multiple nonlinear resonances into account
for a correct description of dynamical tunneling.
If one asociates the $\hbar$ variation with the related density of states
in a molecular system then it is tempting to claim that such a mechanism
must be the correct semiclassical limit of the Stuchebrukhov-Marcus
vibrational superexchange theory. 
The discussions in the next section indicate that such a claim
is indeed reasonable.

\subsection{Resonance-assisted tunneling (RAT)}
\label{rat}

The early investigations of dynamical tunneling
made it clear that nonlinear resonances played an
important role in the phenomenon of dynamical tunneling. 
However in the majority of the studies one had a single resonance of
low order dominating the tunneling process
and hence an effectively integrable classical dynamics. Very few
attempts were made for systems which had multiple resonances
and thus capable of exhibiting near-integrable to mixed phase space
dynamics. Moreover studying the
tunneling process with varying $\hbar$ is necessary to implicate the
nonlinear resonances and other phase space structures without ambiguity.
Definitive answers in this regard were provided by Brodier, Schlagheck, and
Ullmo in their study\cite{bsu01} of time-dependent Hamiltonians. 
In this section we provide a brief outline of their work and refer the
reader to some of the recent reviews\cite{bsu02,seu05,Schl06} 
which provide details of the theory
and applications to several other systems.

\begin{figure} [htbp]
\begin{center}
\includegraphics[width=80mm]{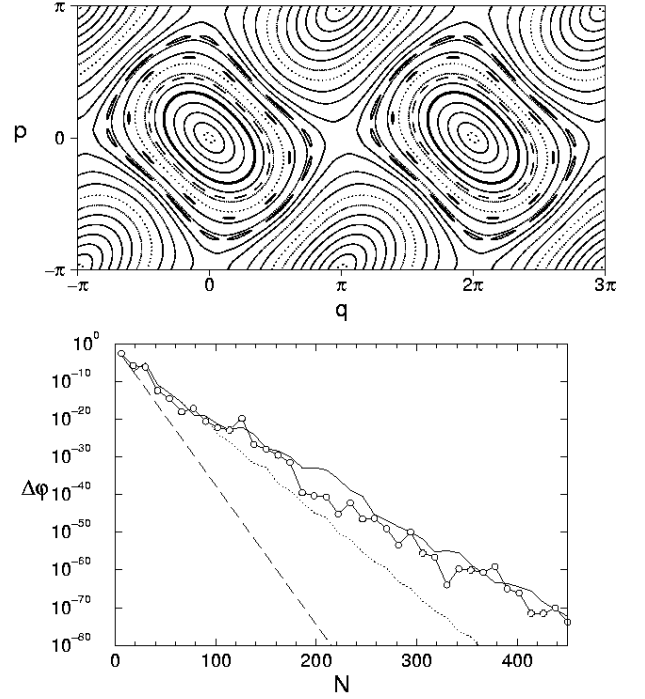}
\caption{The classical phase space (upper panel) and quantum eigenphase
splittings (lower panel) of the kicked Harper system with $\tau=1$. The
splittings are calculated for the $n$-th excited state at 
$N \equiv 2\pi/\hbar = 6(2n+1)$. Variation with $N$ is 
equivalent to fixing the state localized on the torus with action
$I = \pi/6$ and varying $\hbar$. Exact quantum splittings (circles)
are well reproduced by semiclassical prediction (thick line)
based on the $16$:$2$, the $10$:$1$, and the $14$:$1$ resonances.
The dashed line is the integrable
estimate obtained using the Hamiltonian in Eq.~(\ref{kharpinteg}) and
clearly inaccurate. Note that multiple precision arithmetics was
used to compute eigenphase splittings below the ordinary
machine precision limit. Figure courtesy of P. Schlagheck.}
\label{fig9}
\end{center}
\end{figure}

In order to highlight the salient features of RAT consider a one degree
of freedom system that evolves under a periodic time-dependent Hamiltonian
$H(q,p,t)=H(q,p,t+\tau)$. The phase space structure in this case can
be easily visualized by constructing the stroboscopic Poincar\'{e} 
section {\it i.e.,} plotting $(q,p)$ at integer values of the period
$t = n \tau$. We are interested in the generic case wherein the phase space
exhibits two symmetry-related regular islands that are embedded in the
phase space. Typically the phase space has various nonlinear resonances
that arise due to frequency commensurability between the external driving
field frequency $\omega \equiv 2 \pi/\tau$
and the system frequencies. 
Fig.~\ref{fig9} shows an example of a near-integrable phase space
wherein a few resonances are visible.
For the moment assume that
a prominent $r$:$s$, say the $10$:$1$ resonance in Fig.~\ref{fig9}, 
nonlinear resonance manifests in the phase space. The
motion in the vicinity of this $r$:$s$ resonance can be analysed using
secular perturbation theory. In order to do this one introduces a 
time-independent Hamiltonian $H_{0}(q,p)$ that approximately describes the
regular motion in the islands. In particular assume that appropriate
action-angle variables $(I,\theta)$ can be introduced such that
$H_{0}(q,p) \rightarrow H_{0}(I)$ and thus the total Hamiltonian can be
expressed as $H(I,\theta,t) = H_{0}(I) + V(I,\theta,t)$.
The term $V$ represents a weak perturbation in the center of the island.
A $r$:$s$ nonlinear resonance occurs when
\begin{equation}
r \Omega_{r:s} = s \omega_{f} \,\,\, {\rm with}\,\,\, \Omega_{r:s} \equiv
\left[\frac{dH_{0}}{dI}\right]_{I=I_{r:s}}
\end{equation}
The resonant action is denoted by $I_{r:s}$. 
Using techniques\cite{llbook} from 
secular perturbation theory one can show that the dynamics in the vicinity
of the nonlinear resonance is described by the new Hamiltonian\cite{llbook}
\begin{equation}
{\cal H}(I, \vartheta,t) \approx \frac{(I-I_{r:s})^{2}}{2 m_{r:s}}
+ {\cal V}(I,\vartheta,t)
\end{equation}
with ${\cal V}(I,{\cal \vartheta},t) \equiv {\cal V}(I,\vartheta+
\Omega_{r:s}t,t)$, and the ``effective mass" parameter $m_{r:s} \equiv
(d^{2}H_{0}/dI^{2})^{-1}(I_{r:s})$ is related to the anharmonicity of
the system. The new angle $\vartheta \equiv \theta - \Omega_{r:s}t$
varies slowly in time near the resonance and hence, according to
adiabatic perturbation theory, one can replace ${\cal V}$ by
its time average over $r$-periods of the driving field
\begin{equation}
V(I,\vartheta) \equiv \frac{1}{r\tau} 
\int_{0}^{r\tau} dt {\cal V}(I,\vartheta,t)
\label{tavcoup}
\end{equation}
Furthermore, writing the perturbation in terms of a Fourier series and
neglecting the action dependence of the Fourier coefficients {\it i.e.,}
evaluated at $I = I_{r:s}$ the effective integrable Hamiltonian is
obtained as:
\begin{equation}
H_{r:s}(I,\vartheta) = \frac{(I-I_{r:s})^{2}}{2 m_{r:s}} +
\sum_{k=1}^{\infty} V_{k}(I_{r:s}) \cos(kr\vartheta+\phi_{k})
\end{equation}
The above effective Hamiltonian couples the zeroth order
states $|n\rangle$ and $|n+kr\rangle$ with a matrix element 
$V_{k}\exp(i\phi_{k})/2$. Here the quantum numbers refer to the
zeroth-order KAM tori inside the large regular island.
Thus, the eigenstates of $H_{r:s}$ are admixtures of unperturbed states
obeying the ``selection rule" $|n'-n|=kr$ and perturbatively given by
\begin{eqnarray}
|\tilde{n}\rangle &=& |n\rangle + \sum_{k \neq 0} \frac{v_{k}}
{\Delta E_{n+kr}} |n+kr\rangle \nonumber \\
&+& \sum_{k,k' \neq 0} \frac{v_{k} v_{k'}}
{\Delta E_{n+kr}\Delta E_{n+kr+k'r}} |n+kr+k'r\rangle
\end{eqnarray}
where $v_{k} \equiv V_{k} \exp(i\phi_{k})$ and
$\Delta E_{j} \equiv E_{n}-E_{j}$.
Associating the quantized actions
$I_{n} \equiv \hbar(n+1/2)$ with states localized in the regular islands
the energy difference
\begin{equation}
E_{n}-E_{n'} = \frac{1}{2m_{r:s}}(I_{n}-I_{n'})(I_{n}+I_{n'}-2I_{r:s})
\label{symloc}
\end{equation}
becomes small if $I_{n}+I_{n'} \approx 2I_{r:s}$. In other words, the
states $|n\rangle$ and $|n'\rangle$ are coupled strongly
if they straddle the $r$:$s$ resonance
symmetrically. Consequently the nonlinear resonance provides an efficient
route to couple the ground state with $I_{0}<I_{r:s}$ to a highly
excited state with $I_{kr}>I_{r:s}$. 

Once again a crucial issue has to do with the the dominant pathways
that couple $|n\rangle$ and $|n+kr\rangle$. Clearly the two states can
couple in a one-step process via $V_{k}$ or with a $k$-step process
via $V_{1}$. The analysis by Brodier {\it et al.}, 
realizing the rapid decrease of
the Fourier terms $V_{k}$ with $k$, reveals\cite{bsu02} 
that the $k$-step process
is dominant in the semiclassical limit $\hbar \rightarrow 0$.
Note that within the effective Hamiltonian $H_{r:s}$ one
assumes the dominance of a single nonlinear resonance and hence
the focus is on the $r$:$s$ resonance and its higher harmonics. 
Neglecting the Fourier components with $k > 1$ and denoting $2V_{r:s} \equiv
V_{1}$ one obtains the standard pendulum Hamiltonian
\begin{equation}
H_{eff}(I,\vartheta) = \frac{(I-I_{r:s})^{2}}{2m_{r:s}} + 
                        2V_{r:s}\cos(r\vartheta)
\end{equation}
For periodically driven systems the floquet states can be obtained by
diagonalizing the one-period time evolution operator. The near degenerate
floquet states are characterized by
the splitting
\begin{equation}
\Delta \phi_{n} = |\phi_{n}^{+}-\phi_{n}^{-}|
\end{equation}
between the quasienergies, or eigenphases, of the symmetric and antisymmetric
state. Within the perturbation scheme the ground state ($n=0$) splitting is
then obtained as
\begin{equation}
\Delta \phi_{0} = \Delta \phi_{0}^{(0)} + \sum_{k} |{\cal A}_{kr}^{(r:s)}|^{2}
\Delta \phi_{kr}^{(0)}
\end{equation}
with 
\begin{equation}
{\cal A}_{kr}^{(r:s)} \equiv \langle kr|0 \rangle  = \prod_{l=1}^{|k|}
\frac{V_{r:s}}{E_{0}-E_{lr}}
\end{equation}
being a measure of
the contribution of the $kr$-th excited state to the perturbed ground state.
The quantities $\Delta \phi_{n}^{(0)}$ are the unperturbed splittings
resulting from the integrable limit Hamiltonian for which extensive
semiclassical insights exist\cite{Wil86,sc06,tww95}.
Note that the above expression for ${\cal A}_{kr}^{(r:s)}$ is essentially
the same as the one obtained by Stuchebrukhov and Marcus within their
vibrational superexchange approach\cite{sm93}. 
Here one has the generalization for
driven systems whose
phase space exhibits a general $r$:$s$ nonlinear
resonance. 

Brodier {\it et al.} illustrated\cite{bsu01,bsu02} 
the RAT mechanism beautifully by studying
the kicked Harper model
\begin{equation}
H(p,q,t) = \cos(p) + \tau \sum_{n=-\infty}^{\infty} \delta(t-n\tau) \cos(q)
\end{equation}
whose dynamics is equivalent to the symplectic map
\begin{eqnarray}
p_{n+1} &=& p_{n} + \tau \sin(q_{n}) \nonumber \\
q_{n+1} &=& q_{n} - \tau \sin(p_{n+1})
\label{kharp}
\end{eqnarray}
which describes the evolution of $(p,q)$ from time $t=n\tau$ to $t=(n+1)\tau$.
For $\tau=1$ the phase space, shown in Fig.~\ref{fig9},
is near-integrable and exhibits a large
regular island around $(p,q)=(0,0)$. The $2\pi$ periodicity of the system
in both $p$ and $q$ implies the existence of a symmetry related island
at $(p,q)=(0,2\pi)$. The floquet states thus come in nearly degenerate
pairs with eigenphase splittings $\Delta \phi_{n}$
associated with the $n$-th excited state in the regular island.
Since the system is near-integrable for $\tau=1$
it is possible to use classical perturbation theory to construct
a time-independent Hamiltonian
\begin{eqnarray}
H_{0}(p,q) &=& \cos p + \cos q -\frac{\tau}{2} \sin p \sin q \nonumber \\
&-& \frac{\tau^{2}}{12}(\cos p \sin^{2} q + \cos q \sin^{2} p) \nonumber \\
&-& \frac{\tau^{3}}{48} \sin 2p \sin 2q
\label{kharpinteg}
\end{eqnarray}
which is a very good integrable approximation to the kicked Harper map.
Indeed the phase space for the exact time-dependent Hamiltonian
for $\tau=1$ and the integrable approximation in Eq.~(\ref{kharpinteg})
are nearly identical\cite{bsu01}.
The crucial difference, however, is that the integrable approximation
does not have the various nonlinear resonances which are visible
in Fig.~\ref{fig9}. The consequence of negelcting the nonlinear
resonances are catastrophic as far as the splittings are concerned
and from Fig.~\ref{fig9} it is clear that any estimate based on
the integrable approximation is doomed to fail.

Such pendulum-like Hamiltonians have been used in the very early studies
of dynamical tunneling and were also central to the ideas and
conjectures formulated by
Heller\cite{Hel95} in the context of IVR and spectral signatures. However
the importance of the contribution by Brodier, Schlagheck, and Ullmo
lies in the fact that a rigorous semiclassical basis for the RAT mechanism
was established. Although the detailed study was in the context of
a one-dimensional driven system, the insights into the mechanism
are expected to be valid in larger class of systems. These include 
systems with two degrees of freedom and perhaps to some extent systems
with three or more degrees of freedom as well (see discussions in the
introduction and the last section). 
For instance it can be
shown that the RAT mechanism will {\em always} dominate the regular tunneling
process in the semiclassical limit. Furthermore, discussions on the
importance of the higher harmonics of a given nonlinear resonance and the
criteria for identifying and including multiple resonances in the path
between $|n\rangle$ and $|n'\rangle$ are established. In this regard an
important point made by Brodier {\it et al.} is that the condition for
including a $r$:$s$ resonance into the coupling path is not directly related
to the size of the resonance islands - a crucial point that emerges only upon
investigating the analytic continuation of the classical dynamics to complex
phase space\cite{bsu02}. 
Moreover, provided the observations are general enough,
the arguments could
possibly shed some light on the issue of the competition between
low and high order anharmonic resonances 
for IVR in large molecules\cite{pg98,lw96jcp}.

\subsection{RAT precedes CAT?}
\label{catrat}

Up until now the two processes of CAT and RAT have been discussed
separately. Numerous model studies showed that the nonlinear resonances
are important for describing the dynamical tunneling process in the
near-integrable regimes. At the same time specific statistical signatures
could be associated with the influence of chaos on dynamical tunneling. One
sticky issue still remained - the mechanism of the coupling between
a regular state and the chaotic sea. 
As noted earlier, the assumption of a structurless chaos-regular border
and hence their unimportance for CAT process
typically gets violated for sufficiently small $\hbar$.
Many authors\cite{btu93,df98} suspected that the
nonlinear resonances could play an important role in the mixed phase space
scenario as well. However the precise role of the nonlinear resonances
in determining the regular-chaotic coupling was not clear. 
An important first step in this direction was recently taken by
Eltschka and Schlagheck\cite{es05}. 
By extending the resonance-assisted mechanism
Eltschka and Schlagheck argue that the average splittings {\it i.e.,} ignoring
the fluctuations associated with multistate avoided crosings, can
be determined without any adjustable parameters.

\begin{figure}[t]
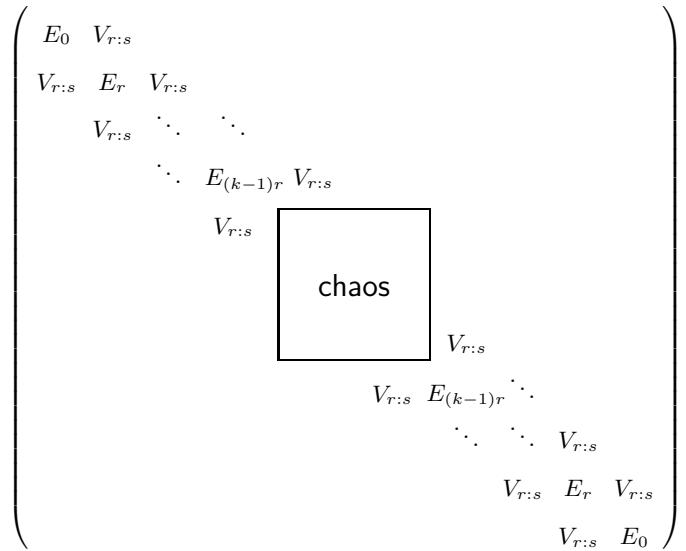

  \begin{displaymath}
    \newlength{\Vrs}
    \settowidth{\Vrs}{$V_{r:s}$}
    \newlength{\csize}
    \setlength{\csize}{5.5em}
    \left(
      \begin{array}{cccclcccc}
        E_0 & V_{r:s} & \phantom{\ddots}\\
        V_{r:s} & E_r & V_{r:s} & \phantom{\ddots}\\
        & V_{r:s} & \ddots & \ddots \\
        & & \ddots & E_{(k-1)r} \hspace*{-0.3cm}
        & \hspace*{0.2cm} V_{r:s} \\[2mm]
        & & & \parbox[c][\csize][t]{\Vrs}{$V_{r:s}$} &
        \framebox{\parbox[c][\csize][c]{\csize}{\centering\large{\sf chaos}}}
        & \parbox[l][\csize][b]{\Vrs}{$V_{r:s}$}\\
        & & & & \parbox{\csize}{\hfill $V_{r:s}$} &
        \hspace*{-0.3cm} E_{(k-1)r} \hspace*{-0.3cm} & \ddots\\
        & & & & & \ddots & \ddots & V_{r:s}\\
        & & & & & \phantom{\ddots} & V_{r:s} & E_r & V_{r:s}\\
        & & & & & & \phantom{\ddots} & V_{r:s} & E_0
      \end{array}
    \right)
  \end{displaymath}
  \caption{Sketch of the effective Hamiltonian matrix that describes tunneling
    between the symmetric quasi-modes in the two separate regular islands.
    The regular parts (upper left and lower right band) includes only
    components that are coupled to the island's ground state by the $r$:$s$
    resonance.
    The chaotic part (central square) consists of a full sub-block with
    equally strong couplings between all basis states with actions beyond the
    outermost invariant torus of the islands. 
    Reproduced from Schlagheck {\it et al.}\cite{seu05}.
    \label{fig10}
  }
\end{figure}

\begin{figure} [htbp]
\begin{center}
\includegraphics[width=80mm]{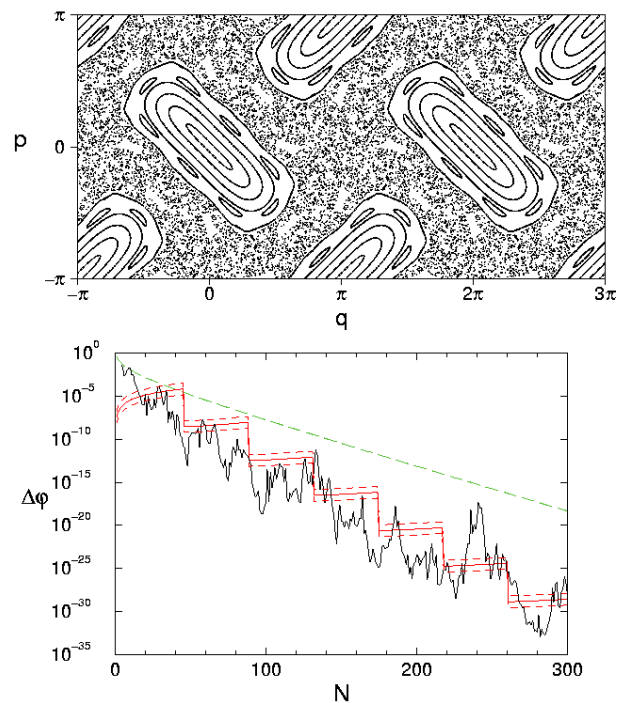}
\caption{(Colour online) A typical mixed regular-chaotic 
phase space obtained for
the kicked Harper map with $\tau=2$ is shown in the upper panel. 
A prominent $8$:$2$ nonlinear resonance is visible.
The lower panel shows the quantum eigenphase splittings calculated
for the ground state within the regular island as a function of
$N \equiv 2\pi/\hbar$. The thick solid line (red)
is the semiclassical prediction
based on the $8$:$2$ resonance for the average splittings along with the
logarithimic standard deviations as dashed lines (red). The long dashed
curve (green) is the Podolskiy-Narimanov\cite{pn03} 
prediction for the same system. Figure
courtesy of P. Schlagheck.}
\label{fig11}
\end{center}
\end{figure}

In order to illustrate the idea note that
in the mixed regular-chaotic regime invariant tori corresponding to the
regular region are embedded into the chaotic sea. A typical case is
shown in Fig.~\ref{fig11} for the kicked Harper map with $\tau=2$.
Therefore only a finite number of invariant tori are supported by
the regular island with the outermost torus corresponding to the
action $I_{c}$. Assuming the existence of
a dominant $r$:$s$ nonlinear resonance implies that the ground state in
the regular region $|0\rangle$
is coupled efficiently to the unperturbed state $|kr\rangle$. 
If $|kr \rangle$ is located beyond the outermost invariant torus of the island
{\it i.e.,} $I_{(k-1)r} < I_{c} < I_{kr}$ then one expects 
a significant coupling
to the unperturbed states located in the chaotic sea. 
Within this scheme the effective Hamiltonian can be thought of as having
the form shown in Fig.~\ref{fig10}.

The effective coupling between the ground state and the chaos obtained
by eliminating the intermediate states within the regular island is
given by
\begin{equation}
V_{\rm eff} = V_{r:s} \prod_{l=1}^{k-1} \frac{V_{r:s}}{E_{0}-E_{lr}}
\label{cateffcoup}
\end{equation}
and follows from the near-integrable resonance-assisted theory.
The chaotic part of the effective Hamiltonian 
is modeled\cite{es05} by a random hermitian
matrix from the Gaussian orthogonal ensemble.
In particular, assuming the validity of the truncated Cauchy distribution
for the splittings, the mean eigenphase splittings for periodically
driven systems assumes the simple form
\begin{equation}
\langle \Delta \phi_{0} \rangle = 
\left(\frac{\tau V_{\rm eff}}{\hbar}\right)^{2}
\label{msplit}
\end{equation}
with a universal prediction for the logarithmic variance of the splittings
\begin{equation}
\left \langle [\ln(\Delta \phi_{0})-\langle \ln(\Delta \phi_{0})\rangle]^{2}
\right \rangle = \frac{\pi^{2}}{4}
\end{equation}
In other words the actual splittings might be enhanced or suppressed as
compared to the mean by about $\exp(\pi/2)$ independent of
the value of $\hbar$ and other external parameters.
In Fig.~\ref{fig11} the expression for the 
mean splitting in Eq.~(\ref{msplit})
is compared to the exact quantum results in the case of the kicked Harper
at $\tau=2$. One immediately observes that the agreement is relatively good
and the approximate plateau like structures are also reproduced.
The plateaus are related to the number of perturbative steps
in Eq.~(\ref{cateffcoup})
that are required to connect the ground state with the chaotic domain.
In order to understand the plateaus note that increasing $N$ in
Fig.~\ref{fig11} corresponds to decreasing $\hbar$ since the state
is fixed. As $\hbar$ decreases there comes a point when the state
$|kr\rangle$ localizes on, rather than being located beyond,  
the outermost invariant torus of the
regular island. Thus the coupling to the chaotic domain via
Eq.~(\ref{cateffcoup}) now requires $(k+1)$ steps rather than $k$ steps.

The mechanism for CAT just outlined has been confirmed for a variety of
systems. Eltschka and Schlagheck confirmed that the plateau structures
appear for the kicked rotor case as well. Wimberger {\it et al.} very
recently showed\cite{wseb06}
that the decay of the nondispersive electronic wavepackets
in driven ionizing Rydberg systems is also controlled by the RAT mechanism.
Mouchet {\it et al.} performed a fairly detailed analysis of an effective
Hamiltonian that can be realized in experiments on cold atoms and provide
clear evidence for the important role played by the various 
nonlinear resonances\cite{mes06}. 
A extensive study\cite{sfgr06}, by Sheinman {\it et al.},
of the lifetimes associated with the
decay of the quantum accelerator modes revealed regimes wherein
the RAT mechanism is crucial.
In this context it is interesting to ask wether the mechanism would also
explain the CAT process
in the Bohigas-Tomsovic-Ullmo system described by the Hamiltonian in
Eq.~(\ref{btuham}). 
It is evident from 
the inset in Fig.~\ref{fig8}b that nonlinear resonances do show up around
the chaos-regular border but a systematic study of the original system
from the recent viewpoints is yet to be done. Similar questions are
being addressed\cite{ss07} 
in the Lin-Ballentine system (cf. Eq.~(\ref{lbham})) {\it i.e.,} to what extent
is the decay of a coherent state localized in the center of the regular island
controlled by the RAT mechanism.
Naturally, in cases wherein the regular islands are ``resonance-free'' the
RAT mechanism is absent. A recent example comes from the work of
Feist {\it et al.} on their studies\cite{fbkrhb06} 
on the conductance through nanowires with
surface disorder which is controlled by dynamical tunneling in a mixed
phase space.

Thus based on the numerous systems studied there are clear indications
that the nonlinear resonances are indeed important to understand the CAT 
mechanism. Nevertheless certain observations reveal that one does not
have a full understanding as of yet. For instance it is clear from
Fig.~\ref{fig11} that the semiclassical theory underestimates the
exact quantum splittings in the regimes of large $\hbar$. This could be due
to the fact that the key nonlinear resonance, on which the estimate is based, 
is not clearly resolved. It could also be due to the fact that within the
pendulum approximation the action dependence of the Fourier coefficients
are neglected. Alternatively it is possible that a {\em different} mechanism
is operative at large $\hbar$ and in particular the Podolskiy-Narimanov
theory might be relevant in this context. However, as evident from the
results reported in the recent work\cite{mes06} by Mouchet {\it et al.}, the
Podolskiy-Narimanov estimate tends to overestimate in certain systems.
In this context note that Sheinman has 
recently derived\cite{sheinthesis}, using assumptions identical
to those of Podolskiy and Narimanov, an expression
for $\Delta E$ which differs from the result in Eq.~(\ref{podnari}).
Moreover the assumption of neglecting the action dependencies of
the Fourier coefficients in Eq.~(\ref{tavcoup}) needs to be carefully studied
especially in cases wherein the nonlinear resonances are rather large.

A more crucial issue pertains to the role played by the partial barriers like
cantori at the regular-chaotic border. At present the RAT based mechanism
still assumes the border to be devoid of such partial barriers and
thus models the chaotic part by a random hermitian matrix. 
As briefly mentioned in the introduction the partial barriers do play
an important role in the molecular systems as well. In this context
the studies\cite{bw861,bw862} by Brown and Wyatt 
on the driven Morse oscillator are
highly relevant. 
The driven Morse system, a model for the dissociation dynamics of
a diatomic molecule, clearly points to the crucial
role of the cantori 
in the quantum dynamics\cite{bw861,bw862}. Interestingly
the dissociation dynamics of the
driven Morse system can be analysed from a dynamical tunneling
perspective to understand the interplay of nonlinear resonances, chaos,
and cantori in the underlying phase space. Investigations along these
lines are in progress.
Currently there is no theory which combines the
RAT viewpoint with the effect of the hierarchical structures located in the
border regions of the phase space\cite{psprivate}.

\section{Consequences for IVR}
\label{ivrnew}

Notwithstanding the issues raised in the previous sections, 
the key point in the context of
this review has to do with the relevance of the RAT and CAT phenomena
to IVR. Heller in his stimulating review\cite{Hel95} 
raised several questions regarding
the manifestation of dynamical tunneling in molecular spectra.
It is thus interesting to ask as to what extent has the recent progress
contributed towards our understanding of quantum mechanisms for IVR. 
This section therefore summarizes the recent progress from
the molecular standpoint.

The Hamiltonians used in this section are effective spectroscopic
Hamiltonians of the form given in Eq.~(\ref{specham}). The origin
and utility of the effective Hamiltonians have already been discussed
in the introduction. Three reasons are worth stating at this juncture.
Firstly, the classical limit of the spectroscopic Hamiltonians are
very easily constructed. Secondly, the various nonlinear resonances
are specified and hence it is possible to perform a careful analysis
of the role of the resonances and chaos in dynamical tunneling. A third
reason has to with the fact that effective Hamiltonians of similar
form arise in different contexts like electron transfer through molecular
bridges\cite{Nitz01}, 
coupled Bose-Einstein condensates, 
and discrete quantum breathers\cite{ffo01,df02,pf06}.
Interestingly the notion of local modes has close connections with the
existence of discrete breathers or intrinsic localized modes which have
been, and continue to be, 
studied by a number of researchers\cite{Fle03,cfk04}. Wether
the intrinsic localized modes can exist quantum mechanically is nothing
but the issue of dynamical tunneling in such discrete systems.
Therefore one anticipates that techniques of the previous sections should
be useful in studying the breather-breather interactions as well. Indeed
a recent work by Dorignac {\it et al.} invokes\cite{dess04} 
the appropriate off-resonant
states to couple the degenerate states.

\subsection{Local mode doublets}

To illustrate the RAT and CAT phenomena recent studies
focus on the effective local mode Hamiltonian\cite{Bag88}
\begin{equation}
H = H_{0} + g' V_{1:1}^{(12)} + \gamma V_{2:2}^{(12)} + \frac{\beta}{2\sqrt{2}}
(V_{2:1}^{(1b)} + V_{2:1}^{(2b)})
\label{bagham}
\end{equation}
with $g' = g+\lambda'(n_{1}+n_{2}+1)+\lambda''n_{b}$,
describing the highly excited vibrational states of a ABA triatomic. 
In the above the local mode stretches are denoted by $(1,2)$ and the
bending mode is denoted by $b$.
As stated previously such systems were one of the first ones
to highlight the manifestation of dynamical tunneling in molecular spectra.
Here, however, the bending mode is included and thus allows for several
anharmonic resonances and a mixed phase space.
The zeroth-order anharmonic Hamiltonian is
\begin{eqnarray}
H_{0}&=&\omega_{s}(n_{1}+n_{2}) + \omega_{b} n_{b} + x_{s}(n_{1}^{2}+n_{2}^{2})
+ x_{b} n_{b}^{2} \nonumber \\
&+& x_{sb} n_{b}(n_{1}+n_{2}) + x_{ss} n_{1} n_{2}
\end{eqnarray}
with $n_{j} = a_{j}^{\dagger}a_{j}$ being the occupation number of the
$j$-th mode written in terms of the harmonic creation $(a_{j}^{\dagger})$
and the destruction $(a_{j})$ operators.
The zeroth-order states $|n_{1},n_{2},n_{b}\rangle$ are coupled by
the anharmonic resonances of the form
\begin{equation}
V_{p:q}^{(ij)} = (a_{i}^{\dagger})^{q} (a_{j})^{p} + (a_{j}^{\dagger})^{p}
(a_{i})^{q}
\label{respert}
\end{equation}
Thus $V_{p:q}^{(ij)}$ couples states $|{\bf n}\rangle$ and $|{\bf n}'\rangle$
with $|n_{i}'-n_{i}|=q$ and $|n_{j}'-n_{j}|=p$.
Due to the structure of the Hamiltonian the polyad number $P=(n_{1}+n_{2})+
n_{b}/2$ is exactly conserved and thus the system has effectively two
degrees of freedom.

\begin{table}
\caption{Local mode splittings in cm$^{-1}$ for H$_{2}$O}
\label{table1}
\begin{ruledtabular}
\begin{tabular}{ccc}
State\footnote{Symmetrized local mode $(nm^{\pm})v_{b}$ labels}
& $\Delta E_{expt}$\footnote{Taken from ref.~\onlinecite{tenny}} 
& $\Delta E_{b}$\footnote{Calculated using the Baggott Hamiltonian.} \\ \colrule
$(10^{\pm})0$ & 98.876  & 100.26 \\
$(20^{\pm})0$ & 48.278  & 48.407 \\
$(30^{\pm})0$ & 13.669  & 12.919 \\
$(40^{\pm})0$ & 2.661   & 2.334 \\
$(50^{\pm})0$ & 0.442   & 0.514 \\
$(60^{\pm})0$ & 0.105   & 0.134 \\
$(70^{\pm})0$ & 0.145   & 0.013 \\
\end{tabular}
\end{ruledtabular}
\end{table}

The Hamiltonian has been proposed by Baggott\cite{Bag88} to fit the vibrational
states of H$_{2}$O and D$_{2}$O for specific choice of the parameters of $H$.
A comparison of the local mode splittings predicted using the Baggott
Hamiltonian and the recent experimental data\cite{tenny} 
of Tennyson {\it et al.}
is shown in the table ~\ref{table1}. The agreement is quite good considering
the fact that the polyad $P$ perhaps breaks down at higher levels of
excitations. Another reason for the choice of the
system had to do with the fact that the 
classical-quantum correspondence for the Hamiltonian
in Eq.~(\ref{bagham}) was already well established 
from several earlier works\cite{ke95,ke97,lk97}.
Additional motivation has to do with the possibility of 
studying the RAT and the
CAT mechanisms in a controlled fashion by analysing the following 
sub-Hamiltonians\cite{Kes05}:
\begin{eqnarray}
H_{A}&=&H_{0} + \frac{\beta}{2\sqrt{2}}(V_{2:1}^{(1b)}+V_{2:1}^{(2b)}) \\
H_{B}&=&H_{A} + \gamma V_{2:2}^{(12)}
\end{eqnarray}
The Hamiltonian $H_{A}$ for smaller polyads exhibits near-integrable
dynamics. The same is true for $H_{B}$ but with an additional weak
$2$:$2$ resonance and hence
it is possible to study the influence of multiple nonlinear
resonances. On the
other hand the full Hamiltonian $H$ has a mixed regular-chaotic phase space
and thus on going from $H_{A} \rightarrow H_{B} \rightarrow H$ 
a systematic study of the influence of phase space structures on
dynamical tunneling is possible. 
The classical limit can be obtained using the 
correspondence
\begin{equation}
a_{j}^{\dagger} \leftrightarrow \sqrt{I_{j}} \exp(i\theta_{j}) \,\,\,\,;
\,\,\,\, a_{j} \leftrightarrow \sqrt{I_{j}} \exp(-i\theta_{j})
\end{equation}
and yields a nonlinear multiresonant Hamiltonian in terms of the
action-angle variables of the zeroth-order $H_{0}$.
For instance the resonant perturbation Eq.~(\ref{respert}) has the
classical limit
\begin{equation}
V_{p:q}^{(ij)} \rightarrow V_{p:q}({\bf I},{\bm \theta}) = 
2(I_{i}^{q} I_{j}^{p})^{1/2} \cos(q\theta_{i}-p\theta_{j})
\end{equation}
with the obvious correspondence ${\bf I} \rightarrow ({\bf n}+1/2)\hbar$.

\begin{figure} [htbp]
\begin{center}
\includegraphics[width=80mm]{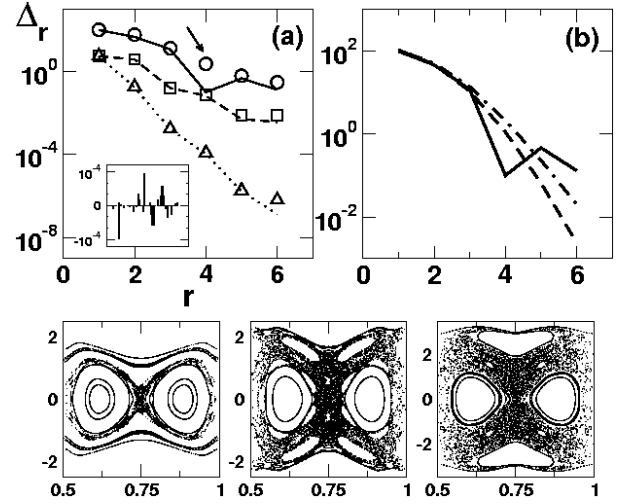}
\caption{(a) Local mode doublet splitting $\Delta_{r}$ 
in cm$^{-1}$ with increasing
excitation quantum $r$ for polyad $P=6$. The exact quantum (lines)
are compared to the vibrational superexchange (symbols) at minimal
order. Note that the agreement is excellent except for the $r=4$ case
for the full Hamiltonian (solid line).
Inset shows the contribution from $130$ different coupling chains towards
obtaining $\Delta_{6}$ in case A. (b) $\Delta_{r}$ estimated from the
$1$:$1$ nonlinear resonance alone (dashed) 
and the $1$:$1$+$2$:$2$ resonances
(dot-dashed) compared with the full case (solid line). Note that the
Hamiltonian with $1$:$1$ and $1$:$1$+$2$:$2$ are classically integrable.
The three lower panels show the phase space $(K_{1}/2,\phi_{1}/4\pi)$
at increasing
energies corresponding
to $r=2,3$, and $4$ for Eq.~(\ref{bagham}). Significant amount of chaos
and the $2$:$1$ stretch-bend nonlinear resonances set in for $r=4$.}
\label{fig12}
\end{center}
\end{figure}

The zeroth-order states $|n_{1}=r,n_{2}=0,n_{b}=2(P-r)\rangle \equiv
|r,0\rangle$ and $|0,r\rangle$ are degenerate in the absence of the
resonant couplings. However, in the presence of the couplings the
degeneracy is broken resulting in the local mode doublets with
a splitting $\Delta_{r}$. In other words if an initial state $|r,0\rangle$
is prepared then on a time scale of $T=2\pi/\Delta_{r}$ the population
is completely trasferred to the symmetric partner state $|0,r\rangle$.
This corresponds to a dynamical tunneling process since there are no
energetic barriers to be blamed in this case\cite{Kes05}. 
Quantum mechanically for a given polyad $P$
the splittings can be obtained easily by diagonalizing a 
$N \times N$ matrix with $N=(P+2)(P+3)/2$. The superexchange calculation
in this multiresonant case is more involved than in the single resonance
case. One important consequence of the multiresonant couplings is that
the number of coupling chains connecting $|r,0\rangle$ and
$|0,r\rangle$ in the state space for a given $P$ is infinite. 
It turns out\cite{Kes03} that the minimal
order estimate
\begin{equation}
\Delta_{r} = \sum_{abc} \beta^{a} g'^{b} \gamma^{c} \sum_{\mu} 
\Delta_{r}(\Gamma_{abc}^{(\mu)})
\label{minimal}
\end{equation}
provides a very good approximation to the exact quantum splittings. 
The index $\mu$ labels all possible coupling chains $\Gamma_{abc}$ for
a specific choice of $a,b$, and $c$ satisfying the constraint
$a+2b+4c=2r$. The contribution to the overall splitting from a specific
coupling chain, which has the form as in Eq.~(\ref{coup1dchain}), is
denoted by $\Delta_{r}(\Gamma_{abc}^{(\mu)})$. The expression in
Eq.~(\ref{minimal}) is nothing but the lowest order of perturbation that
would connect the two degenerate states involving a net change
$\Delta {\bf n} \equiv |{\bf n}-{\bf n}'|=2r$.
The excellent agreement between 
the exact quantum and the superexchange calculation for $P=6$
is shown in Fig.~\ref{fig12}a for the actual system and the various 
subsystems\cite{Kes05}. 
Note that in case of $H_{A}$ a superexchange calculation
of $\Delta_{6}$ involves twelfth-order perturbation with $130$ coupling
chains of varying signs, some as large as $10^{-4}$, conspiring to
yield a fairly accurate value. In Fig.~\ref{fig12}b the splittings
are calculated by retaining only 
the $1$:$1$ resonance or the $1$:$1$+$2$:$2$
resonances. The $\Delta_{r}$ monotonically decrease in both cases with
increasing $r$ {\it i.e.,} OH-stretch excitations as observed earlier.
However starting with $r=4$ the integrable subsystem cannot account
for the exact splittings. In particular the exact case exhibits nonmonotonic
behaviour which neither the integrable limit cases nor the superexchange
calculation can predict\cite{Kes03,Kes05}. 
On the other hand the
monotonic trend seen in Fig.~\ref{fig12}a 
for the near-integrable subsystems is not true in general. For instance
in an earlier work\cite{Kes03} strong fluctuations 
were seen in the near-integrable
case for $P=8$. It is well known that such fluctuations have to with the
avoided crossings of the doublets with one or more states. The avoided
crossings can occur in near-integrable (see Fig.~\ref{fig9} for example)
as well as in the mixed phase space
situations. 
The phase space for the full Baggott Hamiltonian is
shown in Fig.~\ref{fig12} with increasing energy. The surface
of sections are shown in the canonically conjugate\cite{ke97}
variables $(K_{1}/2,\phi_{1}) \equiv
((I_{1}-I_{2})/2,(\theta_{1}-\theta_{2})/2)$ at fixed sectioning
angle $\phi=\theta_{1}+\theta_{2}-4\theta_{b}$ with $\dot{\phi} > 0$.
The normal mode regions appear around $I_{1} \sim I_{2}$.
The appearance of significant amount of chaos and the $2$:$1$ stretch-bend
resonances in the phase space at higher energies
suggests that the RAT and, to some extent CAT, mechanisms could explain
the various results.

\begin{figure} [htbp]
\begin{center}
\includegraphics[width=80mm]{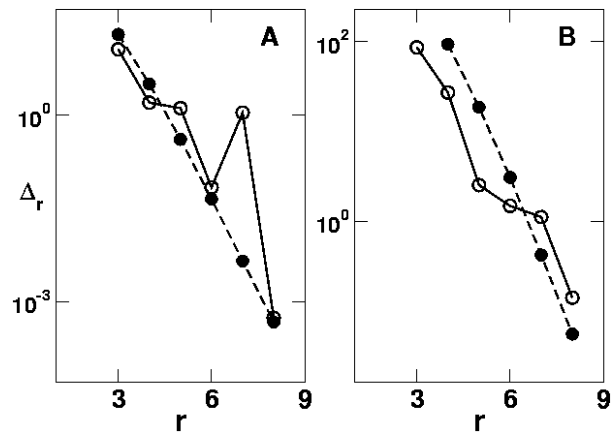}
\caption{Local mode splittings in cm$^{-1}$ for the 
polyad $P=8$ for (A) H$_{2}$O
and (B) D$_{2}$O using the appropriate parameters in the full Hamiltonian
of Eq.~(\ref{bagham}). The open circles denote the exact quantum values
whereas the filled circles represent a semiclassical calculation using
an integrable approximation. Note that the integrable approximation
decays monotonically in both cases and fails to account for the
fluctuations. The phase space is mixed regular-chaotic for both the
molecules but the fluctutations in $\Delta_{r}$ are muted in D$_{2}$O.}
\label{fig13}
\end{center}
\end{figure}

As another example, in Fig.~\ref{fig13} the local mode splittings
computed using Eq.~(\ref{bagham}) for parameters appropriate
to H$_{2}$O and D$_{2}$O and a higher polyad $P=8$ are shown.
It is clear that the splittings for a given local mode
state are typically larger for D$_{2}$O as compared to H$_{2}$O.
Note that this immediately establishes the dynamical nature of the
tunneling since usual arguments of isotopic substitution invoked for
the potential barrier tunneling cases would predict the opposite.
This, of course, is not surprising since the dynamical tunneling
involves transfer of vibrational quanta as opposed to the transfer
of atoms. In addition the enhanced splittings for D$_{2}$O is 
consistent with the fact that H$_{2}$O is a better local mode
system than D$_{2}$O.
The phase space for both systems are mixed regular-chaotic and
similar to that shown in Fig.~\ref{fig12}. Does the nonintegrability
of the system dynamics, and hence the chaos, have any impact on the
local mode splittings? In order to address the question 
the splittings are computed for an integrable approximation
to the Baggott Hamiltonian. The splittings computed using the
integrable approximation are compared to the exact splittings in
Fig.~\ref{fig13} for both the molecules. 
The integrable limit splittings, as expected, decay monotonically with
increasing excitation. Although the comparison looks favorable, except
for the nonmonotonicity, for H$_{2}$O it is clear from Fig.~\ref{fig13}
that similar levels of agreement are not seen for D$_{2}$O.
More importantly, the fluctuations in $\Delta_{r}$ are completely
missed and thus suggest that the exact splittings are
determined by a subtle interplay of the nonlinear resonances and chaos.

\begin{figure} [htbp]
\begin{center}
\includegraphics[width=80mm]{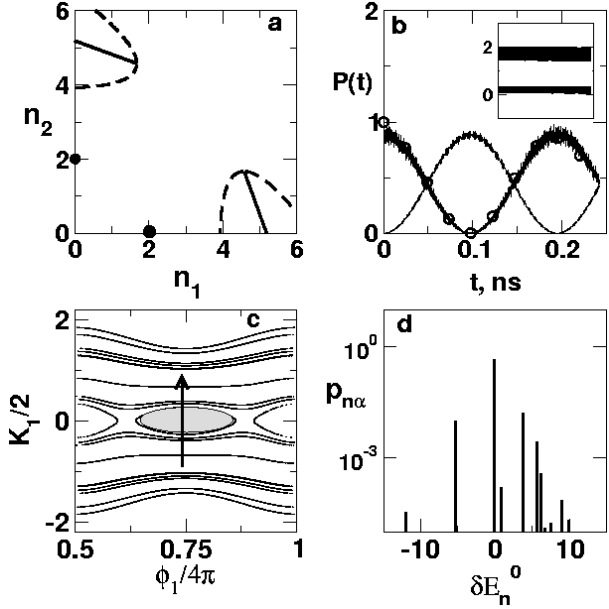}
\caption{(a) State space $(n_{1},n_{2})$ 
for $P=6$ with the location of the 
$2$:$1$ stretch-bend nonlinear resonance zones (dashed regions) 
shown for reference. (b) Quantum
survival probability for the state $|2,0\rangle$ (circles) computed using
$H_{A}$ shows two-level like mixing with 
the symmetric partner $|0,2\rangle$
despite trapping
of classical trajectory on the same timescale (inset). Thus the quantum energy
transfer is due to dynamical tunneling. (c) RAT mechanism due to the
$1$:$1$ nonlinear resonance (shaded) in the phase space induced by the
remote $2$:$1$s explains the energy flow. (d) Overlap intensities
versus $\delta E_{\bf n}^{0} \equiv (E_{\alpha}-E_{\bf n}^{0})$ (in
units of mean level spacing)
shows involvement of off-resonant states.}
\label{fig14}
\end{center}
\end{figure}

To illustrate RAT consider the mixing between the zeroth-order states
$|2,0\rangle$ and $|0,2\rangle$ due to the stretch-bend resonance in
$H_{A}$. The states are not coupled directly by the resonances as can be
inferred by the state space shown in Fig.~\ref{fig14}a. The resonance
zones, computed using the Chirikov approximation\cite{Chi79}, 
are quite far from the
states. Indeed Fig.~\ref{fig14}b shows that the
classical dynamics exhibits trapping, 
however the quantum survival probability of $|2,0\rangle$ shows
complete mixing with $|0,2\rangle$ on a time scale of about $0.1$ nanoseconds.
As before off-resonant states participate in vibrational superexchange, 
clear from Fig.~\ref{fig14}d, resulting in the observed mixing. There
are now several coupling chains at minimal order and the perturbative
calculation yields $\Delta_{2} \approx 0.18$ cm$^{-1}$ in excellent
agreement with the exact value (cf. Fig.~\ref{fig12}). 
What is the role of the stretch-bend resonances? This becomes clear on looking
at the phase space for $H_{A}$ at $E=E_{2,0,8}^{0}$ shown in
Fig.~\ref{fig14}c. One immediately notices a nonlinear resonance
located between the phase space regions corresponding to the two states.
The resonance however is not the $2$:$1$ stretch-bend resonance but
a weak $1$:$1$ resonance between the local stretches induced by the 
stretch-bend couplings. 
It can be shown\cite{Kes05} using classical perturbation theory\cite{llbook}
that the induced $1$:$1$ is governed by a pendulum Hamiltonian of the form
\begin{equation}
H(K_{1},\phi_{1};K_{2},N) = \frac{K_{1}^{2}}{2M_{11}} + 
2 g_{\rm ind}(K_{2},N) \cos(2\phi_{1})
\label{effpendham}
\end{equation}
with $K_{2} \equiv I_{1}+I_{2}$ ($1$:$1$ subpolyad)
and $N \equiv (I_{1}+I_{2})+I_{b}/2$ (classical analog of the polyad $P$)
being the
approximately and exactly conserved quantities. The parameters appearing
in the pendulum Hamiltonian can be expressed in terms of the parameters
of the original Hamiltonian as\cite{Kes05}
\begin{subequations}
\begin{eqnarray}
M_{11} &=& 2(\alpha_{12}-2\alpha_{ss})^{-1} \\
g_{\rm ind}&=&\frac{\beta^{2}}{4} K_{2} (N-2K_{2}) f(K_{2},N) \\
f(K_{2},N) &=& \left\{
\frac{4(\Omega_{s}+\alpha_{ss}K_{2})+\alpha_{12}N}
{[2(\Omega_{s}+\alpha_{ss}K_{2})+\alpha_{12}K_{2}]^{2}}\right\}
\label{pendparams}
\end{eqnarray}
\end{subequations}
where the parameters $\Omega_{s}, \alpha_{ss}$, and $\alpha_{12}$ are
related to the original parameters of Eq.~(\ref{bagham}).
The induced coupling strength\cite{typojcp05} 
is related to the width of the resonance
island seen in Fig.~\ref{fig14}c. Thus having idenitified the key
resonance the splitting between a general local mode pair can be
estimated as
\begin{equation}
\frac{\Delta_{r}^{\rm sc}}{2} = g_{\rm ind} \prod_{k=-(r-2)}^{(r-2)}
\frac{g_{\rm ind}}{E_{r}^{0}-E_{k}^{0}} = \frac{g_{\rm ind}}{[(r-1)!]^{2}}
\left(\frac{M_{11} g_{\rm ind}}{2}\right)^{r-1}
\label{scsplit}
\end{equation}
with $E_{j}^{0} = j^{2}/2M_{11}$ being the unperturbed energy
from Eq.~(\ref{effpendham}). Using parameters appropriate for
H$_{2}$O in Eq.~(\ref{bagham}) and the expressions in Eq.~(\ref{pendparams})
one obtains $M_{11} \approx 1.32 \times 10^{-2}$ and $g_{ind} \approx
3.93$ cm$^{-1}$. Thus $\Delta_{2}^{sc} \approx 0.20$ cm$^{-1}$ which
is in agreement with the exact and superexchange values. This clearly
supports the RAT mechanism for the quantum mixing seen in Fig.~\ref{fig14}b.

\begin{figure} [htbp]
\begin{center}
\includegraphics[width=80mm]{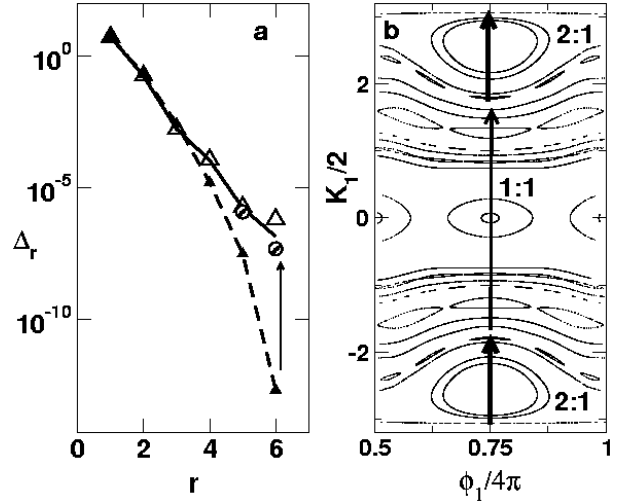}
\caption{(a) Splittings in cm$^{-1}$ for $P=6$ overtone 
states for $H_{A}$. The exact
result (solid line), superexchange calculation (open triangles), and
predictions based on the induced $1$:$1$ stretch-stretch resonance
alone (filed triangles, dashed line) are compared. (b) Phase space
for $H_{A} \approx E_{6,0,0}^{0}$ emphasizes the need for invoking
multiple resonances (arrows). RAT mechanism using the multiple resonances
leads to very good agreement shown in (a) as shaded cricles.}
\label{fig15}
\end{center}
\end{figure}

Although the classical dynamics of $H_{A}$ is near-integrable 
the phase space can be very rich in terms of several resonances. The RAT
mechanism depends on identifying the crucial resonances. In 
Fig.~\ref{fig12} the monotonic decrease of $\Delta_{r}$ with $r$
for the system described by $H_{A}$ and $H_{B}$ might suggest that
the induced resonance alone controls the splittings. However 
Fig.~\ref{fig15}a shows that calculating
the splittings using Eq.~(\ref{effpendham}) yields very poor results for
$r=5,6$. The reason becomes clear on inspecting the phase space
and Fig.~\ref{fig15}b shows the case of $r=6$ as an example. The
Husimi distribution of the state $|6,0,0\rangle$ would be localized
about $K_{1} \approx 6$ and thus the $2$:$1$ stretch-bend and the 
$1$:$1$ induced resonance islands seperate the state from its symmetric
partner $|0,6,0\rangle$ localized around $K_{1} \approx -6$.
Several other resonances are also visible at this 
energy in Fig.~\ref{fig15}b but they are of higher order and perhaps
not relevant for large $\hbar$ values. 
Moreover the Husimi representations indicate\cite{Kes05} 
that the state $|5,0,2\rangle$
is localized inside the $2$:$1$ island (ground state) and the state
$|4,0,4\rangle$ corresponds to the separatrix state. Thus the
$2$:$1$ resonance must be coupling the $|6,0,0\rangle$ and $|4,0,4\rangle$
states. As a consequence the RAT mechanism can be schematically
written as
\begin{equation}
|6,0\rangle \stackrel{\beta_{\rm eff}}{\longrightarrow} |4,0\rangle
\stackrel{g_{\rm ind}} {\longrightarrow} |0,4\rangle \stackrel{\beta_{\rm eff}}
{\longrightarrow} |0,6\rangle
\end{equation}
with $\beta_{\rm eff}$ being the effective coupling across the $2$:$1$ islands
and $g_{\rm ind}$ being the 
effective $1$:$1$ coupling in Eq.~(\ref{pendparams})
for $r=6$. This is illustrated in Fig.~\ref{fig15}b on the phase space
as well. Again approximating the $2$:$1$ 
islands by pendula it is possible\cite{Kes05}
to determine the $\beta_{\rm eff}$ and the computed splittings agree rather
well with the exact results as shown in Fig.~\ref{fig15}a. Thus
Fig.~\ref{fig15} demonstrates the RAT mechanism in that the mixing
between two remote (both in state space and the phase space)
states $|6,0,0\rangle$ and $|0,6,0\rangle$ is characterized by the nonlinear
resonances. Further confirmation of the multiresonant mechanism comes from
the fact that values of $\Delta_{6}$ for $H_{B}$ and the full Hamiltonian
also compare well with the exact results. The subtelty of the mechanism
lies in the fact that all the nonilinear resonances in Fig.~\ref{fig15}b
arise from the two $2$:$1$ resonances and at smaller values of $\hbar$
one expects the higher order resonances to come into play.

\begin{figure} [htbp]
\begin{center}
\includegraphics[width=80mm]{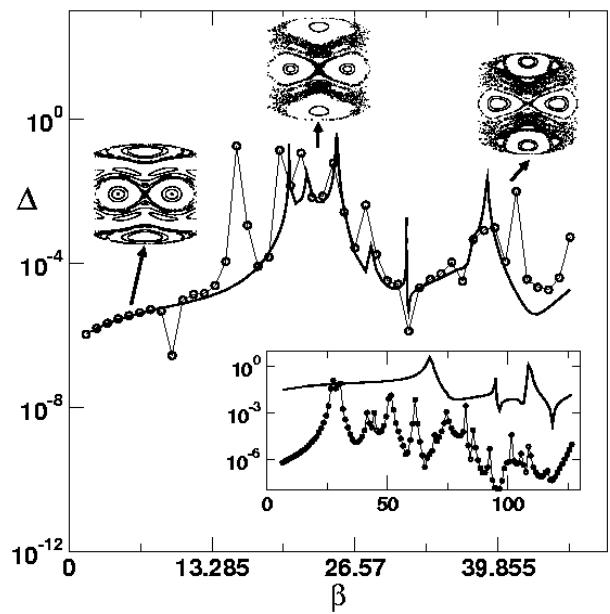}
\caption{Variations in the
splittings for the ground state of the $2$:$1$ island with coupling strength
$\beta$.
(a) Effect of breaking the polyad by adding weak $3$:$1$ resonance
to $H$ with strength $\delta=0.1$ cm$^{-1}$ (circles) contrasted
with the polyad conserved ($P=16$) case (thick line). 
The phase space (with axes as in Fig.~\ref{fig12})
is shown in the latter case for select values of $\beta$.
(Inset) Fluctuations in the splitting for $P=8$ case (thick line) compared
to that with $\hbar$ reduced by half (circles).
All quantities are in units of cm$^{-1}$.} 
\label{fig16}
\end{center}
\end{figure}

The role of RAT in explaining the local mode doublet splittings is obvious
from the preceeding discussion and examples. Model studies
reveal that even very small, on the scale of $\hbar$,
nonlinear resonances mediate dynamical
tunneling in molecular systems\cite{Kes05}. 
However the role of CAT
is far from being established and some of the issues involved were
highlighted before in this review. 
In essence, to implicate CAT atleast
a third chaotic state has to be identified\cite{btu93,tu94}. 
Moreover the system should
be in the deep semiclassical limit {\it i.e.,} small enough $\hbar$ so that
the quantum dynamics can sense the presence of chaos in phase space. These
are fairly severe constraints and experimentally there are not many systems
which respect the constraints. Intutively speaking, highly excited energy
regions of moderate sized molecules might be ideal systems. One possibility
is that the signatures of CAT could be present in
the eigenstates of highly excited molecules and could potentially
complicate the state assignments. One such example can be found in
the work\cite{ke95} by Keshavamurthy and Ezra on the dynamical assignments
of the highly excited states of H$_{2}$O for $P=8$. 
Another example, perhaps, can be found in the work\cite{jjtw99} 
by Jacobson {\it et al.}
wherein the highly excited ($\sim$ 15000 cm$^{-1}$ above the ground state)
vibrational states of acetylene are
dynamically assigned. In their study Jacobson {\it et al.} found that
approximate assignments could be done in an energy range which had
highly chaotic phase space whereas many of the eigenstates
could not be assigned in another energy range despite the existence
of large regular regions in the phase space.
To some extent an indirect
role of CAT is evident from Fig.~\ref{fig12}b wherein the dominant
integrable resonances cannot account for the exact splittings. Some more
hints originate from Fig.~\ref{fig16} where the fluctuations in the
splittings associated with the ground state of the $2$:$1$ island are
shown in two different situations. The first of these investigates the
effect of breaking the polyad constant $P$ by adding weak $3$:$1$ 
stretch-bend resonances to the Hamiltonian in Eq.~(\ref{bagham}).
Thus the three degrees of freedom Hamiltonian is
\begin{equation}
H_{3d} = H + \delta (V_{3:1}^{(1b)} + V_{3:1}^{(2b)})
\label{3dh2o}
\end{equation}
with $\delta=0.1$ cm$^{-1}$ which is three orders of
magnitude smaller than the mean level
spacing of $H$. 
A consequence of $\delta \neq 0$ is that the effective density of states
is increased providing many more states with different
$P$ to interact with the state
of interest.
In Fig.~\ref{fig16} the effect of such
a small perturbation is shown for $P=16$ and clearly there are far more 
fluctuations in the mixed regime. However away from
the fluctuations the inverse participation ratio
of the state is nearly one in the basis of the two degrees of freedom
Hamiltonian $H$ which suggests minor influence of states with $P \neq 16$.
Despite this and the weak coupling strength $\delta$ there are significant
differences between the computed splittings for $H_{3d}$ and $H$. 
Presumably this has to do with the subtle changes in the nature of the
chaotic states upon adding the $3$:$1$ resonances but further studies
are required in order to gain a better understanding. 
The results of a second system shown in the
inset to Fig.~\ref{fig16} reveal the increased flutuations in the
splittings for the state with $P=8$ on decreasing the value of $\hbar$ by
half. Interestingly in this case the average splitting seems to decrease
with increasing $\beta$ and thus increasing stochasticity in the phase space.

Most of the detailed phase space studies have focused on the role played
by classical phase space structures for vibrational modes alone. The
involvement of rotations, torsions, and other large amplitude modes in
molecular systems and their implications for IVR\cite{kmp00} 
and dynamical tunneling
have received relatively less attention. 
Early pioneering studies\cite{hp84} by Harter and Patterson 
showed that the splitting of the doubly
degenerate $K$-levels of a symmetric top on
the introduction of asymmetry could be associated with tunneling across
a separatrix on the so called rotational energy surface.
Would the coupling of vibrations
to these other modes enhance or suppress the splittings?
Harter has extended the rotational energy surfaces idea to
describe and interpret the dynamics and spectra for a class of anharmonic
coriolis coupled vibrational modes\cite{Har86}.
Lehmann studied\cite{Leh91,Leh92} the
coupling of rotations to the local mode vibrations in XH$_{n}$ systems.
In particular Lehmann, and independently Child and Zhu\cite{cz91}, 
considered the modification of the local mode
dynamical tunneling due to the interactions with rotations. Depending on
the local mode tunneling time the rotation motion would
reflect the extent of dynamical symmetry of the local mode state.
Interestingly Lehmann's analysis reveals that it is possible for the rotation
of a molecule to significantly reduce the tunneling rate {\it i.e.,} decrease
the doublet splitting\cite{Leh91}.
Such rotational quenching of the local mode tunneling is due to the fact
that in addition to the transfer of vibrational quanta it is also necessary to
reorient the angular momentum in the body fixed frame. Semiclassical
insights arise by looking at the motion on the rotational energy surfaces.
It is not clear as to how much of a role does the classical chaos plays in
the studies of Harter and Lehmann. However,
the local mode splitting is determined by RAT and CAT mechanisms and hence
they also determine the timescales relative to rotations.
It remains to be seen if the different mechanisms are reflected in
the nature of the rovibrational eigenstates.
A possible signature of CAT was
suggested\cite{Ort96} by Ortigoso nearly a decade
ago. Ortigoso found anomalously large $K$-splittings for some
of the asymmetry doublets occuring in the
rotation-torsion energy levels of acetaldehyde.
Further
studies involving coupling with vibrations have not been undertaken
although, based on time scale separations one can argue that the
observed enhancements might lead to increased vibration-rotation
couplings with important consequences for IVR.

\subsection{State mixing in multidimensions}

Every single example in this section
up until now has involved systems with discrete symmetry
and two degrees of freedom. 
The requirement of discrete symmetries for dynamical tunneling is not
very restrictive. As discussed before in this review this aspect has been
emphasized in the early work as well. Tomsovic\cite{Tom98}
has also discussed the possibility of 
observing CAT in the absence of reflection
symmetries.
On the other hand studies on dynamical tunneling
in systems with large degrees of freedom 
are crucial. 
Given that there are important differences\cite{llbook}, briefly mentioned
in the introduction, 
between the classical dynamics in two and more than two degrees of
freedom it is natural to ask if the RAT and CAT mechanisms hold in general.
Especially from the IVR perspective the crucial question is wether
dynamical tunneling can provide a route for mixing between near-degenerate
states and hence energy flow between phase space regions supporting
qualitatively different types of motion. Thus is it possible that an excited
CH-stretch can, over long times, evolve into a CC-stretch despite the lack
of any obvious coupling? Heller and Davis conjectured\cite{dh812} that 
it should be possible and the agent would be dynamical tunneling. 
However support for the conjecture from a phase space viewpoint have not
been forthcoming.  

\begin{figure} [htbp]
\begin{center}
\includegraphics[width=80mm]{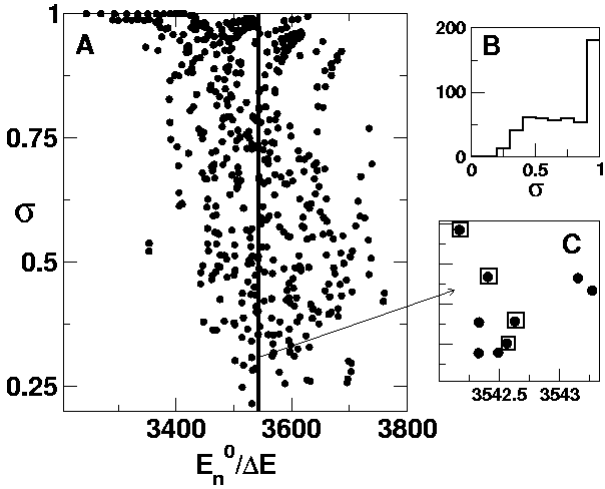}
\caption{(A) Dilution factors $\sigma$ of the
zeroth-order states $|{\bf n}\rangle$
with $H_{0} |{\bf n}\rangle = E_{\bf n}^{0} |{\bf n}\rangle$ versus energy
for the Hamiltonian in Eq.~(\ref{3dham}). 
The polyad value is $P=8$ and couplings
$0.5 K_{{\bf m}_{1}}=K_{{\bf m}_{2}}=K_{{\bf m}_{3}}\approx 0.2 \Delta E$
with mean level spacing $\Delta E \approx 4.4$ cm$^{-1}$. (B) Histogram
showing the distribution of $\sigma$ in (A). Note that a large number of states
have unit dilutions but there are also a number of
states with $\sigma < 1$ indicating mixing of the zeroth-order states. Is the
mixing due to dynamical tunneling? In (C) an expanded view of the states
around $\bar{E} \approx 3542.5 \Delta E$ in (A) is shown. Four states
of interest are highlighted by squares.}
\label{fig17}
\end{center}
\end{figure}

A first step towards such goals was recently taken\cite{Kes051} by studying the
model Hamiltonian 
\begin{equation}
H = H_{0} + \sum_{r} K_{{\bf m}_{r}} \left[(a_{1}^{\dagger})^{\alpha_{r}}
(a_{2})^{\beta_{r}} (a_{3})^{\gamma_{r}} (a_{4})^{\delta_{r}} + h.c.\right]
\label{3dham}
\end{equation}
describing four coupled modes $j=1,2,3,4$ with the zeroth-order part
\begin{equation}
H_{0} = \sum_{j} (\omega_{j} n_{j} + x_{jj} n_{j}^{2}) 
+ \sum_{j<k} x_{jk} n_{j} n_{k}
\end{equation}
The zeroth-order states $|{\bf n}\rangle$, eigenstates of $H_{0}$ with
eigenvalues $E_{\bf n}^{0}$, get coupled and hence mixed due to the
anharmonic resonances chracterized by ${\bf m}_{r} = (\alpha_{r},-\beta_{r},
-\gamma_{r},-\delta_{r})$ with strengths $K_{{\bf m}_{r}}$. 
In order to model a system without any symmetries the parameters of $H_{0}$
were taken from the work\cite{bhmqs00} 
of Beil {\it et al.} corresponding to the
four high frequency modes of the molecule CDBrClF. Three perturbations
${\bf m}_{1}=(1,-2,0,0), 
{\bf m}_{2}=(1,-1,-1,0)$, and ${\bf m}_{3}=(1,-1,0,-1)$ with very
weak strengths $K_{{\bf m}_{r}}/\Delta E \equiv k_{{\bf m}_{r}} < 1$,
relative to the mean level spacing $\Delta E$,
are selected. Note that the coupling structure of the Hamiltonian implies the
existence of a good polyad quantum number $P=n_{1}+(n_{2}+n_{3}+n_{4})/2$.
Thus the model Hamiltonian has effectively three degrees of freedom.

A few words about the choice of the Hamiltonian is perhaps appropriate
at this stage. Firstly, the form of $H_{0}$ is quite generic and the
parameters were chosen from an experimental fit\cite{bhmqs00} in order to
test the ideas on a real molecular system. Secondly the Hamiltonian
in Eq.~(\ref{3dham}) can be considered to be in the intrinsic resonance
representation\cite{chm97}. 
In other words the Hamiltonian models the weak residual
couplings between the KAM tori corresponding to $H_{0}$. Finally the
choice of the resonant couplings is also arbitrary; a different set
of couplings would generate mixings among different zeroth-order states.

\begin{figure} [htbp]
\begin{center}
\includegraphics[width=80mm]{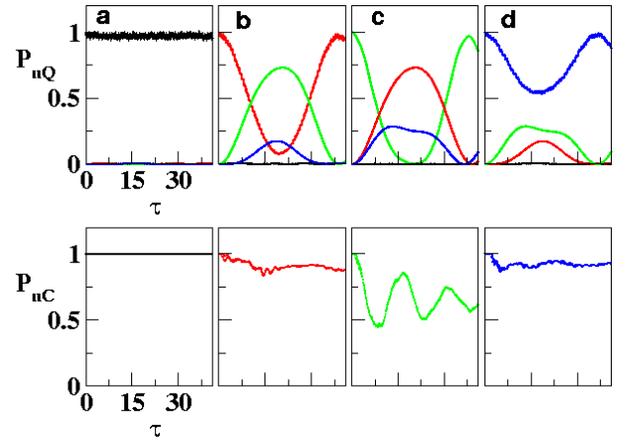}
\caption{(Colour online) Quantum (top panel) and 
classical (bottom panel) survival 
probabilities of the four zeroth-order states shown in Fig.~\ref{fig11}C.
Parameters chosen are as in Fig.~\ref{fig11}. Time $\tau$ is measured
in units of the Heisenberg time $\tau_{H} = (2\pi c \Delta E)^{-1}$.
State $|a\rangle$ (black) shows long time trapping in both cases. States
$|b\rangle$ (red), $|c\rangle$ (green), and $|d\rangle$ (blue) mix amongst
each other quantum meachanically in contrast to their classical behaviour.
The quantum cross probabilities $|\langle {\bf n}'|{\bf n}(t)\rangle|^{2}$ 
are also shown with consistent color coding.
Adapted from ref.~\onlinecite{Kes051}.}
\label{fig18}
\end{center}
\end{figure}

\begin{figure} [htbp]
\begin{center}
\includegraphics[width=80mm]{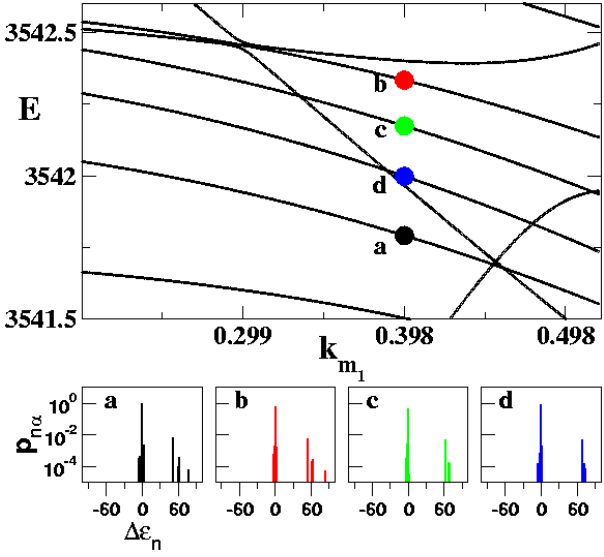}
\caption{(Colour online) The top panel shows variation of the eigenvalues 
$E \equiv E_{\alpha}/\Delta E$ with the coupling parameter $k_{{\bf m}_{1}}$
with the other couplings fixed as in Fig.~\ref{fig11}. Eigenstates having
the largest contribution from the specific zeroth-order states are indicated.
Bottom panel shows the `fractionation' pattern {\it i.e.,} overlap
intensities $p_{{\bf n}\alpha}$ versus $\Delta \epsilon_{\bf n} \equiv
(E_{\bf n}^{0}-E_{\alpha})/\Delta E$ on log scale. The cluster of states
around $\Delta \epsilon_{\bf n} \approx 60$ are the off-resonant states that
mediate the mixings seen in Fig.~\ref{fig17} and Fig.~\ref{fig18}
via a vibrational superexchange mechanism. 
Adapted from ref.~\onlinecite{Kes051}.} 
\label{fig19}
\end{center}
\end{figure}

The relevant issue here has to do with the fate of zeroth-order states
$|{\bf n}\rangle, |{\bf n}'\rangle,\ldots,$ which are near-degenerate
$E_{\bf n}^{0} \approx E_{{\bf n}'}^{0} \approx,\ldots,$
in the presence of such weak perturbations. In Fig.~\ref{fig17}A the
dilution factors for the various zeroth-order states are shown and
it is clear that many zeroth-order states are mixed. In particular within
one mean level spacing several states have varying levels of mixing and
a typical example is shown in Fig.~\ref{fig17}C. Is the mixing observed
in Fig.~\ref{fig17}C due to dynamical tunneling? In order to check four
states are picked and correspond to $|a\rangle=|0,11,1,4\rangle,
|b\rangle=|0,11,2,3\rangle, |c\rangle=|0,12,2,2\rangle$, and
$|d\rangle=|0,13,1,2\rangle$. The choice of the energy range and the states
is fairly arbitrary as evident from Fig.~\ref{fig17}. 
First it can be confirmed by evaluating the quantum survival probability
$P_{{\bf n}Q}$ and its classical analog $P_{{{\bf n}C}}$, shown
in Fig.~\ref{fig18}, that the mixing
of the states of interest is indeed a quantum process {\it i.e.,} classically
forbidden. It is also apparent from Fig.~\ref{fig18} that states $|b\rangle,
|c\rangle$, and $|d\rangle$ mix amongst each other over long time scales.
State $|a\rangle$ on the other hand shows trapping both classically and
quantum mechanically. 
Explanation for the observed mixings in Fig.~\ref{fig17} can be given
in terms of the vibrational superexchange mechanism. 
Indeed the IVR `fractionation'
pattern, shown in Fig.~\ref{fig19}, indicates a a clump of virtual or
off-resonant states about $60 \Delta E$ away from the corresponding
`bright' states. Such fractionation patterns are similar to those
observed in the experiments\cite{cpcesgls03} 
of Callegari {\it et al.} for example.
The off-resonant states have $\sigma \approx 1$ and hence do not 
mix significantly. However it is important to note that the
fractionation pattern looks the same in case of every state and hence
there must be subtle phase cancellation effect to explain the trapping
exhibited by $|a\rangle$. At the same time Fig.~\ref{fig19} also shows that
the vastly different IVR from the four states cannot be obviously related
to any avoided crossing or multistate interactions.

\begin{figure} [htbp]
\begin{center}
\includegraphics[width=80mm]{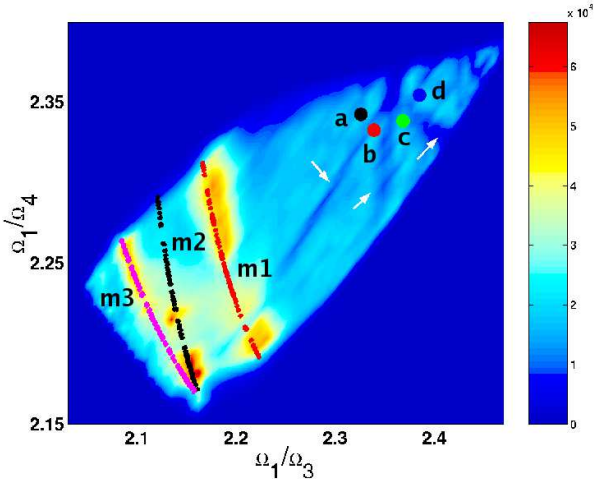}
\caption{(Colour online) The dynamical Arnol'd 
web at $E=\bar{E}$ generated by propagating 25000
classical trajectories for $\tau \approx 40$. The primary resonances
${\bf m}_{1}$ (red), ${\bf m}_{2}$ (black), and ${\bf m}_{3}$ (magenta),
as predicted by ${\cal H}_{0}$ are superimposed for comparison. The
zeroth-order states of interest however are located far away from the
primary resonances and close to the junction of three weak induced
resonances ${\bf m}_{i1}=(0,0,1,-1),{\bf m}_{i2}=(0,1,-1,0)$, and
${\bf m}_{i3}=(0,1,0,-1)$ indicated by arrows. The induced resonances
lead to the observed state mixings in Fig.~\ref{fig17} and Fig.~\ref{fig19}
via the mechanism of RAT. The data for this figure 
were generated by A. Semparithi. Reproduced from ref.~\onlinecite{Kes051}.}
\label{fig20}
\end{center}
\end{figure}

In light of the conjecture of 
Davis and Heller\cite{dh812} and the recent progress
in our understanding of the mechanistic aspects of dynamical tunneling
it is but natural to associate the near-degenerate state mixings with
RAT. However how does one identify the specific resonances at work in
this case? Given that the system has three degrees of freedom 
it is not possible to visualize the Poincar\'{e} surface of section 
at $H \approx \bar{E}$. In the present case the cruicial object to
analyse is the Arnol'd web\cite{llbook} {\it i.e.,} 
the network of nonlinear resonances
and the location of the zeroth-order states therein. Fortunately the
classical limit Hamiltonian
\begin{eqnarray}
{\cal H}({\bf I},{\bm \theta})&=&{\cal H}_{0}({\bf I}) + 2\epsilon
\sum_{r} K_{{\bf m}_{r}} \sqrt{I_{1}^{\alpha_{r}} I_{2}^{\beta_{r}}
I_{3}^{\gamma_{r}} I_{4}^{\delta_{r}}} 
\cos({\bf m}_{r} \cdot {\bm \theta}) \nonumber \\
&\equiv& H_{0}({\bf I}) + \epsilon \sum_{r} K_{{\bf m}_{r}}
f_{r}({\bf I}) \cos({\bf m}_{r} \cdot {\bm \theta}) 
\label{3dhamc}
\end{eqnarray}
from Eq.~(\ref{3dham}) is easily obtained, as 
explained earlier, in terms of the action-angle 
variables $({\bf I},{\bm \theta})$ of $H_{0}$.
In the above equation the variable $\epsilon$ is introduced 
for convenience during a perturbation analysis of the Hamiltonian.
The `static' Arnol'd web at $E \approx \bar{E}$ and fixed polyad
$P_{c} \equiv I_{1}+(I_{2}+I_{3}+I_{4})/2$, classical analog of the
quantum $P$, 
can then be constructed via the intersection of the various resonance
planes ${\bf m}_{r} \cdot \partial {\cal H}_{0}({\bf I})/\partial {\bf I}=0$
with the energy shell ${\cal H}_{0}({\bf I}) \approx \bar{E}$.
The static web involves all the nonlinear resonances restricted to, say,
some maximum order 
$|\alpha_{r}|+|\beta_{r}|+|\gamma_{r}|+|\delta_{r}| \leq M$.
The reason for calling such a construction as static has to do with the
fact that being based on ${\cal H}_{0}$ it is possible that many of the
resonances do not have any dynamical consequence. Thus,
although the static web provides useful information on the location
of the various nonlinear resonances on the energy shell
it is nevertheless critical to determine the `dynamical' Arnol'd web since
one needs to know as to what part of the static web is actually 
relevant to the dynamics. Further discussions on this point can be
found in the paper by Laskar\cite{Las93}
wherein time-frequency analysis is used
to study transport in the four dimensional standard map.
In order to construct the dynamical Arnol'd web it is necessary to be
able to determine the system frequencies as a function of time.
Several techniques\cite{mde87,Las93} have 
been suggested in the literature for 
performing time-frequency analysis of dynamical systems.
An early example comes from the work\cite{mdu96} of
Milczewski, Diercksen, and Uzer wherein 
the Arnol'd web for the Hydrogen atom in crossed electric
and magnetic fields has been computed.
A critical review of the various techniques is clearly outside the scope of
this work and hence, without going into detailed discussion of
the advantages and disadvantages, 
the wavelet based approach\cite{aw01,cwu03} 
developed by Wiggins and coworkers
is utilized for constructing the web. 
There is, however, ample evidence that the wavelet based local frequency
analysis is ideally suited for this purpose and therefore
a brief description of
the method follows.
Classical trajectories with initial conditions
satisfying ${\cal H}({\bf I},{\bm \theta}) \approx \bar{E}$ are generated and
the nonlinear frequencies $\Omega_{k}(t)$ are computed by performing
the continuous wavelet transform of the time series $z_{k}(t)=\sqrt{2I_{k}(t)} 
\exp(i\theta_{k}(t))$: 
\begin{equation}
L_{g}z_{k}(a,b) = a^{-1/2} 
\int_{-\infty}^{\infty} z_{k}(t) g^{*}\left(\frac{t-b}{a}\right) dt
\label{wavelet}
\end{equation}
with $a>0$ and real $b$. Various choices can be made for the mother wavelet
$g(t)$ and in this work it is chosen to be the Morlet-Grossman function
\begin{equation}
g(t) = \frac{1}{\sqrt{2 \pi \sigma^{2}}} \exp\left(2 \pi i \lambda t -
\frac{t^{2}}{2\sigma^{2}}\right)
\end{equation}
and the parameter values $\lambda=1$, $\sigma=2$.
Note that Eq.~(\ref{wavelet}) yields the frequency content of $z_{k}$
within a time window around $t=b$. 
In many instances one is interested in the dominant frequency and hence
the required local frequency is extracted by determining the scale
($a$, inversely proportional to frequency) which maximizes the modulus
of the wavelet transform\cite{aw01} {\it i.e.,}
$\Omega_{k}(t=b)={\rm max}_{a}|L_{g}z_{k}(a,b)|$.
The trajectories are followed in the frequency
ratio space $(\Omega_{1}/\Omega_{3},\Omega_{1}/\Omega_{4})$ since the
energy shell, resonance zones, and the location of the zeroth-order states
can be projected onto the space of two independent frequency ratios. Such
`tune' spaces have been constructed and analysed before, for instance, by
Martens, Davis, and Ezra in their seminal work\cite{mde87} 
on IVR in the OCS molecule.
A density plot is then created by recording the total number of visits
by the trajectories to a given region of the ratio space. Quite naturally
the density plot highlights the dynamically significant portions of the
Arnol'd web at a given energy and polyad. The computation of such a dynamical
web is shown in Fig.~\ref{fig20} at an energy $\bar{E}$ corresponding
to the zeroth-order states of interest. One immediately observes that
apart from highlighting the primary resonances Fig.~\ref{fig20} shows 
the existence of higher order induced resonances. More importantly the
zeroth-order states are located far away from the primary resonances and thus
the state mixings cannot be ascribed to the direct couplings in the
Hamiltonian. However, the states are located in proximity to the junction
formed by the weak induced resonances denoted as ${\bf m}_{i1}, {\bf m}_{i2}$,
and ${\bf m}_{i3}$. The nature of these induced resonances make it very clear
that the state mixings are due to RAT. 

How can one be sure that it is indeed the RAT mechanism that is at work
here? The simplest answer is to try and interfere with the RAT mechanism
by removing or altering\cite{Kes051} the relevant resonances 
as seen in Fig.~\ref{fig20}.
This was precisely what was done in the recent work\cite{Kes051} and
classical perturbation theory is again used to
remove the primary resonances in Eq.~(\ref{3dhamc}) to $O(\epsilon)$. 
Since the methodology is well known only a brief description of the
perturbative analysis follows. The first step involves reduction
of the Hamiltonian in Eq.~(\ref{3dhamc}), noting the exactly
conserved polyad $P_{c}$, to an effective three
degrees of freedom system. This is achieved by performing a canonical
transformation $({\bf I},{\bm \theta}) \rightarrow ({\bf J},{\bm \psi};P_{c})$
via the generating function
\begin{equation}
F({\bf J},{\bm \psi};P_{c}) = \sum_{r} ({\bf m}_{r} \cdot {\bm \theta}) J_{r} 
             + \theta_{1} P_{c}
\end{equation}
The new variables can be obtained in terms of the old variables by using
the generating function properties ${\bm \psi}=\partial_{\bf J}F$
and ${\bf I}=\partial_{\bm \theta}F$. The reduced Hamiltonian takes the
form
\begin{equation}
H({\bf J},{\bm \psi};P_{c}) = H_{0}({\bf J};P_{c}) +
          \epsilon \sum_{r} K_{{\bf m}_{r}} g_{r}({\bf J};P_{c}) \cos \psi_{r}
\end{equation}
with $g_{r}({\bf J};P_{c})$ being determined from the functions
$f_{r}({\bf I})$. The resonant angles are related to the original angle
variables by $\psi_{r}={\bf m}_{r} \cdot {\bm \theta}$.
The dynamical web in Fig.~\ref{fig20} clearly
shows that the zeroth-order states of interest are far away from
the primary resonances. Thus the three primary resonances are removed
to $O(\epsilon)$ by making use of the generating function
\begin{equation}
G = \sum_{r} \psi_{r} \bar{J}_{r} +
\epsilon \sum_{r} d_{r}(\bar{{\bf J}}) \sin \psi_{r}
\end{equation}
where the unknown functions $d_{r}(\bar{{\bf J}})$ are to be chosen such
that the reduced Hamiltonian does not contain the primary resonances
to $O(\epsilon)$. Using the standard generating function relations
one obtains
\begin{eqnarray}
J_{r} &=& \bar{J}_{r} + \epsilon d_{r}(\bar{{\bf J}}) \cos \psi_{r} \\
\bar{\psi}_{r} &=& \psi_{r} + 
\epsilon \sum_{r'} d_{rr'}(\bar{{\bf J}}) \sin \psi_{r'} 
\end{eqnarray}
with $r=1,2,3$ and
$d_{rr'}(\bar{{\bf J}})\equiv \partial d_{r}(\bar{{\bf J}})/\partial 
\bar{J}_{r'}$. Writing the reduced Hamiltonian
in terms of the variables $(\bar{{\bf J}},\bar{{\bm \psi}})$ and
using the identities involving the Bessel functions 
${\mathcal J}_{n}$
\begin{eqnarray}
\cos(\epsilon a \sin b) &=& {\mathcal J}_{0}(\epsilon a) +
            2\sum_{l \geq 1} {\mathcal J}_{2l}(\epsilon a) \cos(2lb) \\
\sin(\epsilon a \sin b) &=& 2\sum_{l \geq 0} {\mathcal J}_{2l+1} 
     \sin((2l+1)b) 
\end{eqnarray}
one determines the choice for the unknown functions to be
\begin{equation}
d_{r}(\bar{{\bf J}}) = -K_{{\bf m}_{r}} 
              \frac{g_{r}(\bar{{\bf J}};P_{c})}{\bar{\Omega}_{r}}
\end{equation}
Consequently, at $O(\epsilon^{2})$ the effective Hamiltonian 
\begin{equation}
{\cal H}(\bar{{\bf J}},\bar{{\bm \psi}};P_{c}) \approx
{\cal H}_{0}(\bar{{\bf J}};P_{c}) + \epsilon^{2}
{\cal H}_{2}(\bar{{\bf J}},\bar{{\bm \psi}};P_{c})
\end{equation}
containing the
induced resonances is obtained. 

Following the procedure for RAT the
induced resonances are approximated by pendulums. For example in the
case of ${\bf m}_{i3}$ one obtains
\begin{equation}
{\cal H}_{eff}^{(24)} = \frac{(K_{24}-K_{24}^{r})^{2}}{2M_{24}} +
2V_{24} \cos(2\phi_{24})
\label{3dbar}
\end{equation}
with $K_{24} \sim 2I_{4}$ and $2\phi_{4} \sim (\theta_{2}-\theta_{4})$. The
resonance center is $K_{24}^{r}$ and the effective coupling can now be
expressed in terms of the conserved quantities $I_{3},P_{c}$, and 
$P_{24}=I_{2}+I_{4}$. Note that $P_{24}$ is the appropriate polyad for
the induced resonance $\Omega_{2}$:$\Omega_{4}$=$1$:$1$. Thus it is possible
to explicitly identify (cf. Eq.~(\ref{3dbar})) the 
barrier for dynamical tunneling in this three
degrees of freedom case. Observe that the barrier is parametrized by an
exact constant of the motion $P_{c}$ and two approximate constants
of the motion $I_{3}$, and $P_{24}$.
The effective couplings can be translated back to
effective quantum strengths $\lambda_{\bf m} \approx V_{\bf m}/2$ quite easily.
The perturbative analysis, for parameters relevant to Fig.~\ref{fig18},
yields 
\begin{equation}
\lambda_{{\bf m}_{i1}} \ll |\lambda_{{\bf m}_{i2}}|
\approx |\lambda_{{\bf m}_{i3}}| = 0.07 K_{{\bf m}_{2}} 
\label{resstr}
\end{equation}
and thus the
induced resonances are more than an order of magnitude smaller in strength
as compared to the primary resonances. Based on the RAT theory it is
possible to provide an explanation for the drastically different
behaviour of 
state $|a\rangle$ when compared to the other three states as seen in
Fig.~\ref{fig19}. The state $|a\rangle$ is not symmetrically located,
refering to the arguments following Eq.~(\ref{symloc}),
with respect to $|b\rangle$ and thus, combined with the smallness of
$\lambda_{{\bf m}_{i1}}$, does not show significant mixing. 

\begin{figure} [htbp]
\begin{center}
\includegraphics[width=80mm]{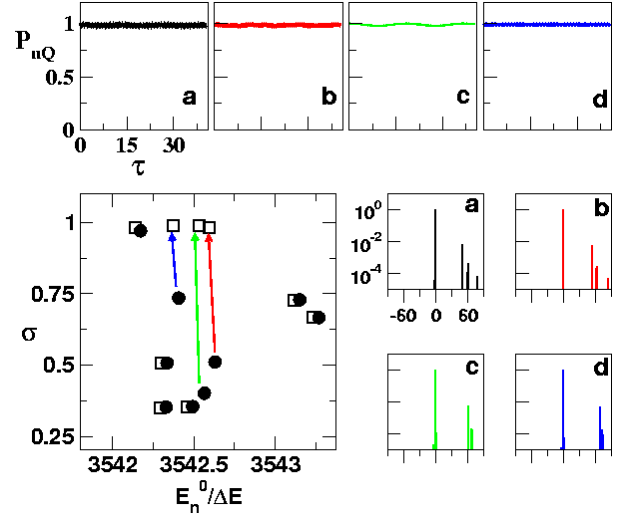}
\caption{(Colour online) The top panel shows the
survival probabilities for the states of
interest using the modified Hamiltonian in Eq.~(\ref{3dhammod}). Note that
the counter resonant terms have essentially shut off the dynamical tunneling
proving the RAT mechanism. The bottom left figure shows the change
in the dilution factors. Solid symbols refer to the $\sigma$ calculated
from the original Hamiltonian in Eq.~(\ref{3dham}) (cf. Fig.~\ref{fig17}C)
and the open squares are the $\sigma$ calculated using Eq.~(\ref{3dhammod}).
Only the states influenced by the induced resonances in Fig.~\ref{fig20}
have their $\sigma \rightarrow 1$ whereas nearby states show very little
change. The four panels on the right show the overlap intensity for
the modified Hamiltonian. Notice the clump of states near $\approx 60
\Delta E$ as in Fig.~\ref{fig19}.}
\label{fig21}
\end{center}
\end{figure}

Given that
the key resonances and their strengths are known (cf. Eq.~(\ref{resstr}))
is it possible to interfere with the mixing between the states?
In other words one is exploring the possibility of controlling the
dynamical tunneling, a quantum phenomenon, by modifying the
local phase space structures.
Consider modifying the quantum Hamiltonian (based on classical information!)
of Eq.~(\ref{3dham}) as follows:
\begin{equation}
H'=H + |\lambda_{{\bf m}_{i2}}| (a_{2}^{\dagger} a_{3} + H.c.)+
|\lambda_{{\bf m}_{i3}}| (a_{2}^{\dagger} a_{4} + H.c.)
\label{3dhammod}
\end{equation}
In the above terms have been added to counter the induced resonances. Due to
the weakness of the induced couplings quantities like mean level spacings,
eigenvalue variations shown in Fig.~\ref{fig19}, and spectral intensities
show very little change as compared to the original system. However 
Fig.~\ref{fig21} shows that the survival probabilities of all the states
exhibit long time trapping. The dilution factors shown in Fig.~\ref{fig21}
also indicate almost no mixing between the near-degenerate states. Thus
dynamical tunneling has been essentially shut down for these states.
The fractionation pattern for the states with the modified Hamiltonian
is quite similar to the ones seen in Fig.~\ref{fig19}. One again observes
a clump of off-resonant states around the same region. Surely the
vibrational superexchange calculation would now predict the absence of
mixing due to subtle cancellations but it is clear that a more transparent
explanation comes from the RAT mechanism.
More significantly Fig.~\ref{fig21} shows that other nearby zeorth-order
states are unaffected by the counter resonant terms. Thus this is an
example of local control of dynamical tunneling or equivalently
IVR. 

\begin{figure} [htbp]
\begin{center}
\includegraphics[width=80mm]{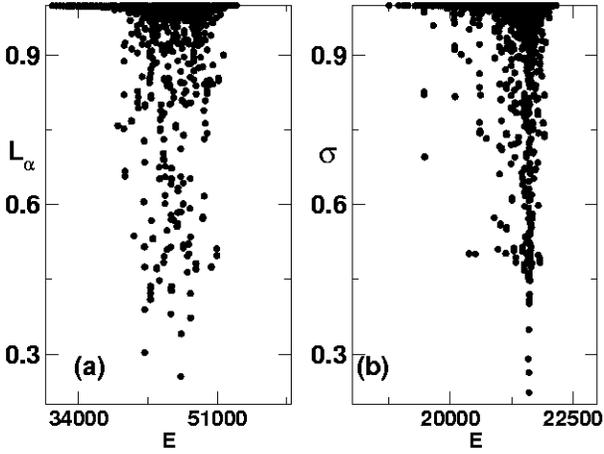}
\caption{Additional examples for state mixings due to
weak couplings. (a) Interpolyad mixing between the eigenstates of the Baggott
Hamiltonian due to the polyad breaking Hamiltonian in Eq.~(\ref{3dh2o})
with $\delta=0.1$ cm$^{-1}$. The inverse participation ratios of the
various eigenstates of $H_{3d}$ in the Baggott eigenbasis are shown
versus the eigenvalues $E$ in cm$^{-1}$.
(b) Dilution factors of the zeroth-order states for a Hamiltonian of
the form as in Eq.~(\ref{3dham}) but with a different $H_{0}$. The polyad
chosen is $P=16$ and the three resonant couplings have 
strengths $0.5 K_{{\bf m}_{1}} = K_{{\bf m}_{2}} = K_{{\bf m}_{3}}
\approx 0.05$ in units of the mean level spacing, 
$\Delta E \approx 1.8$ cm$^{-1}$. The zeroth-order energies $E$ are also in
units of $\Delta E$.}
\label{fig22}
\end{center}
\end{figure}

The model Hamiltonian results in this section correspond to the near-integrable
limit and thus RAT accounts for the state mixings. It would be interesting to
study the mixed phase space limit for three degrees of freedom systems
where classical transport mechanisms can compete as well. 
This however
requires a careful study by varying the effective $\hbar$ of the system
in order to distinguish between the classical and quantum mechanisms.
Note that such zeroth-order state mixing due to weak couplings
occurs in other model systems as well. 
In particular the present analysis and insight suggests that the
mixing of edge and interior zeroth-order states in cyanoacetylene\cite{Hut84}
must be due to the RAT mechanism.
Two more examples in Fig.~\ref{fig22}
highlight the mixing between near-degenerate states due to
very weak couplings. The case
in Fig.~\ref{fig22}a corresponds to the mixings induced by the
weak $3$:$1$ resonance in the case of the model Hamiltonian of 
Eq.~(\ref{3dh2o}) with $\delta=0.1$ cm$^{-1}$. The inverse 
participation ratios of the eigenstates $|\alpha_{3d}\rangle$ of $H_{3d}$  
\begin{equation}
L_{\alpha}^{(3d)} \equiv \sum_{P} |\langle \alpha_{P}|\alpha_{3d}\rangle|^{4}
\end{equation}
are calculated in the Baggott eigenbasis $\{|\alpha_{P}\rangle\}$
spanning several polyads
centered about $P=16$. By construction the participation ratio
of every eigenstate would be close to unity if the small $\delta$
value did not induce any interpolyad mixing. However it is clear from
Fig.~\ref{fig22}a that several eigenstates do get mixed.
The other example shown in Fig.~\ref{fig22}b
pertains to a case where the Hamiltonian is of the same form as 
in Eq.~(\ref{3dham}) but with the parametrs of $H_{0}$ that corresponds
to the CF$_{3}$CHFI molecule\cite{pqsw00}. 
The dilution factors of the various 
zeroth-order states in the polyad $P=16$ with extremely small couplings
exhibit mixing. In both the examples it would be interesting to
ascertain the percentage of states that are mixed due to CAT/RAT. 

\section{Summary}

\begin{figure} [htbp]
\begin{center}
\includegraphics[width=80mm]{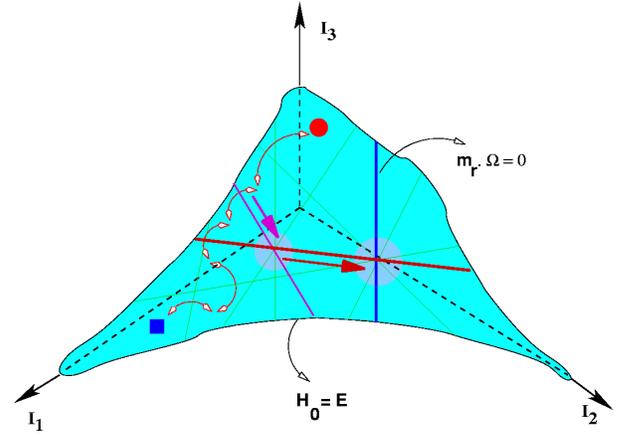}
\caption{(Colour online) A sketch of the resonance web
in three degrees of freedom in
the $(I_{1},I_{2},I_{3})$ action (state) space.
The energy surface $H_{0}=E$ is intersected by
several resonances which form the web and control the IVR out of a state
located on the energy surface.
The strength of the resonances is indicated by the thickness and
shaded regions highlight some of the resonance junctions.
Mixing {\it i.e.,} energy flow between
the state on the top (red circle) and a
far away state (blue square) is due
to RAT, CAT (a possible tunneling sequence is shown by red arrows),
classical across and along (thick arrows) transport.
The competition between the tunneling and classical transport mechanism
depends on the effective $\hbar$ of the system. Barriers to classical
transport do exist and are expected to play
an important role\cite{mde87}
but they have not been indicated in this simple sketch.}
\label{fig23}
\end{center}
\end{figure}

The central message of this review is that {\em both classical and quantum
routes to IVR are controlled by the network of nonlinear resonances}.
A simple sketch of IVR in state space via classical diffusion and
dynamical tunneling is shown in Fig.~\ref{fig23} and
should be contrasted with the sketch shown in Fig.~\ref{fig3}.
The vibrational superexchange picture of Fig.~\ref{fig3} explaining
IVR through the participation of virtual off-resonant states
is equivalent to the picture shown in Fig.~\ref{fig23} which explains
IVR in terms of the Arnol'd web and states on the energy shell.
Furthermore, as sketched in Fig.~\ref{fig23}, a given zeroth-order state
is always close to some nonlinear resonance or the other. Thus it becomes
important to know the dynamically relevant regions of the web. 
Classically it is possible to drift along the resonance lines and come
close to the junction where several resonances meet. At a junction
the actions can move along an entirely different resonance line. In this 
fashion, loosely referred to as Arnol'd diffusion\cite{Arno64,Viva84,Cin02}, 
the classical dynamics
can lead to energy flow between a given point in Fig.~\ref{fig23}
and another distant point. Arnol'd diffusion, however, is exponentially
slow and its existence requires several constraints to be satisfied by
the system\cite{Shur76,Lochasi}. 
Therefore dynamical tunneling between two
far removed points shown in Fig.~\ref{fig23} is expected to be more
probable and faster than the classical Arnol'd diffusion.
On the other hand, for generic Hamiltonian systems, a specific primary
resonance will be intersected by infinitely 
many weaker (higher order than the primary) resonances but finitely
many stronger resonances. The stronger resonances can lead to
a characteristic cross-resonance diffusion near the resonance junction.
The cross-resonance diffusion, in contrast to the Arnol'd diffusion,
takes place on time scales much shorter than exponential\cite{Hallbook}. 
In Fig.~\ref{fig23}
one can imagine a scenario wherein a sequence of dynamical tunneling
events finally lands up close to a resonance junction. Would the 
classical erratic motion near the junction interfere with the dynamical
tunneling process? 
Is it then possible that an edge state might compete with an interior
state due to CAT and/or RAT?
Note that the four zeroth-order states in
Fig.~\ref{fig20} are in fact close to one such resonance junction. 
The classical survival probabilities in Fig.~\ref{fig18} indicate
some transport but further studies are required in order to 
characterize the classical transport.
The issues involved are subtle since
the competition
between the coexisting classical and quantum routes
has hardly been studied up until now and warrants further attention.
Few of the studies that have been performed\cite{lw971,dim02,dim021,Tan02} 
to date suggest that
quantum effects tend to localize the along-resonance diffusion on the web.

Nearly a decade ago Leitner and Wolynes
analysed\cite{lw96} the Hamiltonian 
in Eq.~(\ref{specham}) to 
understand the role of the vibrational superexchange at low energies.
Within the state space approach
it was argued that the superexchange mechanisms contribute significantly
to the energy flow from the edge states to the interior of
the state space. More importantly the superexchange
contributions modify the IVR threshold as quantified
by the transition parameter $T(E)$ defined in Eq.~(\ref{lwqet}). 
From the phase
space perspective as in Fig.~\ref{fig23} the direct and superexchange
processes correspond to classical and quantum dynamical tunneling mechanisms
respectively which are determined by the topology of the web. 
In contrast, note that the present studies on the model Hamiltonian
in Eq.~(\ref{3dham}) actually shows dynamical tunneling at fairly high
energies and between interior states in the state space. 
Further studies on such model systems would lead to better insights
into the nature of $T(E)$.
It would certainly be interesting to see if the criteria $T(E) \approx 1$
of Leitner and Wolynes 
is related to the criteria for
efficient CAT and RAT in the phase space. 
Needless to say, such studies combined with the idea of local
control demonstrated in the pevious section can provide important
insights towards controlling IVR. The effective $\hbar$ is
expected to play a central role in deciding the importance of
various phase space structures. However intutive notions based entirely
on this `geometrical' criteria can be misleading.
For instance, Gong and Brumer 
in their studies\cite{gb05} on the modified kicked rotor
found the quantum dynamics to be strongly influenced by regular islands
in the phase space whose areas were at least an order of magnitude smaller
than the effective Planck constant. Another example comes from the
work by Sirko and Koch\cite{sko02} wherein it was 
observed that the ionization of
bichromatically driven hydrogen atoms is mediated by nonlinear common
resonances\cite{How91}
with phase space areas of the order of the Planck constant.
Further examples can be found in the recent studies\cite{ss07,Kes05}.

Although for large molecules the vibrational superexchange viewpoint
is perhaps the only choice for computations,
for a mechanistic understanding the phase
space perspective is inevitable. Furthermore, as noted earlier, 
there exists ample evidence for the notion
that even in large molecules the actual effective
dimensionality of the IVR manifold in the state space is far less than
the theoretical maximum of $(3N-6)$. Hence detailed classical-quantum
correspondence studies of systems with three or four degrees of freedom 
can provide useful insights into the IVR processes in large molecules.  
At the same time a proper understanding 
of and including the effect of partial barriers
into the theory of dynamical tunneling is necessary for further progress. 
Some hints to the delicate competition between quantum and classical
transport through cantori can be found in the fairly detailed
studies by Maitra and Heller\cite{mh00} 
on the dynamics generated by the whisker map\cite{Chi79}.
Such studies are required to gain insights into the
method of controlling quantum dynamics via suitable modifications of
the local phase space structures\cite{cbcflvpfg04,hcu06}.  
Extending the local phase space control idea to systems with higher
degrees of freedom poses a difficult challenge. 
If the sketch in Fig.~\ref{fig23} is reasonable then it suggests
that the real origins of the hierarchical nature of IVR, due to
both quantum and classical routes, is intricately tied to the geometry
of the Arnol'd web. Significant efforts are
needed to characterize the resonance
web dynamically in order to have a clear understanding of how the specific
features of the web influence the IVR. 
Recent works\cite{Kes051,sblkt06,sk06,bhc05,wbw04,sclu04,acp130}
are beginning to focus on extracting such details in
systems with three or more degrees of freedom which promises to
provide valuable insights into the classical and quantum mechanisms
of IVR in polyatomic molecules. 
Perhaps such an understanding
would allow one to locally perturb specific regions of the resonance web and
hence ultimately lead to control over the classical and quantum dynamics.

\section{Acknowledgements}

It is a
pleasure to acknowledge several useful
discussions with Arul Lakshminarayan and Peter Schlagheck on the topic
of dynamical tunneling.
I am grateful to Prof. Steve Wiggins
for his hospitality at Bristol where parts of this review were written
in addition to discussions on the wavelet technique. 
Financial support for the author's research reported here
came from the Department of Science and Technology and
the Council for Scientific and Industrial Research, India.


\begin{thebibliography}{99}
\bibitem{psprivate}{Recent work from the group of P. Schlagheck has
managed to include the cantorus structure into the framework of RAT
and substantial improvements are seen between semiclassical and
quantum results for the splittings in the kicked Harper model. (Private
communication).}
\bibitem{typojcp05}{We note here a typographical error in Eq.(19a) of
ref.~\onlinecite{Kes05} involving the expression for $g_{\rm ind}$. The
corrected version appears in this review as Eq.~(\ref{pendparams}).}
\end{thebibliography}
\end{document}